\documentclass{aa}

\usepackage[varg]{txfonts}

\usepackage{hyperref}
\usepackage{graphicx}
\usepackage{grffile} 
\usepackage[dvipsnames]{xcolor}
\usepackage{amsmath}
\usepackage{amssymb}
\usepackage{commath}
\usepackage{enumitem}
\usepackage{upgreek}
\usepackage{breakcites}
\usepackage[utf8]{inputenc}

\usepackage[capitalise]{cleveref} 
\crefname{figure}{Fig.}{Figs.}
\crefname{section}{Sect.}{Sects.}

\DeclareRobustCommand{\VAN}[3]{#2}
\let\VANthebibliography\thebibliography
\def\thebibliography{\DeclareRobustCommand{\VAN}[3]{##3}\VANthebibliography}

\bibpunct[]{(}{)}{;}{a}{}{,} 
\defcitealias{2012ApJ...746..125H}{HM12}
\defcitealias{2019MNRAS.485...47P}{P19}

\newcommand{\program}{\textsc}
\newcommand{\ssim}{\sim \!}
\newcommand{\lymana}{{Lyman-\ensuremath{\upalpha}}}
\newcommand{\lymanatext}{Lyman-α}
\newcommand{\lya}{{Ly\ensuremath{\upalpha}}}
\newcommand{\lyatext}{Lyα}
\newcommand{\Angstrom}{\ensuremath{\AA}}

\hypersetup{
    unicode=true,
    final=true,
    plainpages=false,
    pdfstartview=FitV,
    pdftoolbar=false,
    pdfmenubar=true,
    bookmarksopen=true,
    bookmarksnumbered=true,
    breaklinks=true,
    linktocpage,
    colorlinks=true,
    linkcolor=blue,
    urlcolor=blue,
    citecolor=blue,
    anchorcolor=green
}
\AtBeginDocument{
  \hypersetup{
    pdftitle={Prospects for Observing the low-density Cosmic Web in \texorpdfstring{\lymana}{\lymanatext} Emission},
    pdfauthor={J. Witstok et al.},
    pdfsubject={A simulation study of the brightness of the intergalactic medium in \texorpdfstring{\lymana}{\lymanatext}},
    pdfkeywords={{Intergalactic medium} -- {Large-scale structure of Universe} -- {Diffuse radiation} -- {Cosmology: theory} -- {Methods: numerical}}
  }
}

\begin{document}

\title{Prospects for Observing the low-density Cosmic Web \\ in \texorpdfstring{\lymana}{\lymanatext} Emission}
\titlerunning{Prospects for Observing the low-density Cosmic Web in \texorpdfstring{\lymana}{\lymanatext} Emission}
\subtitle{}

\author{
    Joris Witstok\thanks{E-mail: \href{mailto:jnw30@cam.ac.uk}{jnw30@cam.ac.uk}}\inst{\ref{inst:IoA}, \ref{inst:Kavli}, \ref{inst:Cav}}
    \and Ewald Puchwein\inst{\ref{inst:IoA}, \ref{inst:Kavli}, \ref{inst:IfA}}
    \and Girish Kulkarni\inst{\ref{inst:IoA}, \ref{inst:Kavli}, \ref{inst:DTP}}
    \and Renske Smit\inst{\ref{inst:Kavli}, \ref{inst:Cav}, \ref{inst:ARI}}
    \and Martin G. Haehnelt\inst{\ref{inst:IoA}, \ref{inst:Kavli}}
}
\authorrunning{J. Witstok et al.}

\institute{
    Institute of Astronomy, University of Cambridge, Madingley Road, Cambridge CB3 0HA, UK\label{inst:IoA}
    \and Kavli Institute for Cosmology, University of Cambridge, Madingley Road, Cambridge CB3 0HA, UK\label{inst:Kavli}
    \and Cavendish Laboratory, University of Cambridge, 19 JJ Thomson Avenue, Cambridge CB3 0HE, UK\label{inst:Cav}
    \and Leibniz Institute for Astrophysics, An der Sternwarte 16, 14482 Potsdam, Germany\label{inst:IfA}
    \and Tata Institute of Fundamental Research, Homi Bhabha Road, Mumbai 400005, India\label{inst:DTP}
    \and Astrophysics Research Institute, Liverpool John Moores University, 146 Brownlow Hill, Liverpool L3 5RF, UK\label{inst:ARI}
}


\abstract{
    Mapping the intergalactic medium (IGM) in \lymana\ emission would yield unprecedented tomographic information on the large-scale distribution of baryons and potentially provide new constraints on the UV background and various feedback processes relevant to galaxy formation. Here, we use a cosmological hydrodynamical simulation to examine the \lymana\ emission of the IGM due to collisional excitations and recombinations in the presence of a UV background. We focus on gas in large-scale-structure filaments in which \lymana\ radiative transfer effects are expected to be moderate. At low density the emission is primarily due to fluorescent re-emission of the ionising UV background due to recombinations, while collisional excitations dominate at higher densities. We discuss prospects of current and future observational facilities to detect this emission and find that the emission of filaments of the cosmic web will typically be dominated by the halos and galaxies embedded in them, rather than by the lower density filament gas outside halos. Detecting filament gas directly would require a very long exposure with a MUSE-like instrument on the ELT. Our most robust predictions that act as lower limits indicate this would be slightly less challenging at lower redshifts ($z \lesssim 4$). We also find that there is a large amount of variance between fields in our mock observations. High-redshift protoclusters appear to be the most promising environment to observe the filamentary IGM in \lymana\ emission.
}


\maketitle

\section{Introduction}
\label{sec:Introduction}

As the reservoir of the majority of baryons in the Universe, the intergalactic medium (IGM) presents an invaluable means to understanding the evolution of cosmic structure \citep{2009RvMP...81.1405M}. The IGM has been detected in absorption at a wide range of overdensities out to redshift $z \sim 6$ by using \ion{H}{I} \lymana\ (\lya) absorption lines in the spectra of background quasars. Successively larger numbers of quasars have been targeted for this purpose, resulting in a large data set of \lya\ absorption measurements of the IGM. Before Reionization is completed, understanding the physical state of the IGM is complicated by the rather uncertain details of the emergence of the first stars, black holes and galaxies during the epoch of Reionization, but the post-Reionization ($z \lesssim 5.5$) IGM should be well-described by cosmological hydrodynamical simulations \citep{1994ApJ...437L...9C, 1996ApJ...457L..51H, 1999elss.conf..346W, 2017ApJ...837..106O, 2019MNRAS.486.4075O, 2015MNRAS.446.3697L, 2017MNRAS.464..897B}. In these simulations, the observed properties of the IGM are reproduced by a fluctuating gas density distribution tracing the cosmic structure formation process. The gas is thereby in ionisation equilibrium with a uniform UV background (UVB) created by galaxies and active galactic nuclei (AGN). This has led to constraints on the ionisation and thermal state of the IGM out to $z \sim 6$ \citep{1997ApJ...489....7R, 1999ApJ...511..521D, 2000MNRAS.318..817S, 2003MNRAS.342.1205M, 2008ApJ...688...85F, 2011MNRAS.410.1096B, 2012MNRAS.419.2880B, 2013MNRAS.436.1023B, 2017PhLB..773..258G, 2019ApJ...872...13W, 2019MNRAS.486..769K} derived from \lya\ absorption observations.

In contrast, \lya\ emission from the IGM has received relatively little attention, despite a history of just over half a century of theoretically predicted prospects \citep{1967ApJ...147..868P, 1987MNRAS.225P...1H, 1996ApJ...468..462G, 2001ApJ...562..605F, 2003ApJ...599L...1F, 2005ApJ...622....7F, 2005ApJ...628...61C, 2010ApJ...708.1048K, 2010ApJ...725..633F, 2012MNRAS.423..344R, 2013ApJ...763..132S, 2016MNRAS.462.1961S, 2017ApJ...848...52H, 2019MNRAS.489.2417A, 2020MNRAS.494.5439E}. Observing intergalactic \lya\ emission instead of absorption has distinct advantages. Unlike absorption, the \lya\ emission is directly sensitive to the recombination and collisional physics of the neutral as well as the ionised hydrogen content of the IGM and the circumgalactic medium (CGM) that feeds the formation and evolution of galaxies. Second, observations of the \lya\ emission allow one to homogeneously probe three-dimensional volumes. Although three-dimensional \lya-forest studies have now become possible due to the high number density of observed bright quasars \citep[see e.g.][]{2014MNRAS.440.2599C}, the number of such quasars drops rapidly towards high redshifts \citep{2019MNRAS.488.1035K}. Third, observations of \lya\ emission can potentially provide independent constraints on the IGM temperature and photoionisation rate, particularly at densities higher than those probed by the \lya\ forest ($\Delta \gtrsim 10$).

Using narrowband imaging as well as integral field unit imaging, emission in \lya\ from the CGM/IGM has now been observed as ``giant \lya\ nebulae'' in the proximity ($\ssim 100\,\mathrm{kpc}$) of radio-loud as well as radio-quiet quasars \citep{1985ApJ...299L...1D, 1991ApJ...368...28H, 1991ApJ...370...78H, 1990ApJ...365..487M, 2007A&A...461..823V, 2007MNRAS.378..416V, 2008ApJ...672...48C, 2008MNRAS.390.1505H, 2008ApJ...681..856R, 2009A&A...495..471S, 2011MNRAS.418.1115R, 2012MNRAS.425.1992C, 2013MNRAS.429..429R, 2014Natur.506...63C, 2014ApJ...786..106M, 2014MNRAS.443.3795R, 2015Sci...348..779H, 2016ApJ...829....3A, 2016ApJ...831...39B, 2016MNRAS.462.1978F, 2017ASSL..430..195C}. The circumgalactic hydrogen is strongly affected by ionising radiation from these quasars. Observations suggest that the \lya\ emission is mostly recombination radiation, and that dense ($n > 1 \, \mathrm{cm^{-3}}$), ionised, and relatively cold ($T \sim 10^4 \, \mathrm{K}$) pockets of gas should surround massive galaxies \citep{2017ASSL..430..195C}.

\lya\ emission can also result from fluorescent re-emission of the ionising UV background radiation. In the last two decades, significant progress has been made with detecting extended \lya\ emission around galaxies \citep{1996ApJ...457..490F, 1999MNRAS.305..849F, 1999AJ....118.2547K, 2000ApJ...532..170S, 2004AJ....128.2073H, 2008ApJ...681..856R, 2011ApJ...736..160S, 2012MNRAS.425..878M, 2013ApJ...762...38P, 2014MNRAS.442..110M, 2016ApJ...832...37G, 2016A&A...587A..98W, 2017ApJ...837...71C, 2017A&A...608A...8L, 2017MNRAS.465.3803V, 2018ApJ...856...72O, 2018Natur.562..229W, 2019MNRAS.482.3162A}. Using deep ($\ssim 30 \, \mathrm{h}$ exposure time) VLT/MUSE observations of the \textit{Hubble} Deep Field South (HDFS) and \textit{Hubble} Ultra-Deep Field (HUDF) reported in \citet{2015A&A...575A..75B, 2017A&A...608A...1B}, the sensitivity of median-stacked radial profiles of \lya\ emission currently reaches a surface brightness ($\text{SB}$) of $\text{SB} \sim 4 \cdot 10^{-21} \, \mathrm{erg \, s^{-1} \, cm^{-2} \, arcsec^{-2}}$ \citep{2018Natur.562..229W}. This faint signal from \lya\ halos can be traced out to projected (physical) galactic radii of $\ssim 60 \, \mathrm{kpc}$ \citep{2018Natur.562..229W}. Even deeper data sets, like the MUSE Ultra Deep Field \citep[MUDF, described in][]{2019MNRAS.485L..62L} and the MUSE Extremely Deep Field \citep[MXDF, see][]{2021arXiv210205516B} are beginning to be explored. Both will reach a depth on the order of $\ssim 100 \, \mathrm{h}$ (i.e. reaching a sensitivity on the order of a few times $10^{-20} \, \mathrm{erg \, s^{-1} \, cm^{-2} \, arcsec^{-2}}$). The \lya\ emission coming from the intergalactic gas between galaxies is just beginning to be probed and will be the focus of this work.

So far, it has proven very difficult to map the spatial distribution of the IGM beyond the CGM and study its global properties by directly observing the IGM in emission, rather than absorption. In fact, this has so far only been achieved in special cases, e.g., in the vicinity of AGN \citep[e.g.][]{2014Natur.506...63C, 2014ApJ...786..106M, 2015Sci...348..779H, 2016ApJ...831...39B, 2019Sci...366...97U}, by applying statistical image processing techniques \citep{2018MNRAS.475.3854G}\footnote{Here, the circumgalactic medium only showed a preferential direction of extension towards neighbouring galaxies -- no significant signal of filamentary structure in the IGM was found.}, by cross-correlating \lya\ emitters (LAEs) and \lya\ intensity mapping \citep{2019arXiv190600173K}, by observing the thermal Sunyaev-Zel’dovich effect \citep[e.g.][]{2019A&A...624A..48D, 2019MNRAS.483..223T}, or by detection of warm-hot gas in X-ray emission \citep[e.g.][]{1999A&A...341...23K, 2015Natur.528..105E}.

Building on the work of previous studies \citep[such as those by][]{1996ApJ...468..462G, 2003ApJ...599L...1F, 2005ApJ...628...61C, 2013ApJ...763..132S, 2016MNRAS.462.1961S}, this work investigates the possibility of such observations, exploring a simulation run based on the Sherwood simulation project \citep{2017MNRAS.464..897B} incorporating an on-the-fly self-shielding model to predict the properties of \lya\ emission from the cosmic web. The simulation is aimed at accurately modelling the IGM, and employs a modified version of the uniform metagalactic UV background model by \citet[; \citetalias{2012ApJ...746..125H} hereafter]{2012ApJ...746..125H} calibrated to match observations of the \lya\ forest. The large volume and high dynamic range of the simulation allows us to probe the physical environment of the IGM with well resolved under- and overdense regions. Moreover, we can study the prospects of an array of current and future observational facilities aiming to detect this emission. We focus on a future reincarnation of VLT/MUSE on next-generation observatories like the Extremely Large Telescope (ELT) for a more detailed sensitivity analysis.

We describe the simulations used in this work in \cref{sec:Methodology}, together with our model for \lya\ production in the IGM. \cref{sec:Results} presents our results and a discussion of the detection prospects. We summarise our conclusions in \cref{sec:Conclusions}. Throughout this work, we adopt the cosmological parameters $\Omega_\text{m} = 0.308$, $\Omega_\Lambda = 0.692$, $\Omega_\text{b} = 0.0482$, and $h=0.678$ (so $H_0 = 67.8 \, \mathrm{km \, s^{-1} \, Mpc^{-1}}$), taken from the best fitting $\Lambda$CDM model for the combined \textit{Planck+WP+highL+BAO} measurements \citep{2014A&A...571A..16P}. The helium fraction is assumed to be $f_\text{He} = 0.24$.

\section{Methodology}
\label{sec:Methodology}

\lya\ emission from the moderately dense IGM is produced via recombinations and collisional excitations. Recombination is the process where a free electron is captured by an ion, in this case \ion{H}{II}. \lya\ is emitted provided the recombination leaves hydrogen in an excited state, and the last step of the resulting series of energy transition(s) is from energy level $n=2$ to $n=1$. Collisional excitation is the effect in which neutral hydrogen (\ion{H}{I}) is excited through a collision with an electron, which can subsequently lead to the emission of \lya\ in the same way as with recombinations. We use a hydrodynamical simulation calibrated to UV background constraints from the \lya\ forest along with an on-the-fly self-shielding prescription to model these processes.

In the analysis, we focus on low-density gas (below the critical density where self-shielding becomes a dominant process; at $z=4.8$, this corresponds to an overdensity $\Delta \equiv \rho/\bar{\rho} \simeq 100$, see \cref{sssec:Density limits}) as we are primarily interested in detecting emission from the cosmic web. Furthermore, modelling of all relevant feedback and radiative transfer effects becomes increasingly challenging at higher densities.

\begin{figure}
    \centering
    \includegraphics[width=\columnwidth]{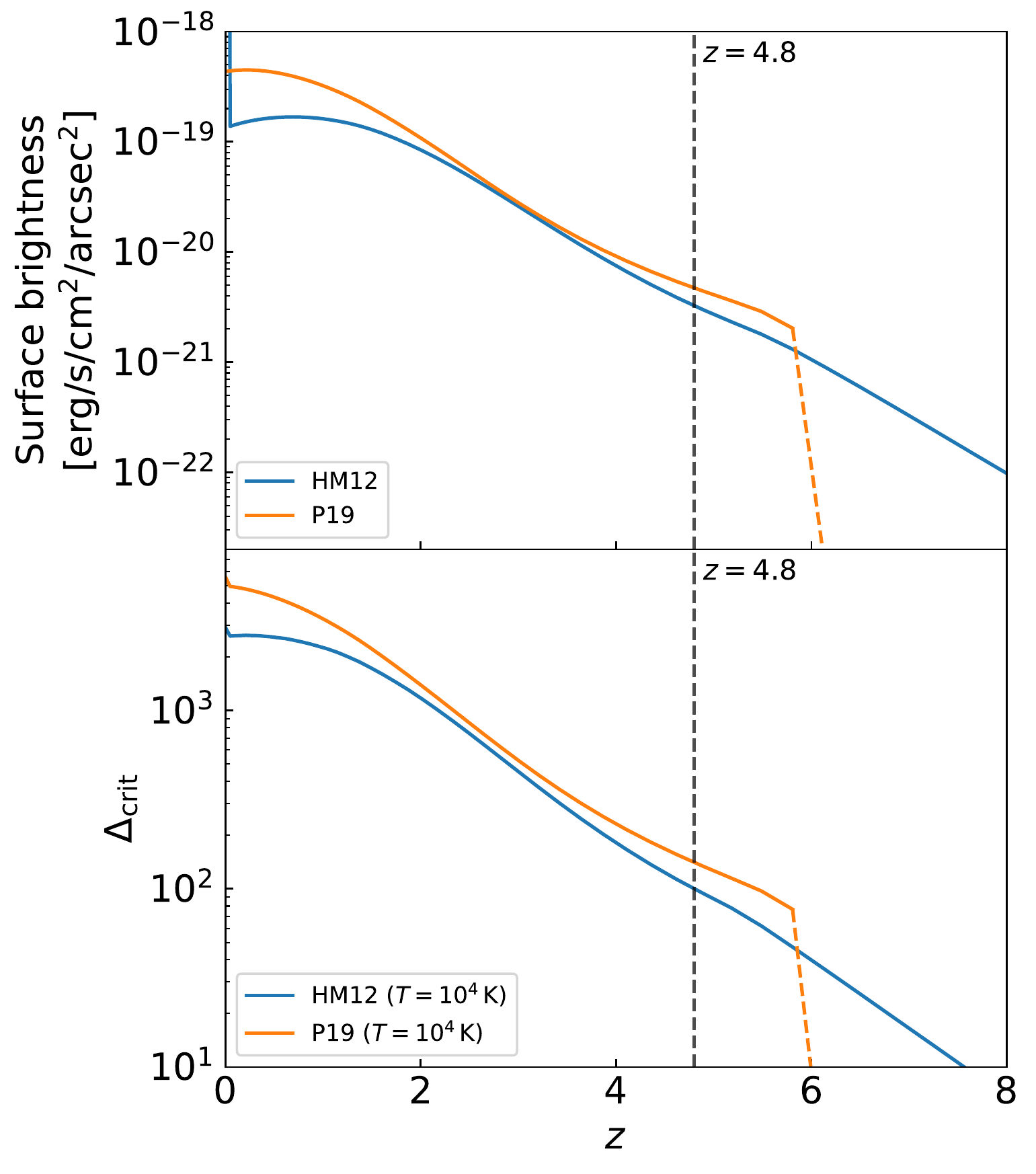}
    \caption[Self-shielding critical overdensity and \lya\ SB mirror limit for recombination emission]
    {Limiting surface brightness of \lya\ in the ``mirror'' assumption (top panel), where $65\%$ of ionising photons in the UV background are reprocessed into \lya\ photons (see text for details), and self-shielding critical density contrast $\Delta_\text{crit}$ (bottom panel). The two different lines correspond to UV backgrounds of \citet[, \citetalias{2012ApJ...746..125H}]{2012ApJ...746..125H} and \citet[, \citetalias{2019MNRAS.485...47P}]{2019MNRAS.485...47P}. Above $z>6$, where the line is dashed, the \citetalias{2019MNRAS.485...47P} limits are not representative for ionised bubbles during patchy Reionization, as the impact of neutral regions on the effective opacity to hydrogen ionising photons is included in the modelling \citepalias[see][]{2019MNRAS.485...47P} and hence a neutral hydrogen-weighted average over both neutral and ionised regions is computed in that model. A redshift of $z=4.8$ is highlighted by the dashed line.}
    \label{fig:UVB_limits}
\end{figure}

\begin{figure}
    \centering
    \includegraphics[width=\columnwidth]{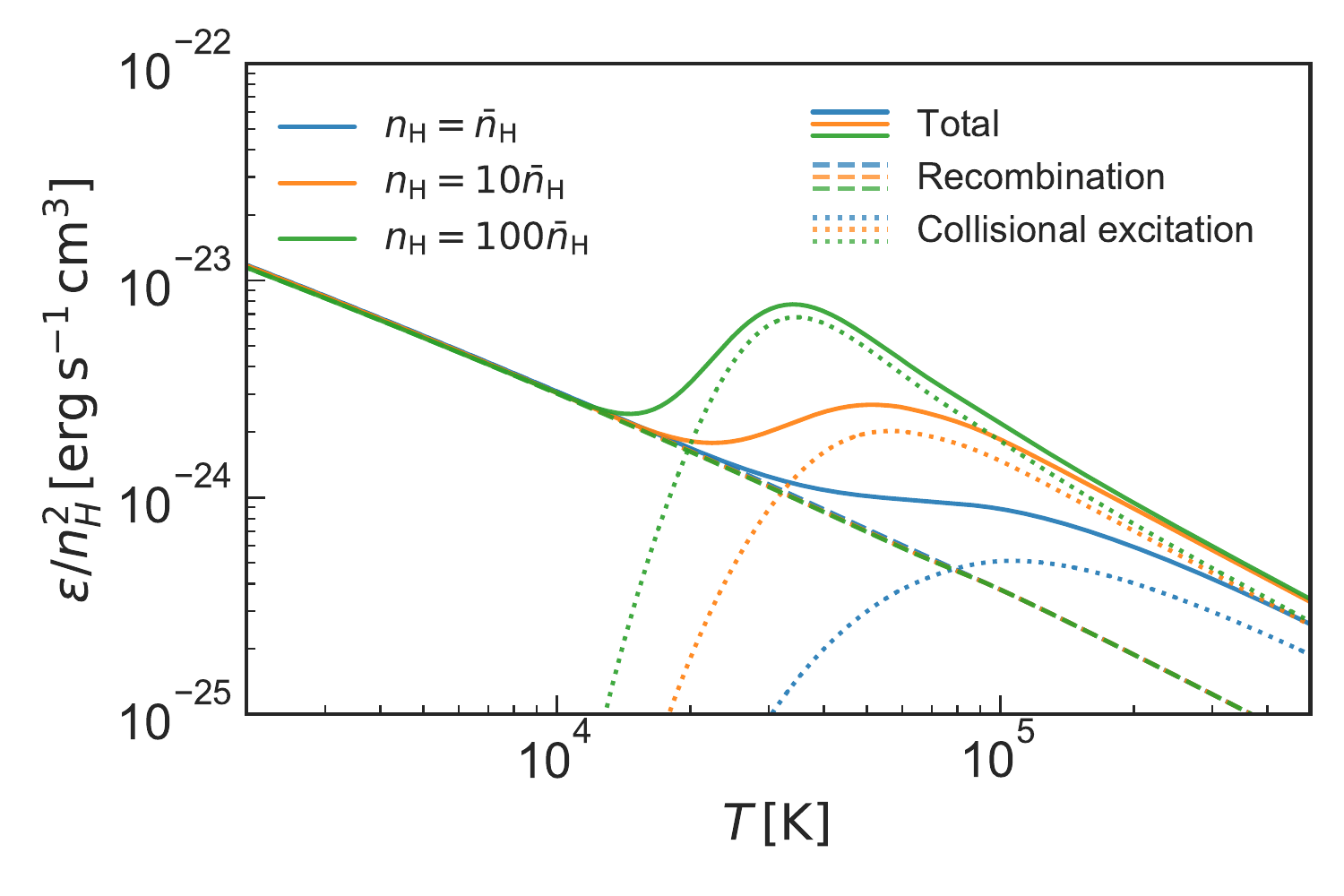}
    \caption[Theoretical \lya\ emissivity as a function of temperature]
    {Normalised emissivity (note the units are $\mathrm{erg \, s^{-1} \, cm^3}$) of the \lya\ line in a cloud of primordial gas at $z=4.8$, due to both recombination and collisional excitation processes, as a function of temperature. There are three values of density, corresponding to overdensities of $1$, $10$, and $100$, respectively (mean cosmological hydrogen density corresponds to $\bar{n}_\text{H} = 3.69 \cdot 10^{-5} \, \mathrm{cm^{-3}}$ at this redshift). The dashed and dotted lines show the contribution from just recombination and collisional excitation, respectively.}
    \label{fig:Emissivity theoretical}
\end{figure}

\subsection{\texorpdfstring{\lya}{\lyatext} emission through recombination}
\label{ssec:Lya recombination emission}

\subsubsection{Emissivity}
\label{sssec:Recombination emissivity}

The underlying equation governing \lya\ emission due to recombination in a gas containing hydrogen is given by \citep[see e.g.][]{2014PASA...31...40D, 2016MNRAS.462.1961S}
\begin{equation}
	\label{eq:Recombination emissivity}
	\epsilon_\text{rec}(T) = f_\text{rec, A/B} (T) \, n_\text{e} \, n_\text{HII} \, \alpha_\text{A/B}(T) \, E_\text{\lya},
\end{equation}
where $\epsilon_\text{rec}$ is the \lya\ luminosity density (in units of $\mathrm{erg \, s^{-1} \, cm^{-3}}$) as a function of the temperature $T$ of the gas. Here, $f_\text{rec, A/B}$ is the fraction of case-A or case-B recombinations that ultimately result in the emission of a \lya\ photon, and the free electron and \ion{H}{II} number densities are denoted by $n_\text{e}$ and $n_\text{HII}$, respectively. Case-A and case-B refer to the way in which recombination occurs: case-A is where all possible recombinations of \ion{H}{II} and a free electron are considered -- this includes any recombination event that take the resulting neutral hydrogen directly to the ground state ($n=1$). In case-B, only recombinations resulting in hydrogen in an excited state are considered. The recombination coefficient, given in unit volume per unit time ($\mathrm{cm^{3} \, s^{-1}}$) for case-A or -B recombination, is denoted by $\alpha_\text{A/B}$, and $E_\text{\lya}$ is the energy of a \lya\ photon.

Since direct recombinations into the ground state do not result in \lya\ emission, an appropriately lower fraction that results in \lya\ emission, $f_\text{rec, A} < f_\text{rec, B}$, has to be used if $\alpha_\text{A}$ rather than $\alpha_\text{B}$ is adopted as the recombination coefficient. The luminosity densities obtained for case-A and -B are then equivalent, except for minor differences due to different fitting functions for the coefficients. We will choose to fix our calculations to use case-B coefficients. We model $f_\text{rec, B}$ using the relations given by \citet{2008ApJ...672...48C} and \citet{2014PASA...31...40D}, whose fitting formulae are presented in \cref{ap:Model parameters}; e.g., at $T = 10\,000 \, \mathrm{K}$, this fraction is $\ssim 0.68$. We elected to use case-B because the model for $f_\text{rec, A} (T)$ from \citet{2014PASA...31...40D} is only valid up to $\ssim 10^{6.5} \, \mathrm{K}$, whereas as we will see below gas temperatures in our simulations range up to $\ssim 10^{7} \, \mathrm{K}$ (\cref{ssec:Surface brightness maps}). For the recombination coefficient, $\alpha_\text{B}(T)$, we adopt the fitting function given in \citet{2011piim.book.....D}. The precise expressions can also be found in \cref{ap:Model parameters}.

\subsubsection{Mirror limit}
\label{sssec:Mirror limit}

In the absence of local ionising UV sources and significant collisional ionisation, the recombination contribution to \lya\ emission should not exceed the surface brightness expected from fully absorbing the external UVB at the boundaries of self-shielded regions and fluorescently re-emitting a corresponding number of \lya\ photons, hence ``mirroring'' the external UVB. In calculating the recombination contribution to \lya\ emission, we will, unless mentioned otherwise, employ this mirror assumption as an upper limit. More precisely, we place an upper $\text{SB}$ limit at the value expected when $65\%$ of the ionising UV background is reprocessed as \lya\ photons \citep[e.g.][]{1996ApJ...468..462G, 2005ApJ...628...61C}, equal to $\text{SB} \simeq 3.29 \cdot 10^{-21} \, \mathrm{erg \, s^{-1} \, cm^{-2} \, arcsec^{-2}}$ for a \citetalias{2012ApJ...746..125H} UV background at $z=4.8$. \cref{fig:UVB_limits} shows the mirror limit for two different UV backgrounds, from \citetalias{2012ApJ...746..125H} and \citet[; \citetalias{2019MNRAS.485...47P} hereafter]{2019MNRAS.485...47P}.

In reality, local ionising sources can boost the recombination emission above the mirror limit. Predicting this reliably is, however, extremely challenging, as it involves modelling the ionizing source populations in galaxies and the escape of ionizing radiation from them in full detail. Our recombination contribution to \lya\ emission computed assuming the mirror limit should hence be considered only as a robust lower limit.

\subsection{\texorpdfstring{\lya}{\lyatext} emission through collisional excitation}
\label{ssec:Lya collisional excitation emission}

\subsubsection{Emissivity}
\label{sssec:Collisional excitation emissivity}

For collisional excitation, the \lya\ luminosity density has a similar form \citep{1990MNRAS.242..692S, 1991ApJ...380..302S, 2014PASA...31...40D, 2016MNRAS.462.1961S}, given by
\begin{equation}
	\label{eq:Collisional excitation emissivity}
	\epsilon_\text{exc}(T) = \gamma_\text{1s2p} (T) \, n_\text{e} \, n_\text{HI} \, E_\text{\lya},
\end{equation}
where $n_\text{HI}$ denotes the number density of neutral hydrogen. We use the fitting functions for the collisional excitation coefficient $\gamma_\text{1s2p}$ given by \citet{1990MNRAS.242..692S} and \citet{1991ApJ...380..302S}. These fitting functions are valid in the temperature range $2 \cdot 10^3 \, \mathrm{K} \leq T \leq 1 \cdot 10^8 \, \mathrm{K}$ (cf. \cref{ap:Model parameters}). The rates are not identical to those applied in the cosmological hydrodynamical simulation (see \cref{ssec:SherwoodSuite}) as these are only given as an ensemble rather than for the specific $2p \rightarrow 1s$ transition in which \lya\ is emitted, but in the relevant temperature regime deviate so little that gas cooling equilibrium would not be appreciably violated.

\subsubsection{Density limits}
\label{sssec:Density limits}

When computing the \lya\ luminosity due to collisional excitation, we will only consider gas well below the critical self-shielding density, derived for the appropriate UV background (the \citetalias{2012ApJ...746..125H} UVB, unless mentioned otherwise). We make use of the critical self-shielding hydrogen number density at $T = 10^4 \, \mathrm{K}$ given in Eq. (13) in \citet{2013MNRAS.430.2427R} for this purpose (shown in the bottom panel of \cref{fig:UVB_limits} as a density contrast), but since this is based on the column density distribution of neutral hydrogen and for the purpose of absorption instead of emission processes, we choose a conservative default density threshold at half this value.

As can be seen in \cref{fig:UVB_limits}, the critical self-shielding overdensity is $\Delta_\text{crit} \simeq 100$ at $z=4.8$. Note that the density \textit{contrast}, $\Delta_\text{crit}$, decreases towards higher redshift, i.e. gas starts to be affected by self-shielding at a lower overdensity at higher redshift. By focusing on gas with densities below this critical threshold, we additionally ensure at this redshift we do not enter the realm of gas densities strongly affected by the detailed baryonic physics of galaxy formation, like feedback processes. For this reason, most of the results presented in this work are chosen to be at $z=4.8$ (and are again a robust lower limit).

\subsection{Emissivity}
\label{ssec:Emissivity}

\cref{fig:Emissivity theoretical} shows the \lya\ luminosity density at $z=4.8$ as a function of gas temperature for a gas of primordial composition at three different overdensities of $1$, $10$, and $100$ (mean cosmological hydrogen density corresponds to $\bar{n}_\text{H} = 3.69 \cdot 10^{-5} \, \mathrm{cm^{-3}}$ at this redshift). In order to derive the corresponding neutral hydrogen densities, we assume that hydrogen is in ionisation equilibrium with the \citetalias{2012ApJ...746..125H} UV background at $z=4.8$. \cref{fig:Emissivity theoretical} also shows the recombination and collisional excitation components of the total \lya\ emission. We find that collisional excitation dominates at high temperatures ($T \gtrsim 2 \times 10^4 \, \mathrm{K}$).

\subsection{The cosmological hydrodynamical simulation}
\label{ssec:SherwoodSuite}

In order to estimate the cosmological \lya\ signal with the theoretical framework above, we make use of a simulation that builds upon the Sherwood simulation project \citep{2017MNRAS.464..897B}. The simulation has been performed with the energy- and entropy-conserving TreePM smoothed particle hydrodynamics (SPH) code \program{p-gadget-3}, which is an updated version of the publicly available \program{gadget-2} code \citep{2001NewA....6...79S, 2005MNRAS.364.1105S}. In this work, we use the same volume as in the 40--1024 simulation of the Sherwood suite. A periodic, cubic volume $40 \, h^{-1} \, \mathrm{cMpc}$ long has been simulated, employing a softening length of $l_\mathrm{soft}=1.56 \, h^{-1} \, \mathrm{ckpc}$, and $1024^3$ dark matter and gas particles. Initial conditions were set up at redshift $z=99$ and the simulation was evolved down to $z=2$. In order to speed up the simulation, star formation was simplified by using the implementation of \citet{2004MNRAS.354..684V} in \program{p-gadget-3}, which converts gas particles with temperature less than $10^5 \, \mathrm{K}$ and density of more than a thousand times the mean baryon density to collisionless stars. This approximation is appropriate for this work as we are not considering the \lya\ emission from the interstellar medium of galaxies, where a complex set of \lya\ radiative transfer processes need to be accounted. The ionisation and thermal state of the gas in the simulation is derived by solving for the ionisation fractions under the assumption of an equilibrium with the metagalactic UV background modelled according to \citetalias{2012ApJ...746..125H}. A small modification to this UV background is applied at $z<3.4$ \citep[see][]{2017MNRAS.464..897B} to result in IGM temperatures that agree with measurements by \citet{2011MNRAS.410.1096B}. We also account for self-shielding of dense gas with an on-the-fly self-shielding prescription based on \citet{2013MNRAS.430.2427R}. For each SPH particle and each time step, our modified \program{p-gadget-3} version computes a suppression factor for the UV background due to self-shielding that is based on the local gas density and uses the parameters given in the first line of table~A1 of \citet{2013MNRAS.430.2427R}. This factor is applied to photoionisation and heating rates before they are used in the chemistry/cooling solver. The solver follows photoionisation, collisional ionisation, recombination and photoheating for gas of a primordial composition of hydrogen and helium, as well as further radiative cooling processes such as collisional excitation, Bremsstrahlung (see \citealt{1996ApJS..105...19K} for the relevant equations), and inverse Compton cooling off the Cosmic Microwave Background \citep{1986ApJ...301..522I}. Metal enrichment and its effect on cooling rates are ignored. We identify dark matter halos in the output snapshots using a friends-of-friends algorithm.

\subsubsection{Narrowband images}
\label{sssec:Narrowband}

When calculating the surface brightness, we construct mock narrowband images of the simulations -- an image that replicates the result of the process of capturing a narrowband image with a telescope -- by taking a thin slice of the simulation in a direction parallel to a face of the simulation box, and converting the emissivity in the simulation to arrive at a surface brightness map, as will be discussed in more detail below. The slice thickness corresponds to an observed wavelength width $\Delta\lambda_\text{obs}$ of the narrowband. Its redshift range is given by
\begin{equation}
    \Delta z = \frac{\Delta \lambda_\text{obs}}{\lambda_\text{\lya}},
\end{equation}
which corresponds to a comoving distance
\begin{equation}
    \Delta d = \frac{c}{H_0} \int_{z}^{z+\Delta z} \frac{1}{\sqrt{\Omega_\text{m} \left( 1 + z' \right)^3 + \Omega_\Lambda}} \dif z'.
\end{equation}

As a reference value for the observed narrowband width, $\Delta \lambda_\text{obs}$, we will use $\Delta \lambda_\text{obs} = 8.75 \, \text{\AA}$ (corresponding to $7$ spectral pixels of the VLT/MUSE instrument; $\Delta \lambda_\text{obs} = 8.75 \, \text{\AA}$ is the median value of narrowband widths in the study by \citealt{2016A&A...587A..98W}; \cref{ssec:Observing facilities} will discuss narrowband imaging in more detail). At a redshift of $z=4.8$, this results in a comoving line-of-sight distance of $\ssim 2.7 \, h^{-1} \, \mathrm{cMpc}$ (see \cref{sssec:Density limits} for an elaboration on the choice of this particular redshift), corresponding to only a small fraction of the total size of the simulation volume. We will discuss the effect of varying the narrowband width on the detectability of \lya\ further in \cref{sssec:Cosmic variance and narrowband widths}.

Using the temperature, density, and ionisation fraction, an emissivity for each individual simulation particle within the narrowband slice can be computed. These emissivities are then converted to luminosities which are projected onto a two-dimensional plane using the SPH kernel of the simulation particles, turning them into a luminosity per unit area, which in turn is converted to a surface brightness.

\subsubsection{Radiative transfer effects}
\label{sssec:Radiative transfer effects}

In the predictions made in this work, \lya\ propagation is always treated in the optically thin limit. For the constructed mock narrowband images, it is assumed that \lya\ photons are emitted in an isotropic manner, and reach the observer without any scattering. The exact effects that scattering would have are difficult to accurately predict (given, e.g., that the effects of dust are poorly constrained), but it is expected that for the filamentary IGM, the difference between our simulations and a model with a physically accurate treatment of radiative transfer will mostly be influenced by two competing effects. First, there might be a broadening of the filamentary structure due to scattering in the nearby IGM, causing the signal to become fainter. Second, however, filaments may also be illuminated by \lya\ radiation coming from nearby dense structures (where additional radiation is likely to be produced in galaxies) that is scattered in the filament, which would cause the filaments to appear brighter. Simulations including radiative transfer indeed show a mixture of these two effects, where the surface brightness of filaments generally is not affected much, or is even boosted \citetext{private communication, Weinberger, 2019}. As the effects of radiative transfer on this work are expected to be moderate (a more detailed discussion on the optical depth of \lya\ is included in \cref{ap:Lya optical depth}), they are assumed not to affect our main findings in a major way. Future work can detail the precise effects of radiative transfer.

We limit the maximum surface brightness from recombinations to what is expected from purely reprocessing or ``mirroring'' the UV background at the boundaries of self-shielded regions (see \cref{sssec:Mirror limit}). This also mitigates the effect where the absence of radiative transfer can bias the surface brightness upwards in cases in which a sightline crosses several dense structures. In reality, however, with the presence of local ionising sources in such dense regions, an amplification with respect to the reprocessed UV background would likely be present as well. This is also suggested by a comparison of our simulation with a post-processing radiative transfer simulation of the same volume using a local source population similar to the one described in \citet{2019MNRAS.485L..24K}. Still, even with an accurate treatment of radiative transfer, the precise effects in the densest regions may rely considerably on the exact baryonic feedback mechanisms that are operating in these regions.

\begin{figure}
    \centering
    \includegraphics[width=\linewidth]{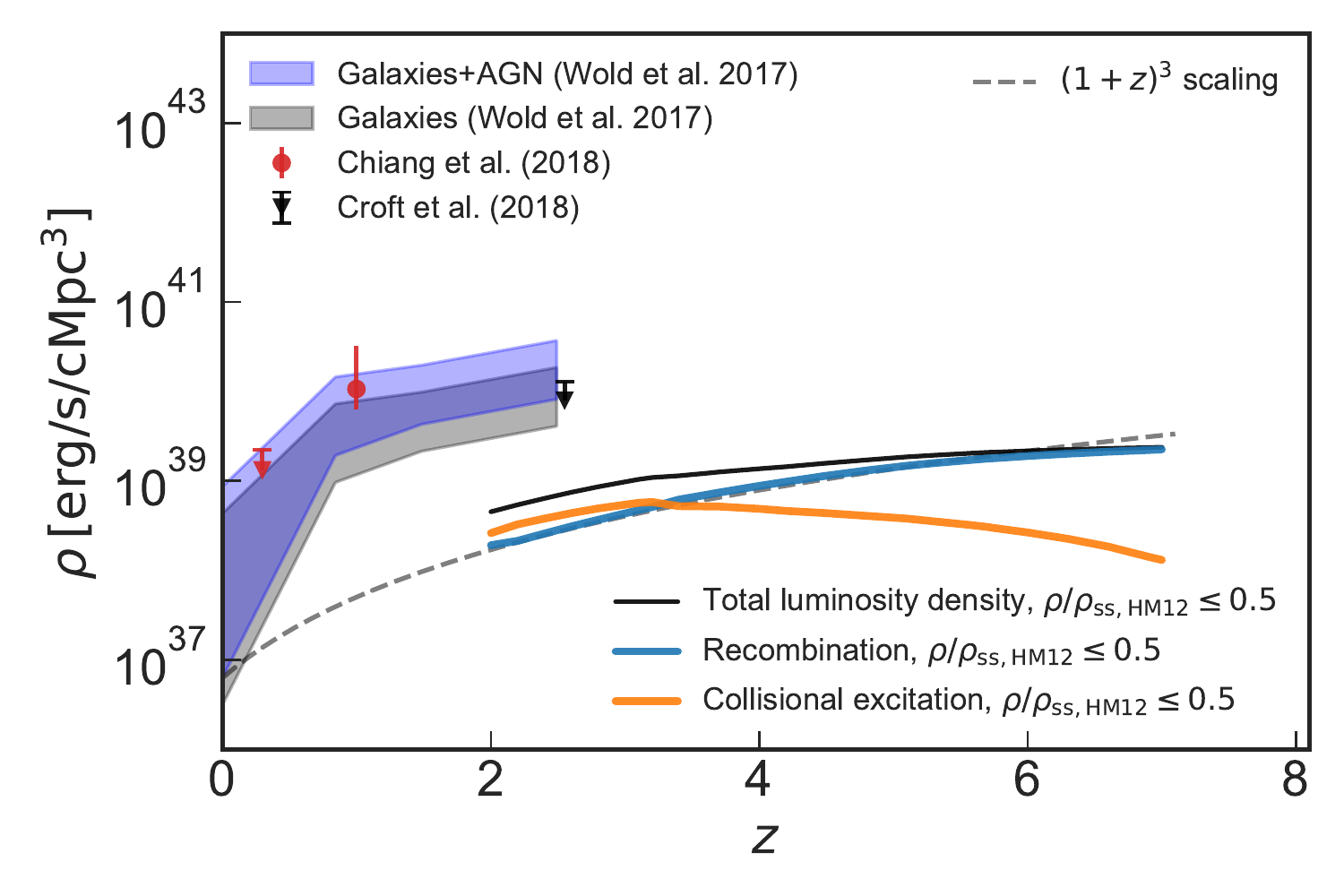}
    \caption[Redshift evolution of the \lya\ luminosity density]
    {Redshift evolution of the comoving \lya\ luminosity density. Blue and orange lines show the results for recombination and collisional excitation emission for gas at densities below half the critical self-shielding density (roughly corresponding to the IGM at an overdensity $\Delta \equiv \rho/\bar{\rho} \lesssim 50$ at $z=4.8$, see \cref{sssec:Density limits}), while the black line shows the total luminosity density for gas below this density threshold; all these follow from the simulation run with a box size of $40 \, h^{-1} \, \mathrm{cMpc}$ and resolution of $2 \times 1024^3$ particles (see \cref{ssec:SherwoodSuite} for more details on the simulation). Observational measurements at low redshift ($z<3$), as presented in \citet{2019ApJ...877..150C}, have been included as a reference. These consist of luminosity densities of just galaxies, and the contribution of both galaxies and AGN (shown as the grey and blue shaded areas, respectively) inferred by \citet{2019ApJ...877..150C} from the intrinsic luminosity density presented in \citet{2017ApJ...848..108W}; furthermore, the measurement and upper limit from \citet{2019ApJ...877..150C} are shown in red, and the upper limit from \citet{2018MNRAS.481.1320C} \citep[converted to a luminosity density by][]{2019ApJ...877..150C} is shown in black (see text for details). Data points are shown as circles, upper limits as downward triangles. Note that the data points should not be directly compared to our predictions as we consider only emission from the low-density gas in the IGM. Also shown is the $(1+z)^3$ scaling relation for recombination emission discussed in the text.}
    \label{fig:z_evolution_lum}
\end{figure}
\begin{figure*}
	\centering
	\includegraphics[width=\linewidth]{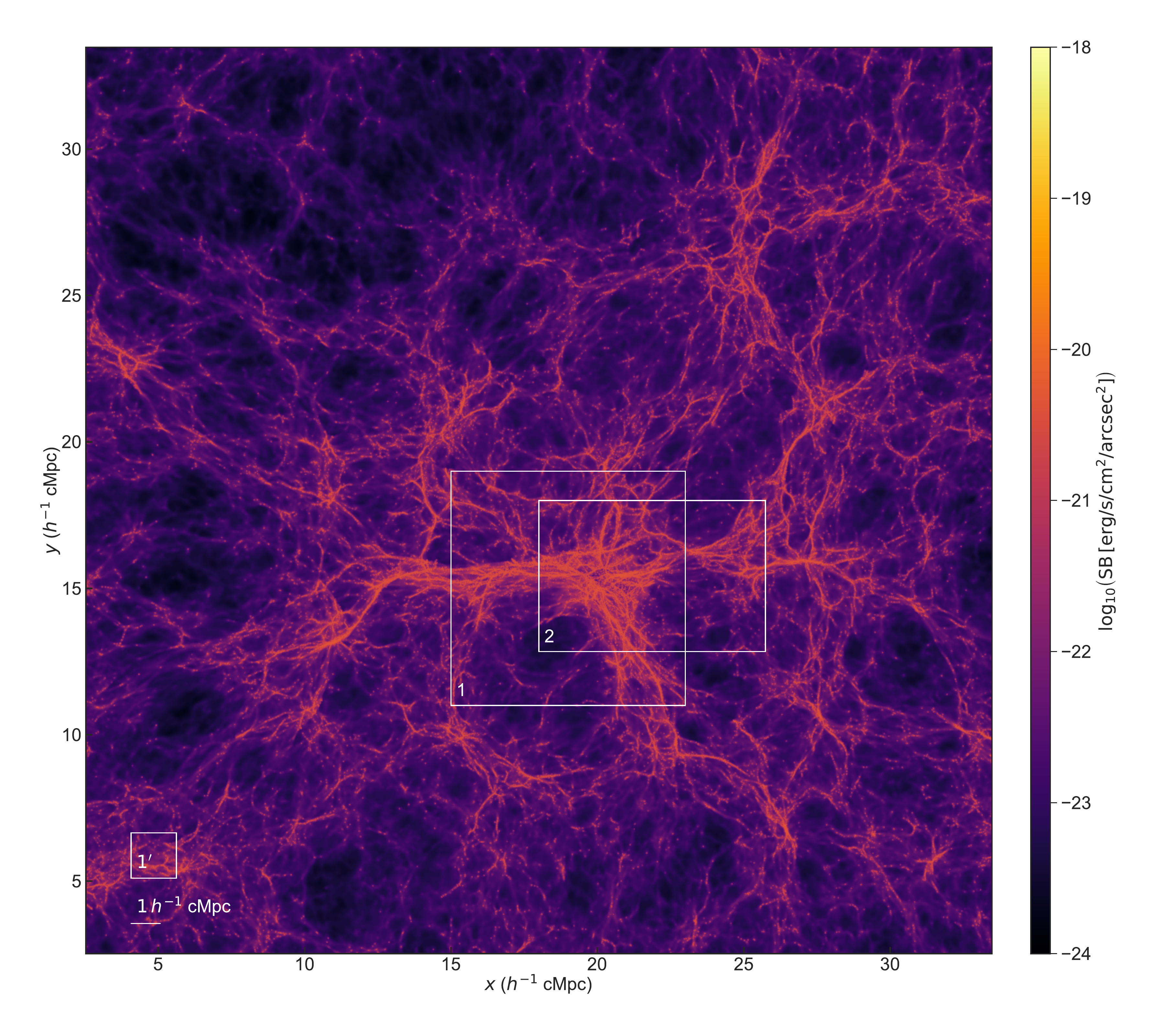}
	\caption[Total \lya\ surface brightness at $z=4.8$]
	{\lya\ surface brightness resulting from the combination of recombination emission (of all gas in the simulation) below the mirror limit, and collisional excitation of gas below half the critical self-shielding density, covering an area of $20 \times 20 \, \mathrm{arcmin}^2$, or $31.0 \times 31.0 \, h^{-2} \, \mathrm{cMpc}^2$, in a narrowband with $\Delta \lambda_\text{obs} = 8.75 \, \text{\AA}$ (corresponding to $\ssim 2.7 \, h^{-1} \, \mathrm{cMpc}$) in a simulation snapshot at $z=4.8$. The images are made by the projection method outlined in the text onto a pixel grid of $6000 \times 6000$ (i.e. the same pixel size as MUSE, making this the equivalent of a mosaic of $20 \times 20$ MUSE pointings -- more details on MUSE will follow in \cref{ssec:Observing facilities}). Regions 1 and 2, indicated by the white rectangles, will be studied in more detail later. Also shown in the bottom left corner are the scales of the MUSE field of view ($1 \times 1 \, \mathrm{arcmin}^2$) and $1 \, h^{-1} \, \mathrm{cMpc}$.}
	\label{fig:SB}
\end{figure*}
\begin{figure*}
    \centering
    \includegraphics[width=\linewidth]{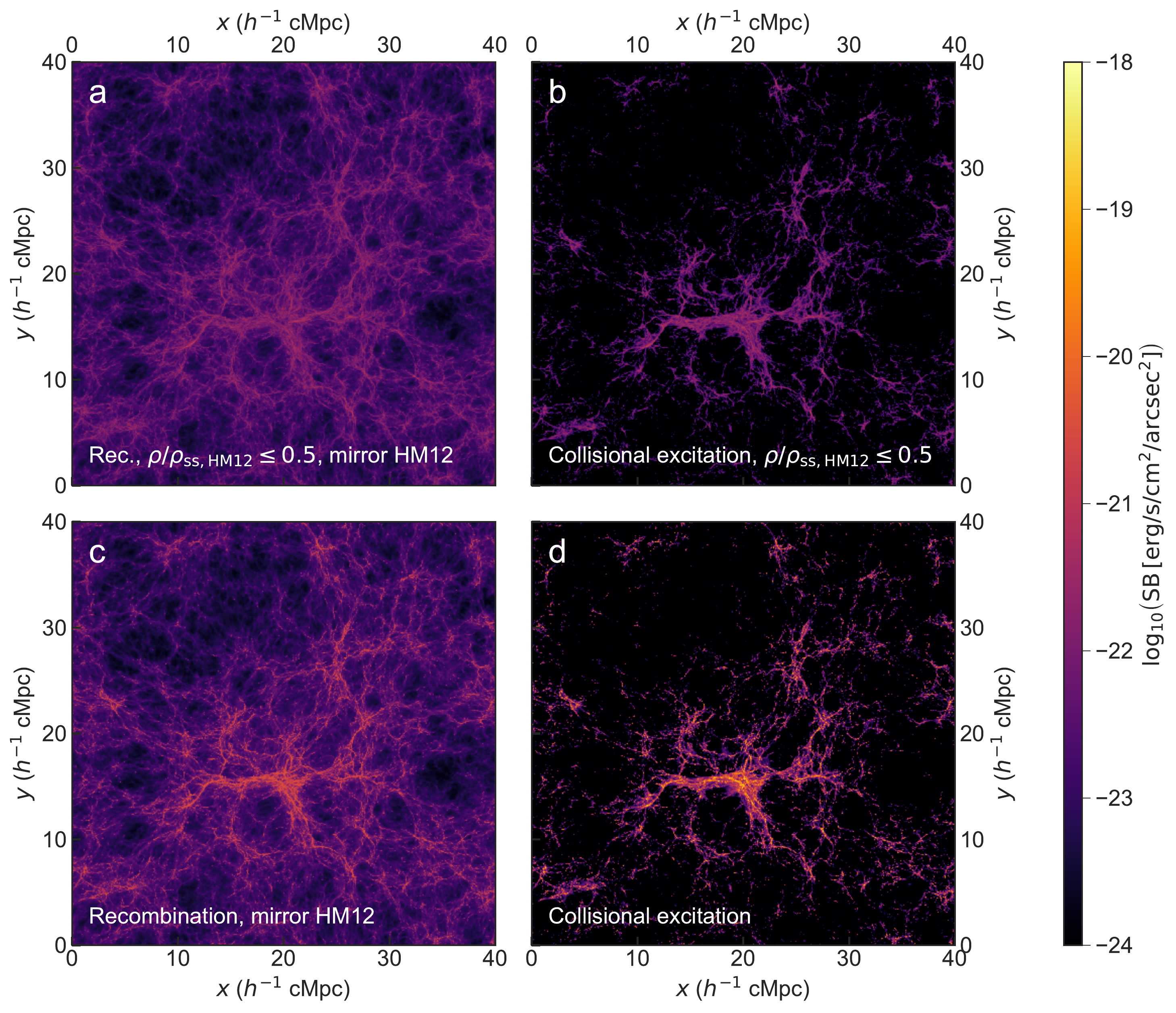}
    \caption[\lya\ surface brightness of recombination and collisional excitation processes at $z=4.8$]
    {\lya\ surface brightness of recombination (panel~\textbf{a}) and collisional excitation (panel~\textbf{b}) processes in a simulation snapshot at $z=4.8$, for the gas at densities below half the critical self-shielding density in a narrowband with $\Delta \lambda_\text{obs} = 8.75 \, \text{\AA}$, or $\ssim 2.7 \, h^{-1} \, \mathrm{cMpc}$ (projections made with pixel grid sizes of $1024 \times 1024$). These images show the entire (two-dimensional) spatial extent of the simulation, $40 \times 40 \, h^{-2} \, \mathrm{cMpc}^2$ ($25.8 \times 25.8 \, \mathrm{arcmin}^2$). Panels~\textbf{c} and \textbf{d} show the same maps, but without a density cut-off.}
    \label{fig:SBrecexc}
\end{figure*}
\begin{figure*}
    \centering
    \includegraphics[width=\linewidth]{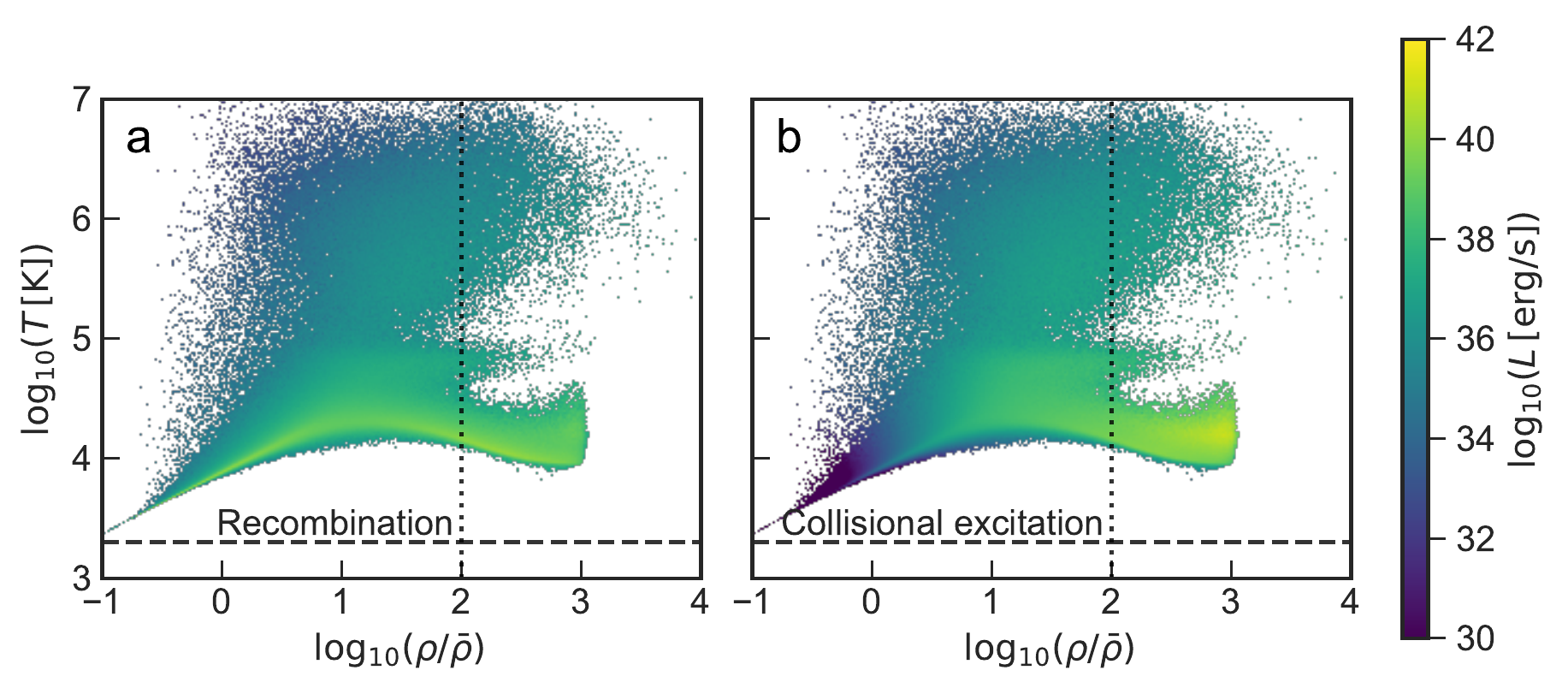}
    \caption[\lya\ luminosities of recombination and collisional excitation processes in phase space at $z=4.8$]
    {Histogram of \lya\ luminosities of recombination (panel~\textbf{a}) and collisional excitation (panel~\textbf{b}) processes in the same region as shown in \cref{fig:SBrecexc} (a narrowband with $\Delta \lambda_\text{obs} = 8.75 \, \text{\AA}$, equivalent to $\ssim 2.7 \, h^{-1} \, \mathrm{cMpc}$) in a simulation snapshot at $z=4.8$ in phase space. The colour represents the total luminosity in the simulation per histogram bin. The horizontal dashed line corresponds to the lower limit above which the fitting function of \citet{1990MNRAS.242..692S, 1991ApJ...380..302S} for collisionally excited \lya\ emission is valid (the upper limit lies above the plotted range), while the vertical dotted line shows the critical self-shielding density threshold at this redshift for the \citetalias{2012ApJ...746..125H} UV background \citep[from Eq. (13) in][]{2013MNRAS.430.2427R}. Densities above the threshold are also more strongly affected by modelling uncertainties.}
    \label{fig:Luminosity phase space}
\end{figure*}

\section{Results}
\label{sec:Results}

\subsection{Luminosity density}
\label{ssec:Luminosity density}

\Cref{fig:z_evolution_lum} shows the redshift evolution of the comoving \lya\ luminosity density in our simulation down to $z=2$. The total luminosity of gas within the entire simulation at densities below half the critical self-shielding density, corresponding to an overdensity $\rho/\bar{\rho} \lesssim 50$ at $z=4.8$ (see \cref{sssec:Density limits}), roughly corresponding to the IGM, is computed. This is also separately done for the recombination and collisional excitation contributions. We then divide by the (comoving) simulation volume to convert the luminosity to a comoving luminosity density.

Observational measurements at low redshift ($z<3$), as compiled by \citet{2019ApJ...877..150C}, have been included as reference. Note that we expect our predictions to be lower than the inference of \citet{2019ApJ...877..150C}, as they also consider emission from high-density gas. The data consist of estimates of the luminosity density of \lya\ emission from galaxies and AGN inferred by \citet{2017ApJ...848..108W} based on a flux-limited sample of \lya\ emitters from GALEX data and scaling the H$\alpha$ galaxy luminosity function measurements \citep{2013MNRAS.428.1128S} out to $z=2$. The measurement and upper limit from \citet{2019ApJ...877..150C} are shown in red. \citet{2019ApJ...877..150C} obtain a constraint on the total \lya\ luminosity density from galaxies and AGN as well as the diffuse IGM by cross-correlating the GALEX UV intensity maps with spectroscopic objects in SDSS. A comparison of the measurements from \citet{2019ApJ...877..150C} and \citet{2017ApJ...848..108W} indicates that at least at $z \lesssim 1$, most \lya\ emission originates in galaxies and AGN. The upper limit from \citet[; converted to a luminosity density by \citealt{2019ApJ...877..150C}]{2018MNRAS.481.1320C} is shown in black in \cref{fig:z_evolution_lum}. \citet{2018MNRAS.481.1320C} fit model spectra to luminous red galaxies in BOSS and cross-correlate the residual \lya\ emission with the \lya\ forest in BOSS quasars to obtain the upper limit from a non-detection shown in \cref{fig:z_evolution_lum}. As such, this procedure places a limit on the component of diffuse \lya\ emission that correlates with the matter distribution \citep{2018MNRAS.481.1320C}.\footnote{An additional measurement, arising from a cross-correlation with BOSS quasars, is restricted to scales within $15 \, h^{-1} \, \mathrm{cMpc}$ of a quasar \citep[equivalent to only $\ssim 3\%$ of space, see][]{2018MNRAS.481.1320C} and is therefore not included as a global luminosity density here.}

Going from redshift $z=2$ to $z=7$, the comoving \lya\ luminosity density increases by just under an order of magnitude (but see \cref{ap:Redshift evolution} for a further discussion of the redshift evolution of surface brightness). As can be seen in the figure, this is mostly due to the increase in recombination emission. Under the simple assumption that the emissivity is produced at a fixed overdensity its emissivity increases like the square of the mean density, which would correspond to a scaling of
\begin{align}
    \label{eq:Recombination emissivity scaling}
    \epsilon_\text{rec} & \sim \Delta^2 (1+z)^6 \text{ (physical luminosity density), or} \\ \nonumber
    \epsilon_\text{rec} & \sim \Delta^2 (1+z)^3 \text{ (comoving luminosity density),}
\end{align}
where $\epsilon_\text{rec}$ is the recombination emissivity and $\Delta \equiv \rho/\bar{\rho}$ the overdensity. As shown by the dashed line in \cref{fig:z_evolution_lum}, the simple scaling for recombination emission in \cref{eq:Recombination emissivity scaling} explains the simulated luminosity density quite well at all redshifts shown.

For collisional excitation, there should be two relevant effects: in the optically thin limit, the neutral fraction in ionisation equilibrium increases proportional to the density, hence $n_\text{HI} \sim n_\text{H}^2$; consequently, the emissivity scales as $\epsilon_\text{exc} \sim n_\text{HI} n_\text{e} \sim n_\text{H}^3$. If the emission were again produced at fixed overdensity, and if there is little evolution in the photoionisation rate, this would hence scale like
\begin{align}
    \label{eq:Collisional emissivity scaling}
    \epsilon_\text{rec} & \sim \Delta^3 (1+z)^9 \text{ (physical luminosity density), or} \\ \nonumber
    \epsilon_\text{rec} & \sim \Delta^3 (1+z)^6 \text{ (comoving luminosity density),}
\end{align}
where $\epsilon_\text{exc}$ is the emissivity from collisional excitation. However, collisional excitation does not follow the predicted $(1+z)^6$ scaling in \cref{eq:Collisional emissivity scaling} (and hence is not shown), even decreasing with redshift at $z \gtrsim 3$. This suggests that it is dominated by emission near the critical self-shielding density (see also \cref{ssec:Surface brightness maps}), and is hence more strongly affected by the density limit at half the critical self-shielding density, which decreases with increasing redshift more strongly than the mean density (i.e. the critical self-shielding \textit{overdensity} decreases towards higher redshift, see \cref{sssec:Density limits}). Still, we note that, depending on the precise distribution of self-shielded regions which is dictated by local ionising sources on a small scale, collisional excitation from dense gas could account for an additional increase of the comoving luminosity density that surpasses the cosmic surface brightness dimming effect, which itself scales as $(1+z)^4$.

\subsection{Surface brightness maps}
\label{ssec:Surface brightness maps}

\cref{fig:SB} shows a surface brightness ($\text{SB}$) map which is the combination of recombination emission (of all gas in the simulation) below the mirror limit, and collisional excitation of gas below half the critical self-shielding density in a simulation snapshot at $z=4.8$, for a narrowband with $\Delta \lambda_\text{obs} = 8.75 \, \text{\AA}$ (at this redshift coinciding with a thickness of the slice of $\ssim 2.7 \, h^{-1} \, \mathrm{cMpc}$). The map shows a region corresponding to $20 \times 20 \, \mathrm{arcmin}^2$. Also shown in the bottom left corner is the size of the MUSE field of view ($1 \times 1 \, \mathrm{arcmin}^2$ -- see \cref{ssec:Observing facilities} for more details). Regions 1 and 2, indicated by the white rectangles, will be studied in more detail later. The values of the surface brightness for this narrowband width are of the order of $\text{SB} \lesssim 10^{-23} \, \mathrm{erg \, s^{-1} \, cm^{-2} \, arcsec^{-2}}$ for the void regions, increasing to typically $\ssim 10^{-21} \, \mathrm{erg \, s^{-1} \, cm^{-2} \, arcsec^{-2}}$ for the IGM filaments. The denser regions have intensity peaks that typically show surface brightness values of $\ssim 10^{-20} \, \mathrm{erg \, s^{-1} \, cm^{-2} \, arcsec^{-2}}$.

\cref{fig:SBrecexc} shows the same narrowband slice as in \cref{fig:SB} (now for the full spatial extent of the simulation box, $40 \times 40 \, h^{-2} \, \mathrm{cMpc}^2$ or $25.8 \times 25.8 \, \mathrm{arcmin}^2$) split into contributions from recombination and collisional excitation processes in the gas. These maps were all made by projection onto a grid of $1024 \times 1024$ pixels. As before, a narrowband slice with $\Delta \lambda_\text{obs} = 8.75 \, \text{\AA}$ ($\ssim 2.7 \, h^{-1} \, \mathrm{cMpc}$) was chosen. Panels~a and b show gas at densities below half the critical self-shielding density, while panels~c and d show all gas. The mirror limit is applied to both panels showing recombination emission (a and c). In this large-scale narrowband image, the total luminosity of recombination processes below half the critical self-shielding density (the total in panel~a before imposing the mirror limit, although no pixels are above the limit) is $\ssim 1.75 \cdot 10^{43} \, \mathrm{erg \, s^{-1}}$, whereas for collisional excitation (the total in panel~b) this is $\ssim 5.45 \cdot 10^{42} \, \mathrm{erg \, s^{-1}}$. Including all gas, the total luminosities are $\ssim 5.02 \cdot 10^{43} \, \mathrm{erg \, s^{-1}}$ for recombination (panel~c; again before imposing the mirror limit, but only $0.37\%$ of pixels are above the limit in this panel), and $\ssim 2.21 \cdot 10^{44} \, \mathrm{erg \, s^{-1}}$ for collisional excitations (panel~d). We note that while collisional excitations dominate over recombinations at high densities, the two processes contribute more equally at the lower densities prevalent in large-scale-structure filaments, with recombination slightly prevailing over collisional excitation. Moreover, gas near or somewhat above the critical self-shielding density contributes significantly to the maximum SB that is reached for both channels; we conclude that the recombination prediction including all gas while having the mirror limit imposed should yield at least a robust lower limit, while the collisional excitation prediction for gas at higher densities is more uncertain, motivating our conservative density limit (\cref{sssec:Density limits}).

While overall these surface brightness maps exhibit the same structure as \cref{fig:SB}, the spatial distribution of emission coming from collisional and recombination processes is different. The degree of clustering in the emission is lower for the emission due to recombination processes, and higher for the component that is due to collisional excitations. Recombination and collisional excitation depend differently on temperature and density, as discussed in \cref{ssec:Luminosity density}. In particular, at fixed temperature and photoionisation rate, recombinations are proportional to the square of the density, $\ssim \rho^2$, while in ionisation equilibrium collisional excitations are proportional to $\ssim \rho^3$. As a consequence, recombinations are more equally spread across the volume, while collisional excitations are clearly more important at higher densities, thus reflecting the filamentary structure of the cosmic web better, and leaving darker voids in between. To understand this in more detail, we now turn to the phase space distribution of the gas in the simulation.

In \cref{fig:Luminosity phase space}, the luminosity in the simulation is shown at the same redshift and the same region as in \cref{fig:SBrecexc} (also in the identical narrowband slice of $\Delta \lambda_\text{obs} = 8.75 \, \text{\AA}$, or $\ssim 2.7 \, h^{-1} \, \mathrm{cMpc}$), now as a luminosity-weighted two-dimensional histogram in temperature and density. This illustrates what was discussed in \cref{ssec:Luminosity density} and seen in \cref{fig:SBrecexc}: collisional excitation is not effective at lower densities, and the most luminous gas particles are located in the upper part of the very high-density cooling branch. Recombination emission, on the other hand, exhibits luminosities that are more comparable at lower and higher densities.

From the phase-space distribution in \cref{fig:Luminosity phase space}, it is clear that very little gas has temperatures outside of the temperature range of $2 \cdot 10^3 \, \mathrm{K} \leq T \leq 1 \cdot 10^8 \, \mathrm{K}$ for which our fitting function for collisionally excited \lya\ is valid (the lower limit of which is indicated by the horizontal dashed line; the upper limit lies above the plotted range and almost all of the gas in the simulation).\footnote{In fact, this is the case for the entire relevant redshift range.} The contribution from gas outside of this temperature range will be very small and we thus neglect it here.

The vertical dotted line shows the critical self-shielding density threshold at this redshift for the \citetalias{2012ApJ...746..125H} UV background \citep[from Eq. (13) in][]{2013MNRAS.430.2427R}, illustrating the limiting density below which gas will not be strongly affected by the details of modelling self-shielding. 

\begin{table*}
    \centering
    \caption[Observational experiments]{
        Overview of a selection of current and future instruments that might be most promising for detecting IGM filaments.
    }
    \begin{tabular}{lcccccc}
        \hline
        Name & Wavelength range & Redshift range & Field of view  & Resolution \\
        & $\lambda$ (\AA)& $z_\mathrm{\lya}$ &  & $R$ \\
        \hline 
        \textit{Current IFU instrumentation} & & & & \\
        \hline 
        KCWI-Blue (Keck) & $3500$-$5600$ & $1.9$-$3.6$ & $20 \times 33 \, \mathrm{arcsec}^2$  & $1000$-$20000$ \\
        MUSE (VLT) & $4650$-$9300$ & $2.8$-$6.7$ & $1 \times 1 \, \mathrm{arcmin}^2$  & $1770$-$3590$ \\
        KMOS (VLT) & $8000$-$25000$ & $5.6$-$19.6$ & $ 65 \times 43 \, \mathrm{arcsec}$  & $2000$-$4200$ \\
        OSIRIS (Keck) & $10000$-$24500$ & $7.2$-$19.1$ & $4.8 \times 6.4  \, \mathrm{arcsec}^2$ & $2000$-$4000$ \\
        SINFONI (VLT) & $11000$-$24500$ & $8.0$-$19.1$ & $8 \times 8 \, \mathrm{arcsec}^2$ & $2000$-$4000$ \\
        \hline
        \textit{Upcoming IFU instrumentation} & & & & & \\
        \hline
        KCRM (KCWI-Red, Keck) & $5300$-$10500$ & $3.4$-$7.6$ & $20 \times 33 \, \mathrm{arcsec}^2$  & $1000$-$20000$ \\
        HARMONI (ELT) & $4700$-$24500$ & $2.9$-$19.1$ & $6.4 \times 9.1 \, \mathrm{arcsec}^2$ & $3000$-$20000$ \\
        BlueMUSE (VLT) & $3500$-$6000$ & $1.9$-$3.9$ & $1.4 \times 1.4 \, \mathrm{arcmin}^2$  & $\ssim 3000$-$5000$ \\
        \hline
        \textit{Upcoming/proposed space missions} & & & & & \\
        \hline
        SPHEREx\tablefootmark{*} & $7500$-$50000$ & $5.2$-$40.1$ & $3.5 \times 11.3 \, \mathrm{deg}^2$  & $41$-$130$ \\
        MESSIER\tablefootmark{*} & $\ssim 2000$-$7000$ & $\ssim 0.5$-$4$ & $2 \times 2 \, \mathrm{deg}^2$ & \dots \\
        WSO-UV & $1150$-$3200$ & $\ssim 0$-$1.5$ & $70 \times 75 \, \mathrm{arcsec}^2$  & $\ssim 500$ \\
    \end{tabular}
    \tablefoot{
        Fields left blank indicate currently unknown or undecided values. All current instruments presented are integral field unit (IFU) spectrographs, upcoming/proposed instruments include several IFU spectrographs and space telescopes (two UV satellites and one IR spectrophotometer). Future experiments are in the development stage, unless marked with an asterisk. \\
        \tablefoottext{*}{Proposed space missions.}
    }
    \label{tab:Experiments}
\end{table*}

\subsection{Observing facilities}
\label{ssec:Observing facilities}

In \cref{tab:Experiments}, an overview of a selection of current and future instruments that could potentially detect \lya\ emission from IGM filaments is shown along with their wavelength and redshift range, field of view (FOV), and resolving power ($R$). Most ground- and space-based instruments that may be considered for detection of the diffuse IGM, will naturally observe in the visible spectrum and the ultraviolet, respectively, given the limitations of ground-based observations due to absorption by Earth's atmosphere. This necessarily restricts the redshift range in which these instruments could observe \lya. For ground-based observations, the typical redshift is $z \gtrsim 2.5$, whereas space-based telescopes observing in the UV can detect \lya\ at lower redshifts: in principle, satellites carrying UV detectors could observe it from $z \sim 0$ up to about $z \sim 1.5$.

Integral field unit (IFU) spectrographs have arguably the best instrument design for directly detecting emission from the cosmic web, owing to the flexibility in extracting pseudo-narrowband images over a wide range of bandwidths and central wavelengths and thereby resolving structures both spatially and spectrally over a large cosmic volume at once. The typical narrowband width extracted from IFU spectrographs to observe \lya\ emission is $< 10 \, \text{\AA}$ \citep[e.g.,][]{2016A&A...587A..98W, 2018Natur.562..229W}, almost an order of magnitude smaller than obtained from photometric narrowband imaging which have typical bandwidths of $\ssim80$-$100 \, \text{\AA}$ \citep{2011ApJ...736..160S, 2018PASJ...70S..13O}. This significantly improves the contrast of IFU emission line maps for observations limited by sky-noise. Despite the limited contrast for individual images, photometric narrowband studies still have detected large-scale \lya\ emission in stacking analyses \citep[e.g.][]{2011ApJ...736..160S, 2012MNRAS.425..878M, 2019arXiv190600173K}, enabled by the wide field of view and large number of sources collected by such cameras. In particular, the recently installed Hyper Suprime-Cam on Subaru is currently obtaining $26 \, \mathrm{deg}^2$ narrowband imaging from redshift $z=2.2$-$6.6$ as part of the Hyper Suprime-Cam Subaru Strategic Program \citep[e.g.][]{2018PASJ...70S..13O}. However, for this work, we will focus on instruments that are most likely to obtain individual detections of \lya\ emission from the cosmic web. Before the appearance of integral field unit imaging, another spectroscopic method used was long-slit spectroscopy \citep[as in e.g.][]{2008ApJ...681..856R}, but with the arrival of integral field spectroscopy, the volume probed by deep observations targeting \lya\ emission could be dramatically increased, rendering long-slit spectroscopy a non-competitive alternative for this purpose.

The Very Large Telescope (VLT) has the widest range of IFU spectrographs. The current near-IR instruments at this facility are SINFONI and KMOS, whose acronyms stand for Spectrograph for INtegral Field Observations in the Near Infrared \citep[SINFONI, see][]{2003SPIE.4841.1548E, 2004Msngr.117...17B}, and the K-band Multi Object Spectrograph \citep[KMOS, see][]{2013Msngr.151...21S}. Due to their spectral range, they are both only able to observe \lya\ at very high redshifts, respectively $z > 8.0$ and $z > 5.6$ (where the partly neutral IGM is expected to absorb most \lya\ emission). Most recently installed (2014) on the VLT is MUSE, the Multi Unit Spectroscopic Explorer, an IFU spectrograph operating in the visible wavelength range \citep[see][]{2010SPIE.7735E..08B}. The combination of its relatively large FOV ($1 \times 1 \, \mathrm{arcmin}^2$) and spectral coverage ($4650$-$9300 \, \text{\AA}$), while maintaining good spectral resolution (ranging between $1770$-$3590$), currently makes it one of the most promising candidates for the purpose of imaging the cosmic web in \lya. BlueMUSE \citep{2019arXiv190601657R} is a proposed second MUSE instrument, optimised for the blue end of the visible wavelength range. Future instruments at the VLT's successor, the Extremely Large Telescope (ELT), include the High Angular Resolution Monolithic Optical and Near-infrared Integral field spectrograph \citep[HARMONI, see][]{2014SPIE.9147E..25T}, which is expected to be operational in 2025.

The blue channel of the Keck Cosmic Web Imager \citep[KCWI, see][]{2018ApJ...864...93M} is an instrument similar to VLT/MUSE at the Keck II telescope. It offers a slightly better spectral sampling, although the FOV and spatial resolution are smaller/lower ($20 \times 33 \, \mathrm{arcsec}^2$ and $1.4 \, \mathrm{arcsec}$). However, since it has only become operational in 2018, no deep-field imaging like the MUSE observations of the \textit{Hubble} Deep Field South and \textit{Hubble} Ultra-Deep Field \citep{2015A&A...575A..75B, 2017A&A...608A...1B} has been released publicly yet. The red channel to KCWI, the Keck Cosmic Reionization Mapper (KCRM), is currently under construction and will complement the blue channel to cover the full wavelength range of $3500$-$10500 \, \text{\AA}$ ($3.4<z_\text{\lya}<7.6$). Similar to SINFONI on the VLT, Keck currently has a near-infrared IFU spectrograph, OSIRIS, with a small FOV that can target \lya\ only above $z>7.2$ (where the considerably neutral IGM is expected to absorb most emission).

For completeness, we also mention several promising space-based experiments: the World Space Observatory-Ultraviolet \citep[WSO-UV, see][]{2018SPIE10699E..3GS}, and MESSIER \citep{2017IAUS..321..199V}, two proposed UV satellites. They are proposed to have large FOVs and high sensitivities, but are limited to the lower redshift range ($z < 1.5$). In this work, we instead focus our attention on the high-redshift regime ($z > 3$). In February 2019, SPHEREx \citep{2018arXiv180505489D} was selected as the next medium-class explorer mission by NASA and is targeted for launch in 2023. SPHEREx will survey the entire sky with a spectrophotometer at very low spectral resolution, sensitive to diffuse \lya\ emission at $z>5.2$.

\begin{figure}
	\centering
	\includegraphics[width=\linewidth]{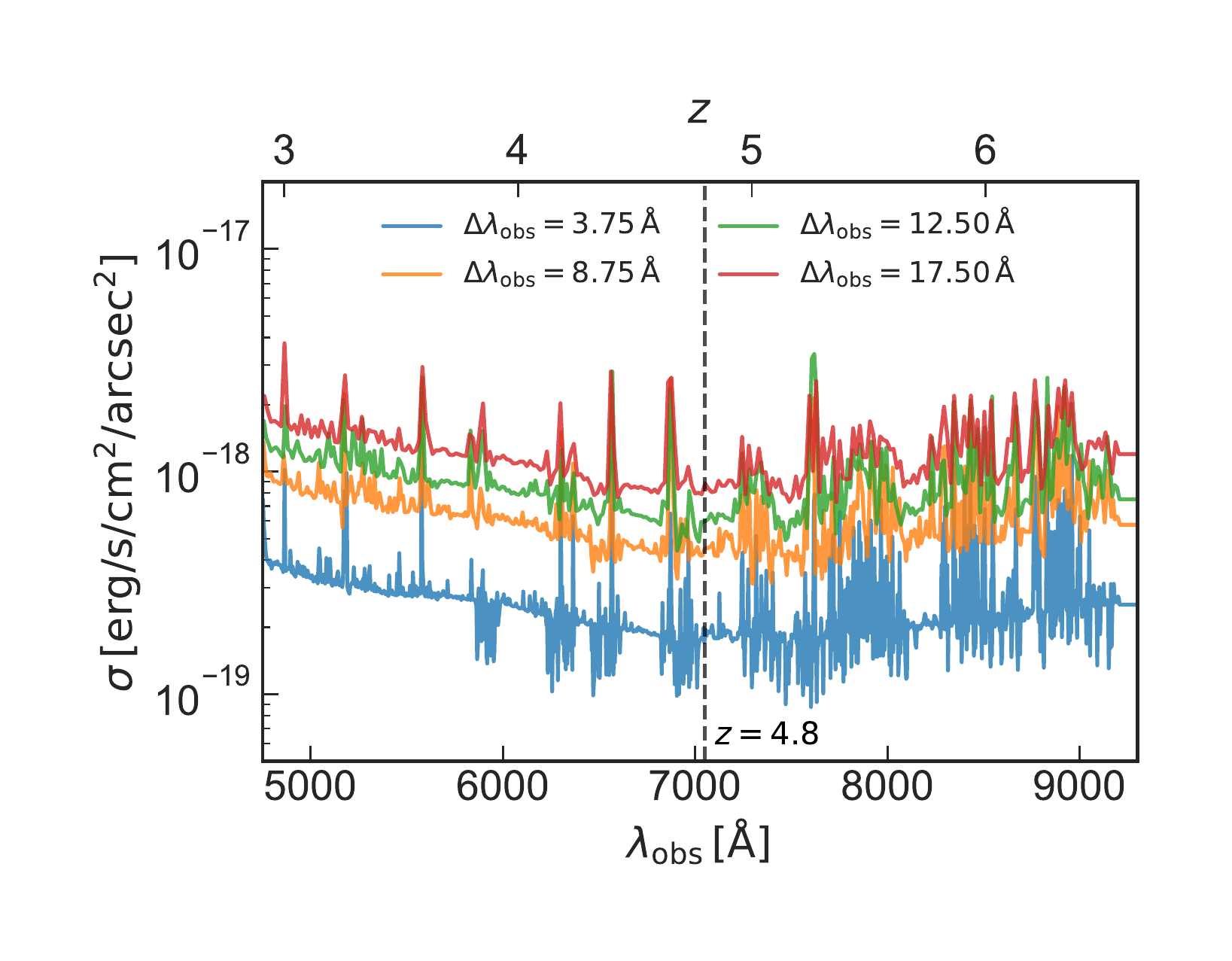}
	\caption[Inferred noise in the MUSE HDFS observation]
	{Inferred noise in the MUSE HDFS observation as a function of observed wavelength or redshift for different pseudo-narrowband widths: $\Delta \lambda_\text{obs} = 3.75 \, \text{\AA}$, $\Delta \lambda_\text{obs} = 8.75 \, \text{\AA}$, $\Delta \lambda_\text{obs} = 12.50 \, \text{\AA}$, and $\Delta \lambda_\text{obs} = 17.50 \, \text{\AA}$. Note that skylines result in increased noise in some spectral ranges. The vertical dashed line indicates the position of \lya\ at $z=4.8$, which is located in a spectral window with lower noise. The throughput of MUSE is at its maximum of $\ssim 40\%$ at $\ssim 7200 \, \mathrm{\Angstrom}$ \citep[e.g.][]{2019arXiv190601657R}.}
	\label{fig:MUSE_sigmas}
\end{figure}

Out of the current instruments, MUSE arguably offers the best compromise of resolution, spectral coverage, and volume surveyed. The combination of its FOV of $1 \times 1 \, \mathrm{arcmin}^2$ and spectral resolution make it a promising instrument to observe the cosmic web in \lya\ emission. As a representative example of what has already been achieved, we now discuss in more detail the MUSE \textit{Hubble} Deep Field South \citep[HFDS; see][]{2015A&A...575A..75B}. This is a $27 \, \mathrm{h}$ integration of the HDFS, reaching a $1\sigma$ surface brightness limit of $1 \cdot 10^{-19} \, \mathrm{erg \, s^{-1} \, cm^{-2} \, arcsec^{-2}}$ for emission lines. In \cref{fig:MUSE_sigmas}, we show the wavelength dependency of the inferred noise from the MUSE HDFS in pseudo-narrowbands of different widths for reference. We will discuss the consideration of different narrowband widths in more detail in \cref{sssec:Cosmic variance and narrowband widths}.

With MUSE, the \lya\ emission can be observed over the redshift range of $2.8$-$6.7$ (see \cref{tab:Experiments}). Hereafter, a redshift of $z=4.8$ is specifically chosen for a more detailed study of our simulations. As already hinted at in \cref{fig:z_evolution_lum}, the diffuse gas in the IGM appears to be denser and potentially intrinsically more luminous in \lya\ at higher redshifts -- however, there are negating effects imposed by self-shielding, with the critical self-shielding overdensity and the mirror limit steadily decreasing towards higher redshifts (\cref{sssec:Mirror limit,sssec:Density limits}). We choose a redshift of $4.8$ that seems to offer a reasonable compromise between these two effects, while also ensuring the results will not be significantly affected by the details of feedback (\cref{sssec:Density limits}). Finally, in reality, there will be an additional component of emission from filaments due to halos and galaxies embedded within them, the exact redshift dependence of which is difficult to predict. The following section will go into more details of the outlook on observations of primarily the diffuse gas with a MUSE-like instrument -- specifically, we will focus on such a wide-field integral-field spectrograph on an ELT-class telescope to explore the most far-reaching observational prospects in the near future -- discussing sensitivity limits, the overall redshift evolution, and optimal observing strategies.

\begin{figure*}
	\centering
	\includegraphics[width=\linewidth]{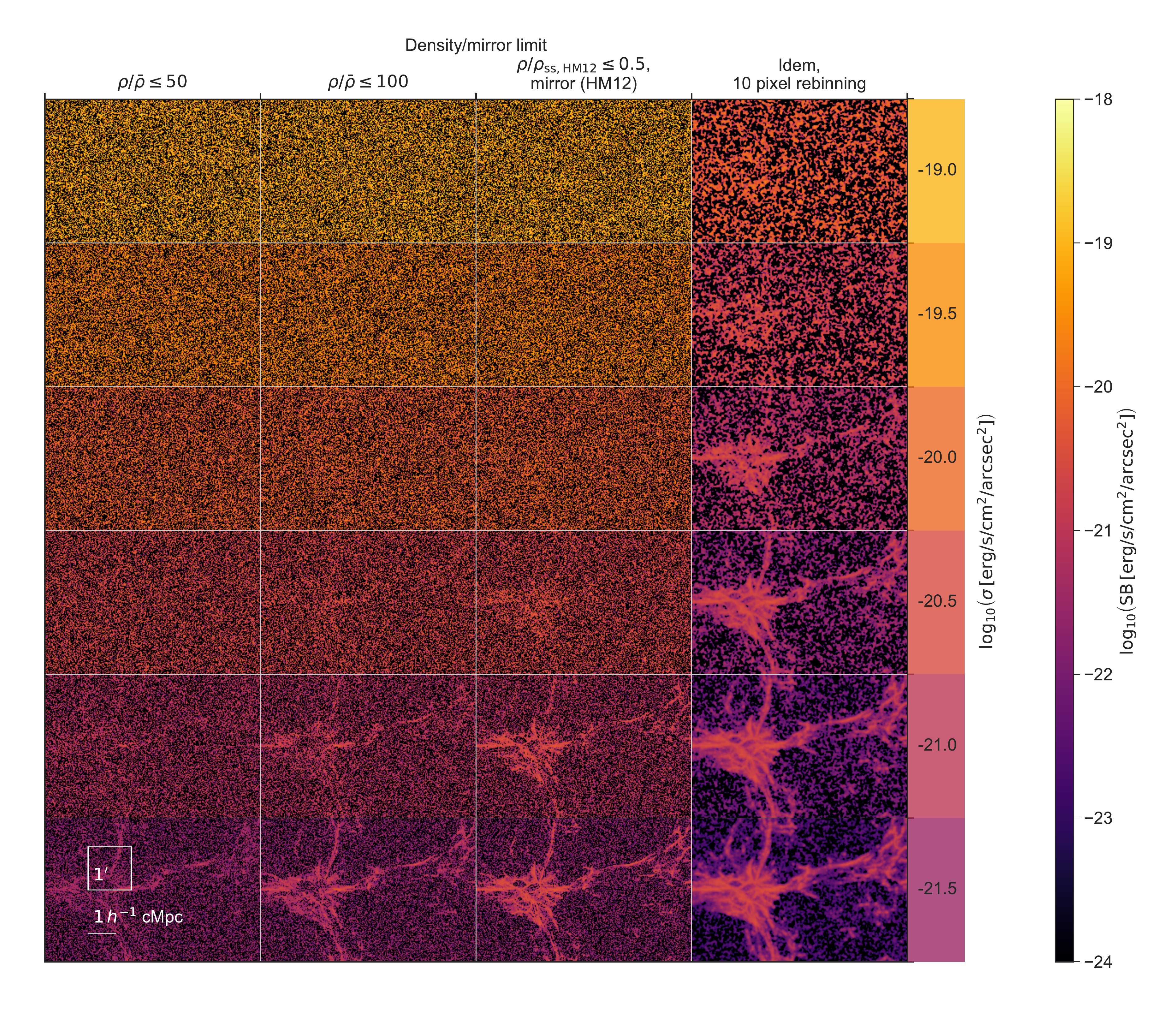}
	\caption[Overview surface brightness map at $z=4.8$ for different noises and density cut-offs]
	{Repeated view of region~2 of the $z=4.8$ surface brightness map in \cref{fig:SB} for different noise levels and assumptions on various limits, with a narrowband with $\Delta \lambda_\text{obs} = 3.75 \, \text{\AA}$ ($\ssim 1.19 \, h^{-1} \, \mathrm{cMpc}$) and convolved with a Gaussian kernel with a FWHM of $0.75 \, \mathrm{arcsec}$ before adding noise \citep[as in the HDFS observation, see][]{2015A&A...575A..75B}. The spatial extent of each panel is $5 \times 3.3 \, \mathrm{arcmin}^2$, or $7.8 \times 5.2 \, h^{-2} \, \mathrm{cMpc}^2$. $1 \sigma$ levels of the Gaussian noise applied per pixel (before rebinning) to each panel in the entire row are indicated directly right of the mosaic, coloured according to the colourbar on the very right, while the density cut-off and mirror limit (if applied) for each column is shown above the mosaic (see text for details). The final column is identical to the column next to it, but has a smoothing of $10 \times 10$ pixels or $2 \times 2 \, \mathrm{arcsec}^2$ applied (see text). Scales of $1 \times 1 \, \mathrm{arcmin}^2$ (the MUSE field of view) and $1 \, h^{-1} \, \mathrm{cMpc}$ are indicated on the lower left. Each panel in the image has $1500 \times 1000$ pixels, again making the pixel size equal to that of MUSE ($0.2 \, \mathrm{arcsec}$ per pixel).}
	\label{fig:4nsobs5}
\end{figure*}

To allow for a more realistic comparison between simulations and observations, some of the surface brightness images hereafter (\cref{fig:4nsobs5,fig:4nsobs_ov150}) are convolved with a Gaussian point spread function (PSF), to mimic the effect of seeing. The PSF full width at half maximum (FWHM) is chosen to be $0.75 \, \mathrm{arcsec}$, corresponding to the most conservative estimate for the MUSE HDFS \citep{2015A&A...575A..75B}. In addition, these figures include noise that is added to the signal predicted from the simulations.

\begin{figure*}
	\centering
	\includegraphics[width=\linewidth]{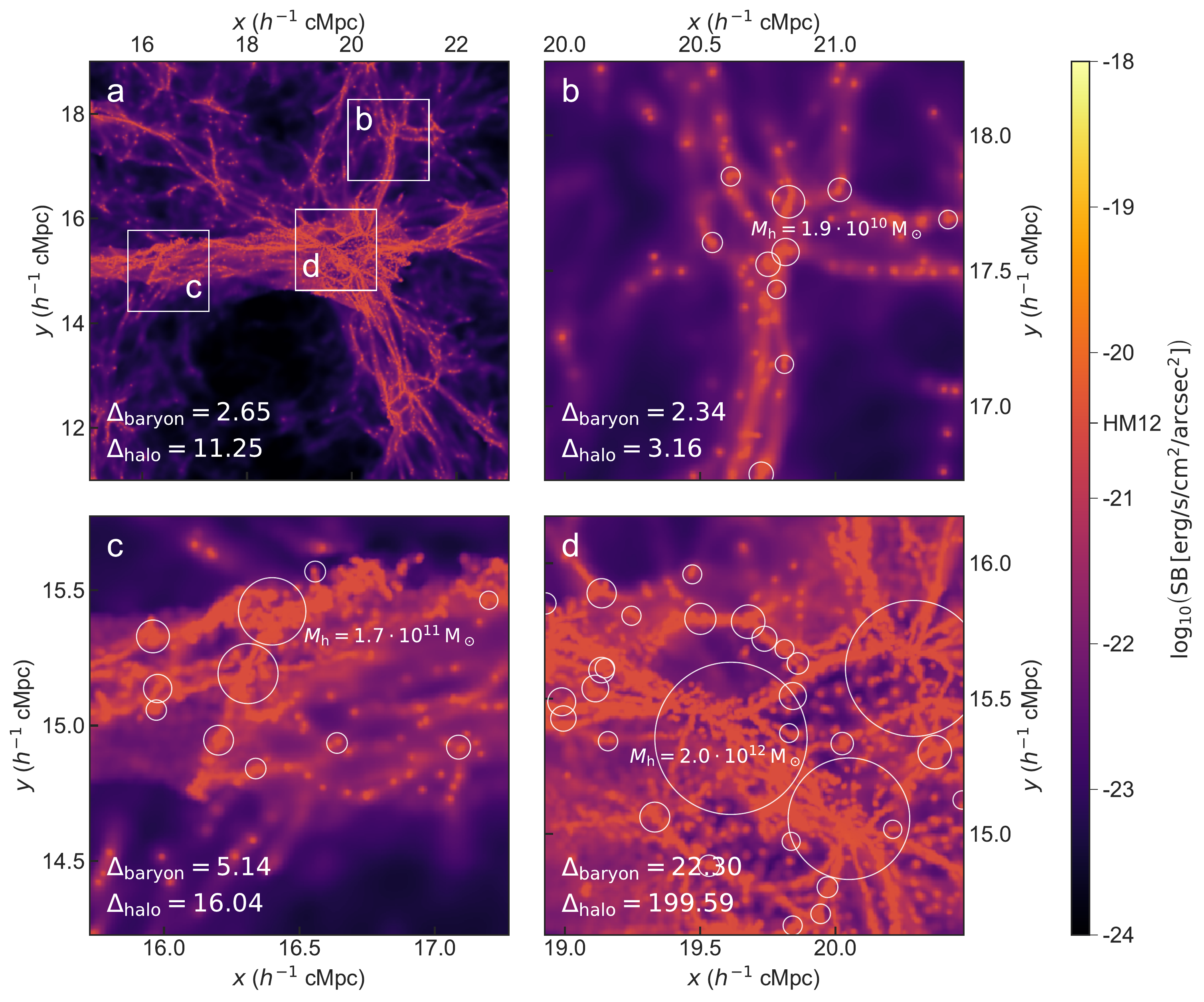}
	\caption[Observed \lya\ surface brightness at $z=4.8$]
	{\lya\ surface brightness for a narrowband with a smaller value of $\Delta \lambda_\text{obs} = 3.75 \, \text{\AA}$ (i.e. $\ssim 1.19 \, h^{-1} \, \mathrm{cMpc}$) in a simulation snapshot at $z=4.8$. As in \cref{fig:4nsobs_ov}, the SB shown is a combination of recombination emission (of all gas in the simulation) below the mirror limit (indicated on the colourbar as \citetalias{2012ApJ...746..125H}), and collisional excitation of gas below half the critical self-shielding density. Panel~\textbf{a}, an overview narrowband image that corresponds to region~1 in \cref{fig:SB} (centred on the same comoving coordinates both spatially and spectrally, but now less extended in wavelength range as the narrowband width has been decreased). This panel shows a region of $8 \times 8 \, h^{-2} \, \mathrm{cMpc}^2$ ($5.2 \times 5.2 \, \mathrm{arcmin}^2$) on a pixel grid of $1024 \times 1024$. Panels~\textbf{b}-\textbf{d}, \lya\ narrowband images the size of $1 \times 1 \, \mathrm{arcmin}^2$ consisting of $300 \times 300$ pixels (as the FOV of MUSE). The volume probed by one of these narrowband images at this redshift is $2.84 \, h^{-3} \, \mathrm{cMpc}^3$. The areas covered by these maps are indicated by the white squares in the overview panel~a. Halos with halo mass of $M_\mathrm{h} > 10^{9.5} \, \mathrm{M_\odot}$ are shown as circles, their size indicating their projected virial radius (see text). The most massive halo in each panel is annotated. In the bottom left corner of each panel, two different measures of the region's overdensity are shown (see text for more details; note that the baryonic value is calculated taking all gas into account, even though only gas below a certain density contributes to the collisional excitation). This figure highlights the importance of cosmic variance for detecting the filamentary structure of the IGM, and therefore demonstrates the instrument pointing is essential to efficiently map the IGM in \lya\ emission.}
	\label{fig:4nsobs_ov}
\end{figure*}

\begin{figure*}
	\centering
	\includegraphics[width=\linewidth]{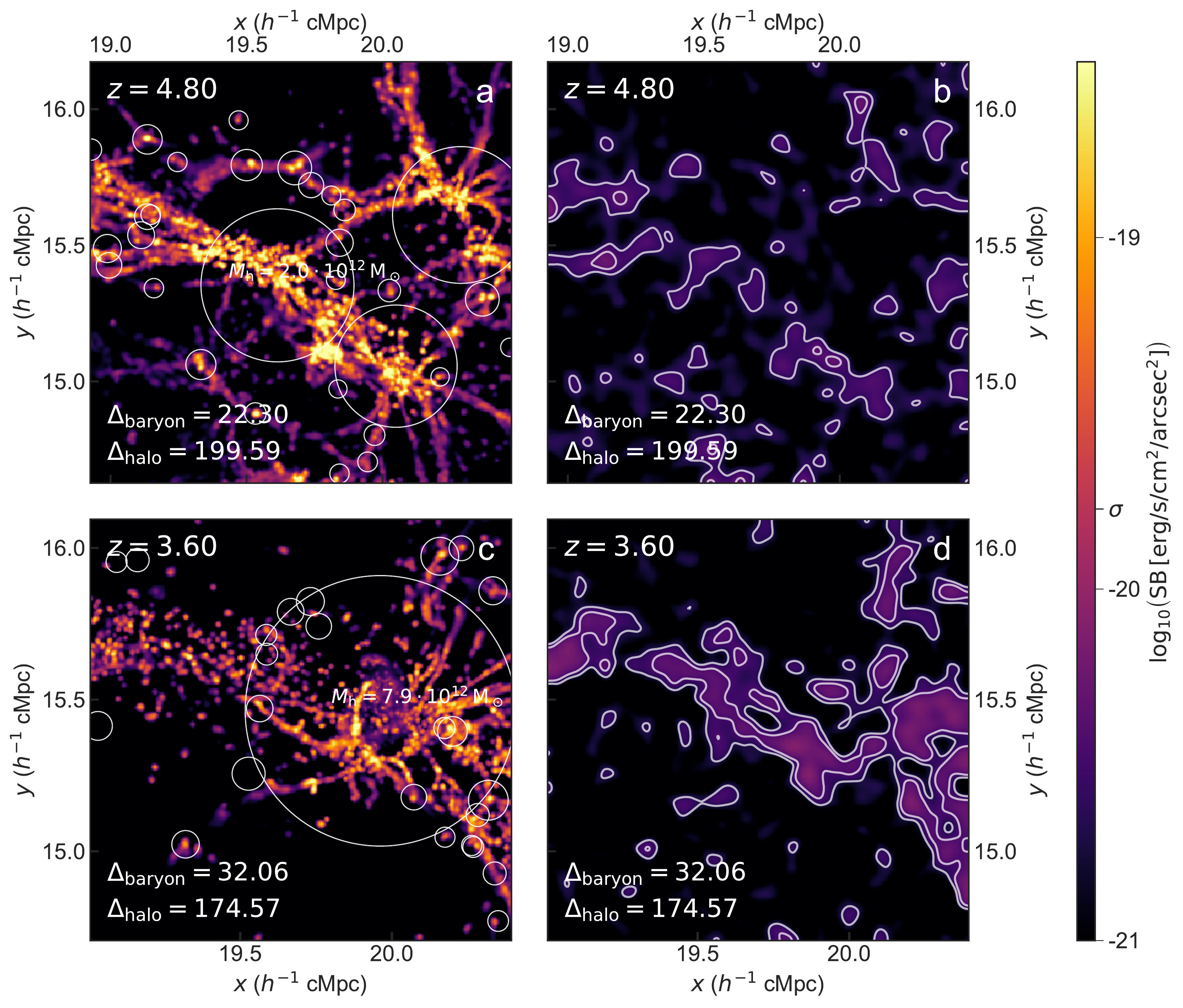}
	\caption[Observed \lya\ surface brightness at $z=4.8$]
	{Mock observations for a MUSE-like wide-field IFU instrument on the ELT covering the same region as panel~d in \cref{fig:4nsobs_ov} at two different redshifts ($z=4.8$ and $z=3.6$ on the top and bottom row, respectively), with no limits imposed and without observational effects versus with mirror and density limits, and modelled noise and seeing applied (left and right column, respectively; see text for details). The smaller narrowband with $\Delta \lambda_\text{obs} = 3.75 \, \text{\AA}$ (i.e. $\ssim 1.19 \, h^{-1} \, \mathrm{cMpc}$) has been used again. Note that these images have a different dynamical range than all other figures, to accentuate the observable \lya\ signal. In panels~b and d, a rebinning of $10 \times 10$ pixels ($2 \times 2 \, \mathrm{arcsec}^2$) was applied, after which the image was smoothed on the same scale, to recover the signal on larger scales. White contours indicate measured $3 \sigma$ and $5 \sigma$ levels. The \lya\ emission of IGM filaments can (marginally) be recovered in such an extremely deep observation, and seems more feasible at low redshift when considering the robust lower limits (i.e. the mirror limit for recombinations and density threshold for collisional excitation, panel~d); however, the predicted full intrinsic luminosity of filaments is notably higher at higher redshift (cf. panel~a and c; see text for further discussion) but quite dependent on the details of the modelling.}
	\label{fig:4nsobs_ov150}
\end{figure*}

\subsection{Simulated observations}
\label{ssec:Simulated observations}

\subsubsection{Cosmic variance and narrowband widths}
\label{sssec:Cosmic variance and narrowband widths}

Before we look in more detail at observational strategies, we introduce two indicators of overdensity in the ``observed'' simulation volume. The reason we introduce these specific characterisations of environment is to provide a quantitative way to distinguish different regions according to the level of their overall overdensity as could be characterised observationally. The first criterium to characterise environment, the baryonic overdensity, $\Delta_\mathrm{baryon}$, is computed by the ratio of baryonic density in the relevant region and the mean baryonic density at the redshift of the simulation. As a second criterion, we use the halo overdensity, $\Delta_\mathrm{halo}$, which is similar but instead of baryons uses halos with halo mass $M_\mathrm{h} > 10^{9.5} \, \mathrm{M_\odot}$: the amount of mass contained in these halos divided by the simulated (sub)volume as a fraction of their mean density (the total mass of all halos with $M_\mathrm{h} > 10^{9.5} \, \mathrm{M_\odot}$ in the simulation box, divided by its total volume). \footnote{Throughout this work, quoted halo masses are the dark matter mass of halos identified in the output snapshots of the simulation by a friends-of-friends algorithm with linking length $0.2$, roughly corresponding to masses measured in spherical regions with a density of $\Delta = 200$ times the mean density of the Universe, i.e. $M_\mathrm{200m}$ \citep[see e.g.][]{2008ApJ...688..709T}.} This particular mass cut-off has been chosen as this is near the resolution limit of the simulation.

Now turning our attention to a MUSE-like instrument specifically, \cref{fig:4nsobs_ov} shows several different surface brightness images of the simulation at $z=4.8$. The region of panel~a has already been shown in \cref{fig:SB} as region~1, while the other three images (panels~b-d) are the angular size of $1 \times 1 \, \mathrm{arcmin}^2$, and have a grid size of $300 \times 300$ pixels (corresponding to the FOV of the current MUSE instrument). The location of the two subregions shown in panels~b-d are marked in panel~a with a white square. In panels~b-d, halos with halo mass of $M_\mathrm{h} > 10^{9.5} \, \mathrm{M_\odot}$ are shown as circles, their size indicating their projected virial radii, $R_\mathrm{vir, \, 200}$ (the radius within which their mass would result in a mean halo density of $200$ times the mean density). Furthermore, the overdensity in each region shown is marked in the lower left corner of each panel in \cref{fig:4nsobs_ov} according to the two different measures that have been introduced above.

Panels~b--d show the signal as predicted from the simulation for three different ``IFU pointings''. The volume probed by each of these images at this redshift is $2.84 \, h^{-3} \, \mathrm{cMpc}^3$. Note that here we have chosen a smaller narrowband with $\Delta \lambda_\text{obs} = 3.75 \, \text{\AA}$ or $\ssim 1.19 \, h^{-1} \, \mathrm{cMpc}$ at this redshift (equivalent to three spectral pixels of MUSE). Filamentary structures are still encapsulated in this width, while a smaller narrowband allows the signal to stand out more clearly from the noise: a wider narrowband, having more pixels in the spectral dimension, increases the overall noise level. The initial value of $\Delta \lambda_\text{obs} = 8.75 \, \text{\AA}$, which we adopted from \citet{2016A&A...587A..98W}, was chosen for the observation of \lya\ halos. Since \lya\ scattering occurs increasingly in high-density regions and in the high-velocity outflowing gas near galaxies \citep[e.g.][]{2006A&A...460..397V}, these structures of high density and high gas velocities cause the \lya\ signal to be spread out over a larger wavelength range.

Filamentary structures, however, have lower densities and peculiar velocities; hence, they will be contained in a narrower wavelength range. Therefore, while on average more individual filaments are present when the chosen narrowband width is larger, the signal from a given filament will tend to get lost in the noise, as illustrated by \cref{fig:MUSE_sigmas}. \Cref{fig:4nsobs_ov} indicates that individual filaments are still abundantly contained within these thin narrowband images with $\Delta \lambda_\text{obs} = 3.75 \, \text{\AA}$, which is getting near the limit of the typical spectral resolution ($\Delta \lambda \approx 2.5 \, \text{\AA}$ for MUSE, see \citealt{2010SPIE.7735E..08B}). Although the precise spectral line width will be determined by the details of radiative transfer (\lya\ photons will be scattered away from the resonance frequency depending on the kinematics of the scattering medium, see \cref{ap:Lya optical depth}), $\Delta \lambda_\text{obs} = 3.75 \, \text{\AA}$ covers a velocity range of $\Delta v = 160 \, \mathrm{km/s}$, which should be large enough to cover the line width for the modest optical depths in filaments \citep[e.g. Eq. (21) in][]{2014PASA...31...40D}.

As expected, regions with a higher signal (like the two bottom panels in \cref{fig:4nsobs_ov}) contain more high-mass ($M_\mathrm{h} > 10^{9.5} \, \mathrm{M_\odot}$) halos compared to low-density regions (e.g. panel~b) and are found to have a higher overdensity, in both our proxies for environment, $\Delta_\mathrm{baryon}$ and $\Delta_\mathrm{halo}$. The \lya\ emission is mainly originating from in and around the virial radii of these halos, but filamentary structures can be seen to extend between them, up to comoving megaparsec scales in panel~d. We note that the panel~c and d are probably the optimal pointings in the entire region shown in panel~a, indicating that with a randomly chosen field, there is only a rather modest chance of observing a filamentary structure with this relatively high surface brightness. This figure therefore highlights the importance of cosmic variance for detecting the filamentary structure of the IGM, and we conclude both the instrument pointing and narrowband width chosen are essential to efficiently map the IGM in \lya\ emission.

In practice, such overdensity candidates at $z \sim 4$ are readily identified at an on-sky number density of $\ssim 1 \, \mathrm{deg^{-2}}$ in broadband surveys (e.g. \citealt{2016ApJ...826..114T, 2018PASJ...70S..12T}; the latter study identified $\ssim 180$ protocluster candidates over $121 \, \mathrm{deg^2}$ at $z \sim 4$). These still require spectroscopic follow-up observations of several individual member galaxies, however, to exclude the possibility of multiple overlapping structures in projection. The feasibility of such campaigns was for example demonstrated by \citet{2016ApJ...826..114T}, who were able to confirm three out of four candidate protoclusters over a $\ssim 4 \, \mathrm{deg^2}$ area at $z \sim 3$-$4$ (in excellent agreement with the expected fraction of true positives from cosmological simulations of more than $76\%$) using just over $\ssim 1 \, \mathrm{h}$ of spectroscopic observations with Subaru/FOCAS per protocluster candidate, thereby reaching a spectral resolution of $\Delta \lambda_\text{obs} \sim 2.5 \, \text{\AA}$.

We note that while a small narrowband width ($\Delta \lambda_\text{obs} = 3.75 \, \text{\AA}$ or $\Delta z \sim 0.003$ as in \cref{fig:4nsobs_ov150}) is optimal for a subsequent deep imaging campaign of extended, filamentary \lya\ emission with a wide-field IFU, not all protocluster members necessarily need to be contained within such a narrow redshift range, since an IFU flexibly allows for the extraction of multiple pseudo-narrowbands along redshift space. Moreover, the IFU observation simultaneously provides the spectroscopic redshift of several galaxies in the protocluster through their \lya\ emission (or even fainter UV metal absorption or emission lines, if the exposure is sufficiently deep), which can help guide the placement of such pseudo-narrowbands.

These recent studies furthermore give rise to a promising outlook for the search of protocluster candidates with extragalactic surveys in the near future. Just over two years into its main survey, the Vera Rubin observatory will already reach a limiting $i$-band AB-magnitude of $\ssim 26$ \citep{2019ApJ...873..111I}, a depth similar to that of the survey used in \citet{2018PASJ...70S..12T}, while the full $10$-year survey (reaching $26.8 \, \mathrm{mag}$) will even approach the depth of the $\ssim 4 \, \mathrm{deg^2}$ field considered by \citet{2016ApJ...826..114T}.

\subsubsection{Sensitivity analysis}
\label{sssec:Sensitivity analysis}

In \cref{fig:4nsobs5}, in all panels, a similar, small section of the main surface brightness map at $z=4.8$ (region~2 in \cref{fig:SB}) is shown in the same narrowband with $\Delta \lambda_\text{obs} = 3.75 \, \text{\AA}$ (i.e. $\ssim 1.19 \, h^{-1} \, \mathrm{cMpc}$), now with a Gaussian smoothing (FWHM of $0.75 \, \mathrm{arcsec}$). The columns show different assumptions on various limits (e.g., the signal from gas below $50$ and $100$ times the mean baryonic density, $\bar{\rho}$), while the overlaid Gaussian noise varies per row (the $1 \sigma$ level applied to the entire row is stated on the right-hand side of the figure). Noise levels quoted are their values per pixel (before rebinning, discussed below), which agrees in size with a MUSE pixel ($0.2 \, \mathrm{arcsec}$). Apart from the different gas density thresholds, the two columns on the right show the expectation in the mirror assumption, where, in addition to the collisional excitation luminosity of gas below a density of half the critical self-shielding density, we calculate the recombination luminosity arising from gas at all densities, but with the surface brightness limited from above by the mirror value (see \cref{sssec:Mirror limit}). At this redshift, the limit is equal to $\text{SB} \simeq 3.29 \cdot 10^{-21} \, \mathrm{erg \, s^{-1} \, cm^{-2} \, arcsec^{-2}}$ for a \citetalias{2012ApJ...746..125H} UV background. Finally, the last column has been rebinned on a scale of $10 \times 10$ pixels ($2 \times 2 \, \mathrm{arcsec}^2$) and subsequently convolved with a Gaussian with FWHM of equal size.

This particular region, chosen for its juxtaposition of both an under- and overdense region, shows that \lya\ emission arising from the less dense components of filamentary structures can only be detected with very high sensitivities (of $\lesssim 10^{-20.5} \, \mathrm{erg \, s^{-1} \, cm^{-2} \, arcsec^{-2}}$ for overdensities of $\rho/\bar{\rho} \leq 100$). Still, with image analysis techniques (e.g. rebinning pixels), the signal of these filaments can stand out at a noise level of $\sigma \sim 10^{-19.5} \, \mathrm{erg \, s^{-1} \, cm^{-2} \, arcsec^{-2}}$. Considering that the sensitivity in recent observations reaches a limiting surface brightness of $\text{SB} \sim 10^{-19} \, \mathrm{erg \, s^{-1} \, cm^{-2} \, arcsec^{-2}}$~\cite[e.g.][]{2015A&A...575A..75B, 2017A&A...608A...1B, 2021arXiv210205516B}, or for median-stacked radial profiles even down to $\text{SB} \sim 4 \cdot 10^{-21} \, \mathrm{erg \, s^{-1} \, cm^{-2} \, arcsec^{-2}}$ (or $\log_{10}{\text{SB}} \simeq -20.4$; see \citealt{2018Natur.562..229W}), this suggests that the very deepest observations are getting close to the detection of such filamentary structures.

Returning to the region shown in panel~d of \cref{fig:4nsobs_ov}, we construct mock observations for a MUSE-like, wide-field integral-field spectrograph on the ELT at two different redshifts, $z=4.8$ and $z=3.6$, in \cref{fig:4nsobs_ov150}. The left panels show emission from all gas without any limits, while the right panels show the combination of recombination emission of all gas in the simulation below the mirror limit, and collisional excitation of gas below half the critical self-shielding density, as before. The panels on the right are convolved with a a Gaussian PSF corresponding to a FWHM of $0.75 \, \mathrm{arcsec}$ \citep[as in the HDFS observation, see][]{2015A&A...575A..75B} and include modelled noise. The noise level has been inferred from a continuum-subtracted pseudo-narrowband image (with the same width) constructed from the $27 \, \mathrm{h}$ MUSE HDFS observation \citep{2015A&A...575A..75B} at $\ssim 7200 \, \mathrm{\Angstrom}$, where the throughput of MUSE is at its maximum of $\ssim 40\%$ \citep[e.g.][; but see also \cref{fig:MUSE_sigmas}]{2019arXiv190601657R}; the $1 \sigma$ level of the inferred noise in this case is $\sigma = 1.72 \cdot 10^{-19} \, \mathrm{erg \, s^{-1} \, cm^{-2} \, arcsec^{-2}}$. Subsequently, the noise level is adjusted to correspond to a MUSE-like instrument on the ELT by scaling the sensitivity by the square root of the ratio of collecting areas between the VLT and ELT ($52 \, \mathrm{m^2}$ and $978 \, \mathrm{m^2}$, respectively\footnote{See for example \url{https://www.eso.org/sci/facilities/paranal/telescopes/ut/m1unit.html} and \url{https://www.eso.org/public/teles-instr/elt/numbers/}.}) and an increased integration time of $t=150 \, \mathrm{h}$ (again assuming a $1/\sqrt{N}$ scaling of the noise level with $N$ the number of collected photons, i.e. a factor $\sqrt{27/150} \simeq 0.42$). The resulting noise level is $\sigma = 1.68 \cdot 10^{-20} \, \mathrm{erg \, s^{-1} \, cm^{-2} \, arcsec^{-2}}$ (indicated on the colourbar).

There are two different evolutions in redshift at play in \cref{fig:4nsobs_ov150}. First of all, we conclude that without conservative limits (not imposing the mirror limit and including gas at higher densities), the \lya\ emission along filaments, originating from dense gas in halos and galaxies embedded in them, is significantly brighter at higher redshift, as is clear from the comparison of the left panels between the two redshifts, $z=4.8$ and $z=3.6$ (panels~a and c); this is an illustration of the cosmic density evolution winning over the increased surface brightness dimming, as discussed in \cref{ssec:Luminosity density}. The modelling of the dense gas dominating the emission is, however, quite uncertain. A robust prediction can be obtained for low-density filamentary gas, for which we find that it can only be marginally detected in an extremely deep observation with an ELT-class telescope (panels~b and d). In our most robust predictions, excluding emission from the dense (and complicated) central regions of halos, \lya\ emission appears brighter at low redshift, where the mirror limit is less affected by surface brightness dimming and self-shielding effects only start to play a role at higher overdensities (SB maps for a larger range of redshifts are shown in \cref{ap:Redshift evolution}). Future work that includes models with more detailed galaxy formation physics, simultaneously capturing the effects of self-shielding and baryonic feedback processes on high-density gas, is needed to investigate how precisely these two effects compete at different redshifts. An accurate treatment of the high-density gas is needed to point out the optimal redshift to observe gas in different environments.

\section{Conclusions}
\label{sec:Conclusions}

We have presented simulation predictions on the properties of \lya\ emission from low-density gas in the IGM at redshifts $2 < z < 7$. Based on our simulations we predict the \lya\ emissivity due to recombinations and collisional excitations in the gas, carefully considering the relevant physical processes. We have employed an on-the-fly self-shielding mechanism and have neglected the effect of \lya\ scattering which is expected to be moderate in the low-density IGM. We impose the ``mirror'' limit for recombination emission and primarily focus on the regime that is not affected strongly by self-shielding for emission produced by collisional excitation, i.e. well below the self-shielding critical density ($\rho/\bar{\rho} \sim 100$ at $z=4.8$).

We found recombination to dominate at lower densities, while collisional excitation becomes the main emission process at higher densities; they contribute approximately equally for the regime we focus on, below half the self-shielding critical overdensity ($\rho/\bar{\rho} \lesssim 50$ at $z=4.8$). Gas near or somewhat above the critical self-shielding density contributes significantly to luminosity produced through both channels; we show that our prediction of recombination emission including all gas, while having the mirror limit imposed, combined with collisional excitation emission of low-density gas should yield a robust lower limit. The prediction for \lya\ emission of collisionally excited gas at higher densities is more uncertain, and we therefore leave this task to future work.

Our predicted values of the surface brightness ($\text{SB}$) at $z=4.8$ for narrowband images with $\Delta \lambda_\text{obs} = 8.75 \, \text{\AA}$ are of the order of $\text{SB} \lesssim 10^{-23} \, \mathrm{erg \, s^{-1} \, cm^{-2} \, arcsec^{-2}}$ for the void regions, increasing to $\ssim 10^{-21} \, \mathrm{erg \, s^{-1} \, cm^{-2} \, arcsec^{-2}}$ for the diffuse gas in filaments. Denser gas within (the halos of) galaxies embedded in the filaments can reach higher values and likely dominates the total emission from filaments. The modelling of this component is, however, very challenging as it depends on the details of the radiative transfer and feedback processes.

We have briefly discussed the prospects of targeting diffuse \lya\ emission with various spectrographs at different telescopes. At this moment, VLT/MUSE is arguably the best option for imaging the \lya\ emission from gas in the filamentary structure of the cosmic web due to its comparably large FOV ($1 \times 1 \, \mathrm{arcmin}^2$) and spectral coverage ($4650$-$9300 \, \text{\AA}$, and thus accessible redshift range of $2.8$-$6.7$ for \lya), while maintaining a high spatial resolution ($0.2 \, \mathrm{arcsec}$ sampling), and good spectral resolution (ranging between $1770$-$3590$). Recent deep observations reaching a limiting \lya\ surface brightness of $\text{SB} \sim 10^{-19} \, \mathrm{erg \, s^{-1} \, cm^{-2} \, arcsec^{-2}}$~\cite[e.g.][]{2015A&A...575A..75B, 2017A&A...608A...1B, 2021arXiv210205516B}, or for median-stacked radial profiles even down to $\text{SB} \sim 4 \cdot 10^{-21} \, \mathrm{erg \, s^{-1} \, cm^{-2} \, arcsec^{-2}}$ (or $\log_{10}{\text{SB}} \simeq -20.4$; see \citealt{2018Natur.562..229W}) suggest that the deepest current observations are already beginning to probe the extended \lya\ radiation emitted by low-density gas ($\rho/\bar{\rho} \lesssim 100$) associated with filamentary structures -- this observed emission, however, is likely dominated by dense gas in halos and galaxies embedded in them.

In our most conservative predictions that should be considered as a lower limit, where we exclude emission from the dense (and complicated) central regions of halos, \lya\ emission appears brighter at low redshift, where the mirror limit is less affected by surface brightness dimming and self-shielding effects only start to play a role at relatively high overdensities. Our mock observations, which aim to simulate observations of regions at different overdensities, show a large amount of variance between fields, making densely populated protoclusters more promising targets for detecting the IGM in \lya\ emission. Our findings suggest an observing strategy exploiting a targeted search of such a distant protocluster could potentially allow deep observations with a wide-field IFU instrument on an ELT-class telescope -- a successor to MUSE -- to directly map the intergalactic, low-density gas in \lya\ emission in detail.

\section*{Acknowledgements}
\label{sec:Acknowledgements}

We are grateful to Sarah Bosman, Elisabeta Lusso, Michael Rauch, and Lutz Wisotzki for useful discussions regarding observational techniques and instruments, and to Lewis Weinberger for his contribution on the effects of radiative transfer. We furthermore thank the anonymous referee for their suggestions. EP acknowledges support by the Kavli Foundation. JW, EP, GK, and MGH gratefully acknowledge support from the ERC Advanced Grant 320596, ``The Emergence of Structure During the Epoch of Reionization''. JW and RS acknowledge support from the ERC Advanced Grant 695671, ``QUENCH'', and the Fondation MERAC. RS acknowledges support from an NWO Rubicon grant, project number 680-50-1518, and an STFC Ernest Rutherford Fellowship (ST/S004831/1). The Sherwood simulations were performed with supercomputer time awarded by the Partnership for Advanced Computing in Europe (PRACE) 8th Call. We acknowledge PRACE for awarding us access to the Curie supercomputer, based in France at the Tr\`es Grand Centre de Calcul (TGCC). This work also made use of the DiRAC Data Analytic system at the University of Cambridge, operated by the University of Cambridge High Performance Computing Service on behalf of the STFC DiRAC HPC Facility (\url{www.dirac.ac.uk}). This equipment was funded by BIS National E-infrastructure capital grant (ST/K001590/1), STFC capital grants ST/H008861/1 and ST/H00887X/1, and STFC DiRAC Operations grant ST/K00333X/1. DiRAC is part of the National e-Infrastructure. This work has also used the following packages in \program{python}: the \program{SciPy} library \citep{Jones2001}, its packages \program{NumPy} \citep{2011CSE....13b..22V} and \program{Matplotlib} \citep{Hunter2007}, and the \program{Astropy} package \citep{2013A&A...558A..33A, 2018AJ....156..123A}.

\bibliographystyle{aa}
\bibliography{AA_paper}

\begin{thebibliography}{124}
\expandafter\ifx\csname natexlab\endcsname\relax\def\natexlab#1{#1}\fi

\bibitem[{{Arrigoni Battaia} {et~al.}(2016){Arrigoni Battaia}, {Hennawi},
  {Cantalupo}, \& {Prochaska}}]{2016ApJ...829....3A}
{Arrigoni Battaia}, F., {Hennawi}, J.~F., {Cantalupo}, S., \& {Prochaska},
  J.~X. 2016, \apj, 829, 3

\bibitem[{{Arrigoni Battaia} {et~al.}(2019){Arrigoni Battaia}, {Hennawi},
  {Prochaska}, {O{\~n}orbe}, {Farina}, {Cantalupo}, \&
  {Lusso}}]{2019MNRAS.482.3162A}
{Arrigoni Battaia}, F., {Hennawi}, J.~F., {Prochaska}, J.~X., {et~al.} 2019,
  \mnras, 482, 3162

\bibitem[{{Astropy Collaboration} {et~al.}(2018){Astropy Collaboration},
  {Price-Whelan}, {Sip{\H{o}}cz}, {G{\"u}nther}, {Lim}, {Crawford}, {Conseil},
  {Shupe}, {Craig}, {Dencheva}, {Ginsburg}, {VanderPlas}, {Bradley},
  {P{\'e}rez-Su{\'a}rez}, {de Val-Borro}, {Aldcroft}, {Cruz}, {Robitaille},
  {Tollerud}, {Ardelean}, {Babej}, {Bach}, {Bachetti}, {Bakanov}, {Bamford},
  {Barentsen}, {Barmby}, {Baumbach}, {Berry}, {Biscani}, {Boquien}, {Bostroem},
  {Bouma}, {Brammer}, {Bray}, {Breytenbach}, {Buddelmeijer}, {Burke},
  {Calderone}, {Cano Rodr{\'\i}guez}, {Cara}, {Cardoso}, {Cheedella}, {Copin},
  {Corrales}, {Crichton}, {D'Avella}, {Deil}, {Depagne}, {Dietrich}, {Donath},
  {Droettboom}, {Earl}, {Erben}, {Fabbro}, {Ferreira}, {Finethy}, {Fox},
  {Garrison}, {Gibbons}, {Goldstein}, {Gommers}, {Greco}, {Greenfield},
  {Groener}, {Grollier}, {Hagen}, {Hirst}, {Homeier}, {Horton}, {Hosseinzadeh},
  {Hu}, {Hunkeler}, {Ivezi{\'c}}, {Jain}, {Jenness}, {Kanarek}, {Kendrew},
  {Kern}, {Kerzendorf}, {Khvalko}, {King}, {Kirkby}, {Kulkarni}, {Kumar},
  {Lee}, {Lenz}, {Littlefair}, {Ma}, {Macleod}, {Mastropietro}, {McCully},
  {Montagnac}, {Morris}, {Mueller}, {Mumford}, {Muna}, {Murphy}, {Nelson},
  {Nguyen}, {Ninan}, {N{\"o}the}, {Ogaz}, {Oh}, {Parejko}, {Parley}, {Pascual},
  {Patil}, {Patil}, {Plunkett}, {Prochaska}, {Rastogi}, {Reddy Janga},
  {Sabater}, {Sakurikar}, {Seifert}, {Sherbert}, {Sherwood-Taylor}, {Shih},
  {Sick}, {Silbiger}, {Singanamalla}, {Singer}, {Sladen}, {Sooley},
  {Sornarajah}, {Streicher}, {Teuben}, {Thomas}, {Tremblay}, {Turner},
  {Terr{\'o}n}, {van Kerkwijk}, {de la Vega}, {Watkins}, {Weaver}, {Whitmore},
  {Woillez}, {Zabalza}, \& {Astropy Contributors}}]{2018AJ....156..123A}
{Astropy Collaboration}, {Price-Whelan}, A.~M., {Sip{\H{o}}cz}, B.~M., {et~al.}
  2018, \aj, 156, 123

\bibitem[{{Astropy Collaboration} {et~al.}(2013){Astropy Collaboration},
  {Robitaille}, {Tollerud}, {Greenfield}, {Droettboom}, {Bray}, {Aldcroft},
  {Davis}, {Ginsburg}, {Price-Whelan}, {Kerzendorf}, {Conley}, {Crighton},
  {Barbary}, {Muna}, {Ferguson}, {Grollier}, {Parikh}, {Nair}, {Unther},
  {Deil}, {Woillez}, {Conseil}, {Kramer}, {Turner}, {Singer}, {Fox}, {Weaver},
  {Zabalza}, {Edwards}, {Azalee Bostroem}, {Burke}, {Casey}, {Crawford},
  {Dencheva}, {Ely}, {Jenness}, {Labrie}, {Lim}, {Pierfederici}, {Pontzen},
  {Ptak}, {Refsdal}, {Servillat}, \& {Streicher}}]{2013A&A...558A..33A}
{Astropy Collaboration}, {Robitaille}, T.~P., {Tollerud}, E.~J., {et~al.} 2013,
  \aap, 558, A33

\bibitem[{{Augustin} {et~al.}(2019){Augustin}, {Quiret}, {Milliard},
  {P{\'e}roux}, {Vibert}, {Blaizot}, {Rasera}, {Teyssier}, {Frank},
  {Deharveng}, {Picouet}, {Martin}, {Hamden}, {Thatte}, {Pereira Santaella},
  {Routledge}, \& {Zieleniewski}}]{2019MNRAS.489.2417A}
{Augustin}, R., {Quiret}, S., {Milliard}, B., {et~al.} 2019, \mnras, 489, 2417

\bibitem[{{Bacon} {et~al.}(2010){Bacon}, {Accardo}, {Adjali}, {Anwand},
  {Bauer}, {Biswas}, {Blaizot}, {Boudon}, {Brau-Nogue}, {Brinchmann},
  {Caillier}, {Capoani}, {Carollo}, {Contini}, {Couderc}, {Daguis{\'e}},
  {Deiries}, {Delabre}, {Dreizler}, {Dubois}, {Dupieux}, {Dupuy}, {Emsellem},
  {Fechner}, {Fleischmann}, {Fran{\c{c}}ois}, {Gallou}, {Gharsa}, {Glindemann},
  {Gojak}, {Guiderdoni}, {Hansali}, {Hahn}, {Jarno}, {Kelz}, {Koehler},
  {Kosmalski}, {Laurent}, {Le Floch}, {Lilly}, {Lizon}, {Loupias}, {Manescau},
  {Monstein}, {Nicklas}, {Olaya}, {Pares}, {Pasquini}, {P{\'e}contal-Rousset},
  {Pell{\'o}}, {Petit}, {Popow}, {Reiss}, {Remillieux}, {Renault}, {Roth},
  {Rupprecht}, {Serre}, {Schaye}, {Soucail}, {Steinmetz}, {Streicher}, {Stuik},
  {Valentin}, {Vernet}, {Weilbacher}, {Wisotzki}, \&
  {Yerle}}]{2010SPIE.7735E..08B}
{Bacon}, R., {Accardo}, M., {Adjali}, L., {et~al.} 2010, in Society of
  Photo-Optical Instrumentation Engineers (SPIE) Conference Series, Vol. 7735,
  Ground-based and Airborne Instrumentation for Astronomy III, ed. I.~S.
  {McLean}, S.~K. {Ramsay}, \& H.~{Takami}, 773508

\bibitem[{{Bacon} {et~al.}(2015){Bacon}, {Brinchmann}, {Richard}, {Contini},
  {Drake}, {Franx}, {Tacchella}, {Vernet}, {Wisotzki}, {Blaizot}, {Bouch{\'e}},
  {Bouwens}, {Cantalupo}, {Carollo}, {Carton}, {Caruana}, {Cl{\'e}ment},
  {Dreizler}, {Epinat}, {Guiderdoni}, {Herenz}, {Husser}, {Kamann}, {Kerutt},
  {Kollatschny}, {Krajnovic}, {Lilly}, {Martinsson}, {Michel-Dansac},
  {Patricio}, {Schaye}, {Shirazi}, {Soto}, {Soucail}, {Steinmetz}, {Urrutia},
  {Weilbacher}, \& {de Zeeuw}}]{2015A&A...575A..75B}
{Bacon}, R., {Brinchmann}, J., {Richard}, J., {et~al.} 2015, \aap, 575, A75

\bibitem[{{Bacon} {et~al.}(2017){Bacon}, {Conseil}, {Mary}, {Brinchmann},
  {Shepherd}, {Akhlaghi}, {Weilbacher}, {Piqueras}, {Wisotzki}, {Lagattuta},
  {Epinat}, {Guerou}, {Inami}, {Cantalupo}, {Courbot}, {Contini}, {Richard},
  {Maseda}, {Bouwens}, {Bouch{\'e}}, {Kollatschny}, {Schaye}, {Marino},
  {Pello}, {Herenz}, {Guiderdoni}, \& {Carollo}}]{2017A&A...608A...1B}
{Bacon}, R., {Conseil}, S., {Mary}, D., {et~al.} 2017, \aap, 608, A1

\bibitem[{{Bacon} {et~al.}(2021){Bacon}, {Mary}, {Garel}, {Blaizot}, {Maseda},
  {Schaye}, {Wisotzki}, {Conseil}, {Brinchmann}, {Leclercq}, {Abril-Melgarejo},
  {Boogaard}, {Bouch{\'e}}, {Contini}, {Feltre}, {Guiderdoni}, {Herenz},
  {Kollatschny}, {Kusakabe}, {Matthee}, {Michel-Dansac}, {Nanayakkara},
  {Richard}, {Roth}, {Schmidt}, {Steinmetz}, {Tresse}, {Urrutia}, {Verhamme},
  {Weilbacher}, {Zabl}, \& {Zoutendijk}}]{2021arXiv210205516B}
{Bacon}, R., {Mary}, D., {Garel}, T., {et~al.} 2021, arXiv e-prints,
  arXiv:2102.05516

\bibitem[{{Becker} \& {Bolton}(2013)}]{2013MNRAS.436.1023B}
{Becker}, G.~D. \& {Bolton}, J.~S. 2013, \mnras, 436, 1023

\bibitem[{{Becker} {et~al.}(2011){Becker}, {Bolton}, {Haehnelt}, \&
  {Sargent}}]{2011MNRAS.410.1096B}
{Becker}, G.~D., {Bolton}, J.~S., {Haehnelt}, M.~G., \& {Sargent}, W. L.~W.
  2011, \mnras, 410, 1096

\bibitem[{{Bolton} {et~al.}(2012){Bolton}, {Becker}, {Raskutti}, {Wyithe},
  {Haehnelt}, \& {Sargent}}]{2012MNRAS.419.2880B}
{Bolton}, J.~S., {Becker}, G.~D., {Raskutti}, S., {et~al.} 2012, \mnras, 419,
  2880

\bibitem[{{Bolton} {et~al.}(2017){Bolton}, {Puchwein}, {Sijacki}, {Haehnelt},
  {Kim}, {Meiksin}, {Regan}, \& {Viel}}]{2017MNRAS.464..897B}
{Bolton}, J.~S., {Puchwein}, E., {Sijacki}, D., {et~al.} 2017, \mnras, 464, 897

\bibitem[{{Bonnet} {et~al.}(2004){Bonnet}, {Abuter}, {Baker}, {Bornemann},
  {Brown}, {Castillo}, {Conzelmann}, {Damster}, {Davies}, {Delabre},
  {Donaldson}, {Dumas}, {Eisenhauer}, {Elswijk}, {Fedrigo}, {Finger},
  {Gemperlein}, {Genzel}, {Gilbert}, {Gillet}, {Goldbrunner}, {Horrobin}, {Ter
  Horst}, {Huber}, {Hubin}, {Iserlohe}, {Kaufer}, {Kissler-Patig}, {Kragt},
  {Kroes}, {Lehnert}, {Lieb}, {Liske}, {Lizon}, {Lutz}, {Modigliani}, {Monnet},
  {Nesvadba}, {Patig}, {Pragt}, {Reunanen}, {R{\"o}hrle}, {Rossi}, {Schmutzer},
  {Schoenmaker}, {Schreiber}, {Stroebele}, {Szeifert}, {Tacconi}, {Tecza},
  {Thatte}, {Tordo}, {\VAN{Werf}{Van der}{van der} Werf}, \&
  {Weisz}}]{2004Msngr.117...17B}
{Bonnet}, H., {Abuter}, R., {Baker}, A., {et~al.} 2004, The Messenger, 117, 17

\bibitem[{{Borisova} {et~al.}(2016){Borisova}, {Cantalupo}, {Lilly}, {Marino},
  {Gallego}, {Bacon}, {Blaizot}, {Bouch{\'e}}, {Brinchmann}, {Carollo},
  {Caruana}, {Finley}, {Herenz}, {Richard}, {Schaye}, {Straka}, {Turner},
  {Urrutia}, {Verhamme}, \& {Wisotzki}}]{2016ApJ...831...39B}
{Borisova}, E., {Cantalupo}, S., {Lilly}, S.~J., {et~al.} 2016, \apj, 831, 39

\bibitem[{{Cai} {et~al.}(2017){Cai}, {Fan}, {Yang}, {Bian}, {Prochaska},
  {Zabludoff}, {McGreer}, {Zheng}, {Green}, {Cantalupo}, {Frye}, {Hamden},
  {Jiang}, {Kashikawa}, \& {Wang}}]{2017ApJ...837...71C}
{Cai}, Z., {Fan}, X., {Yang}, Y., {et~al.} 2017, \apj, 837, 71

\bibitem[{{Cantalupo}(2017)}]{2017ASSL..430..195C}
{Cantalupo}, S. 2017, {Gas Accretion and Giant Ly{\ensuremath{\alpha}}
  Nebulae}, ed. A.~{Fox} \& R.~{Dav{\'e}}, Vol. 430 (Springer International
  Publishing AG), 195

\bibitem[{{Cantalupo} {et~al.}(2014){Cantalupo}, {Arrigoni-Battaia},
  {Prochaska}, {Hennawi}, \& {Madau}}]{2014Natur.506...63C}
{Cantalupo}, S., {Arrigoni-Battaia}, F., {Prochaska}, J.~X., {Hennawi}, J.~F.,
  \& {Madau}, P. 2014, \nat, 506, 63

\bibitem[{{Cantalupo} {et~al.}(2012){Cantalupo}, {Lilly}, \&
  {Haehnelt}}]{2012MNRAS.425.1992C}
{Cantalupo}, S., {Lilly}, S.~J., \& {Haehnelt}, M.~G. 2012, \mnras, 425, 1992

\bibitem[{{Cantalupo} {et~al.}(2008){Cantalupo}, {Porciani}, \&
  {Lilly}}]{2008ApJ...672...48C}
{Cantalupo}, S., {Porciani}, C., \& {Lilly}, S.~J. 2008, \apj, 672, 48

\bibitem[{{Cantalupo} {et~al.}(2005){Cantalupo}, {Porciani}, {Lilly}, \&
  {Miniati}}]{2005ApJ...628...61C}
{Cantalupo}, S., {Porciani}, C., {Lilly}, S.~J., \& {Miniati}, F. 2005, \apj,
  628, 61

\bibitem[{{Cen} {et~al.}(1994){Cen}, {Miralda-Escud{\'e}}, {Ostriker}, \&
  {Rauch}}]{1994ApJ...437L...9C}
{Cen}, R., {Miralda-Escud{\'e}}, J., {Ostriker}, J.~P., \& {Rauch}, M. 1994,
  \apjl, 437, L9

\bibitem[{{Chiang} {et~al.}(2019){Chiang}, {M{\'e}nard}, \&
  {Schiminovich}}]{2019ApJ...877..150C}
{Chiang}, Y.-K., {M{\'e}nard}, B., \& {Schiminovich}, D. 2019, \apj, 877, 150

\bibitem[{{Cisewski} {et~al.}(2014){Cisewski}, {Croft}, {Freeman}, {Genovese},
  {Khandai}, {Ozbek}, \& {Wasserman}}]{2014MNRAS.440.2599C}
{Cisewski}, J., {Croft}, R. A.~C., {Freeman}, P.~E., {et~al.} 2014, \mnras,
  440, 2599

\bibitem[{{Croft} {et~al.}(2018){Croft}, {Miralda-Escud{\'e}}, {Zheng},
  {Blomqvist}, \& {Pieri}}]{2018MNRAS.481.1320C}
{Croft}, R. A.~C., {Miralda-Escud{\'e}}, J., {Zheng}, Z., {Blomqvist}, M., \&
  {Pieri}, M. 2018, \mnras, 481, 1320

\bibitem[{{Dav{\'e}} {et~al.}(1999){Dav{\'e}}, {Hernquist}, {Katz}, \&
  {Weinberg}}]{1999ApJ...511..521D}
{Dav{\'e}}, R., {Hernquist}, L., {Katz}, N., \& {Weinberg}, D.~H. 1999, \apj,
  511, 521

\bibitem[{{de Graaff} {et~al.}(2019){de Graaff}, {Cai}, {Heymans}, \&
  {Peacock}}]{2019A&A...624A..48D}
{de Graaff}, A., {Cai}, Y.-C., {Heymans}, C., \& {Peacock}, J.~A. 2019, \aap,
  624, A48

\bibitem[{{Dijkstra}(2014)}]{2014PASA...31...40D}
{Dijkstra}, M. 2014, \pasa, 31, e040

\bibitem[{{Djorgovski} {et~al.}(1985){Djorgovski}, {Spinrad}, {McCarthy}, \&
  {Strauss}}]{1985ApJ...299L...1D}
{Djorgovski}, S., {Spinrad}, H., {McCarthy}, P., \& {Strauss}, M.~A. 1985,
  \apjl, 299, L1

\bibitem[{{Dor{\'e}} {et~al.}(2018){Dor{\'e}}, {Werner}, {Ashby}, {Bleem},
  {Bock}, {Burt}, {Capak}, {Chang}, {Chaves-Montero}, {Chen}, {Civano},
  {Cleeves}, {Cooray}, {Crill}, {Crossfield}, {Cushing}, {de la Torre},
  {DiMatteo}, {Dvory}, {Dvorkin}, {Espaillat}, {Ferraro}, {Finkbeiner},
  {Greene}, {Hewitt}, {Hogg}, {Huffenberger}, {Jun}, {Ilbert}, {Jeong},
  {Johnson}, {Kim}, {Kirkpatrick}, {Kowalski}, {Korngut}, {Li}, {Lisse},
  {MacGregor}, {Mamajek}, {Mauskopf}, {Melnick}, {M{\'e}nard}, {Neyrinck},
  {{\"O}berg}, {Pisani}, {Rocca}, {Salvato}, {Schaan}, {Scoville}, {Song},
  {Stevens}, {Tenneti}, {Teplitz}, {Tolls}, {Unwin}, {Urry}, {Wandelt},
  {Williams}, {Wilner}, {Windhorst}, {Wolk}, {Yorke}, \&
  {Zemcov}}]{2018arXiv180505489D}
{Dor{\'e}}, O., {Werner}, M.~W., {Ashby}, M. L.~N., {et~al.} 2018, ArXiv
  e-prints [\eprint[arXiv]{1805.05489}]

\bibitem[{{Draine}(2011)}]{2011piim.book.....D}
{Draine}, B.~T. 2011, {Physics of the Interstellar and Intergalactic Medium}
  (Princeton University Press)

\bibitem[{{Eckert} {et~al.}(2015){Eckert}, {Jauzac}, {Shan}, {Kneib}, {Erben},
  {Israel}, {Jullo}, {Klein}, {Massey}, {Richard}, \&
  {Tchernin}}]{2015Natur.528..105E}
{Eckert}, D., {Jauzac}, M., {Shan}, H., {et~al.} 2015, \nat, 528, 105

\bibitem[{{Eisenhauer} {et~al.}(2003){Eisenhauer}, {Abuter}, {Bickert},
  {Biancat-Marchet}, {Bonnet}, {Brynnel}, {Conzelmann}, {Delabre}, {Donaldson},
  {Farinato}, {Fedrigo}, {Genzel}, {Hubin}, {Iserlohe}, {Kasper},
  {Kissler-Patig}, {Monnet}, {Roehrle}, {Schreiber}, {Stroebele}, {Tecza},
  {Thatte}, \& {Weisz}}]{2003SPIE.4841.1548E}
{Eisenhauer}, F., {Abuter}, R., {Bickert}, K., {et~al.} 2003, in Society of
  Photo-Optical Instrumentation Engineers (SPIE) Conference Series, Vol. 4841,
  Instrument Design and Performance for Optical/Infrared Ground-based
  Telescopes, ed. M.~{Iye} \& A.~F.~M. {Moorwood}, 1548--1561

\bibitem[{{Elias} {et~al.}(2020){Elias}, {Genel}, {Sternberg}, {Devriendt},
  {Slyz}, {Visbal}, \& {Bouch{\'e}}}]{2020MNRAS.494.5439E}
{Elias}, L.~M., {Genel}, S., {Sternberg}, A., {et~al.} 2020, \mnras, 494, 5439

\bibitem[{{Fardal} {et~al.}(2001){Fardal}, {Katz}, {Gardner}, {Hernquist},
  {Weinberg}, \& {Dav{\'e}}}]{2001ApJ...562..605F}
{Fardal}, M.~A., {Katz}, N., {Gardner}, J.~P., {et~al.} 2001, \apj, 562, 605

\bibitem[{{Faucher-Gigu{\`e}re} {et~al.}(2010){Faucher-Gigu{\`e}re},
  {Kere{\v{s}}}, {Dijkstra}, {Hernquist}, \&
  {Zaldarriaga}}]{2010ApJ...725..633F}
{Faucher-Gigu{\`e}re}, C.-A., {Kere{\v{s}}}, D., {Dijkstra}, M., {Hernquist},
  L., \& {Zaldarriaga}, M. 2010, \apj, 725, 633

\bibitem[{{Faucher-Gigu{\`e}re} {et~al.}(2008){Faucher-Gigu{\`e}re}, {Lidz},
  {Hernquist}, \& {Zaldarriaga}}]{2008ApJ...688...85F}
{Faucher-Gigu{\`e}re}, C.-A., {Lidz}, A., {Hernquist}, L., \& {Zaldarriaga}, M.
  2008, \apj, 688, 85

\bibitem[{{Francis} {et~al.}(1996){Francis}, {Woodgate}, {Warren}, {M{\o}ller},
  {Mazzolini}, {Bunker}, {Lowenthal}, {Williams}, {Minezaki}, {Kobayashi}, \&
  {Yoshii}}]{1996ApJ...457..490F}
{Francis}, P.~J., {Woodgate}, B.~E., {Warren}, S.~J., {et~al.} 1996, \apj, 457,
  490

\bibitem[{{Fumagalli} {et~al.}(2016){Fumagalli}, {Cantalupo}, {Dekel},
  {Morris}, {O'Meara}, {Prochaska}, \& {Theuns}}]{2016MNRAS.462.1978F}
{Fumagalli}, M., {Cantalupo}, S., {Dekel}, A., {et~al.} 2016, \mnras, 462, 1978

\bibitem[{{Furlanetto} {et~al.}(2003){Furlanetto}, {Schaye}, {Springel}, \&
  {Hernquist}}]{2003ApJ...599L...1F}
{Furlanetto}, S.~R., {Schaye}, J., {Springel}, V., \& {Hernquist}, L. 2003,
  \apjl, 599, L1

\bibitem[{{Furlanetto} {et~al.}(2005){Furlanetto}, {Schaye}, {Springel}, \&
  {Hernquist}}]{2005ApJ...622....7F}
{Furlanetto}, S.~R., {Schaye}, J., {Springel}, V., \& {Hernquist}, L. 2005,
  \apj, 622, 7

\bibitem[{{Fynbo} {et~al.}(1999){Fynbo}, {M{\o}ller}, \&
  {Warren}}]{1999MNRAS.305..849F}
{Fynbo}, J.~U., {M{\o}ller}, P., \& {Warren}, S.~J. 1999, \mnras, 305, 849

\bibitem[{{Gallego} {et~al.}(2018){Gallego}, {Cantalupo}, {Lilly}, {Marino},
  {Pezzulli}, {Schaye}, {Wisotzki}, {Bacon}, {Inami}, {Akhlaghi}, {Tacchella},
  {Richard}, {Bouche}, {Steinmetz}, \& {Carollo}}]{2018MNRAS.475.3854G}
{Gallego}, S.~G., {Cantalupo}, S., {Lilly}, S., {et~al.} 2018, \mnras, 475,
  3854

\bibitem[{{Garzilli} {et~al.}(2017){Garzilli}, {Boyarsky}, \&
  {Ruchayskiy}}]{2017PhLB..773..258G}
{Garzilli}, A., {Boyarsky}, A., \& {Ruchayskiy}, O. 2017, Physics Letters B,
  773, 258

\bibitem[{{Geach} {et~al.}(2016){Geach}, {Narayanan}, {Matsuda}, {Hayes},
  {Mas-Ribas}, {Dijkstra}, {Steidel}, {Chapman}, {Feldmann}, {Avison},
  {Agertz}, {Ao}, {Birkinshaw}, {Bremer}, {Clements}, {Dannerbauer}, {Farrah},
  {Harrison}, {Kubo}, {Micha{\l}owski}, {Scott}, {Smith}, {Spaans}, {Simpson},
  {Swinbank}, {Taniguchi}, {\VAN{Werf}{Van der}{van der} Werf}, {Verma}, \&
  {Yamada}}]{2016ApJ...832...37G}
{Geach}, J.~E., {Narayanan}, D., {Matsuda}, Y., {et~al.} 2016, \apj, 832, 37

\bibitem[{{Gould} \& {Weinberg}(1996)}]{1996ApJ...468..462G}
{Gould}, A. \& {Weinberg}, D.~H. 1996, \apj, 468, 462

\bibitem[{{Haardt} \& {Madau}(2012)}]{2012ApJ...746..125H}
{Haardt}, F. \& {Madau}, P. 2012, \apj, 746, 125 (HM12)

\bibitem[{{Hayashino} {et~al.}(2004){Hayashino}, {Matsuda}, {Tamura},
  {Yamauchi}, {Yamada}, {Ajiki}, {Fujita}, {Murayama}, {Nagao}, {Ohta},
  {Okamura}, {Ouchi}, {Shimasaku}, {Shioya}, \&
  {Taniguchi}}]{2004AJ....128.2073H}
{Hayashino}, T., {Matsuda}, Y., {Tamura}, H., {et~al.} 2004, \aj, 128, 2073

\bibitem[{{Heckman} {et~al.}(1991){Heckman}, {Lehnert}, {van Breugel}, \&
  {Miley}}]{1991ApJ...370...78H}
{Heckman}, T.~M., {Lehnert}, M.~D., {van Breugel}, W., \& {Miley}, G.~K. 1991,
  \apj, 370, 78

\bibitem[{{Heneka} {et~al.}(2017){Heneka}, {Cooray}, \&
  {Feng}}]{2017ApJ...848...52H}
{Heneka}, C., {Cooray}, A., \& {Feng}, C. 2017, \apj, 848, 52

\bibitem[{{Hennawi} {et~al.}(2015){Hennawi}, {Prochaska}, {Cantalupo}, \&
  {Arrigoni-Battaia}}]{2015Sci...348..779H}
{Hennawi}, J.~F., {Prochaska}, J.~X., {Cantalupo}, S., \& {Arrigoni-Battaia},
  F. 2015, Science, 348, 779

\bibitem[{{Hernquist} {et~al.}(1996){Hernquist}, {Katz}, {Weinberg}, \&
  {Miralda-Escud{\'e}}}]{1996ApJ...457L..51H}
{Hernquist}, L., {Katz}, N., {Weinberg}, D.~H., \& {Miralda-Escud{\'e}}, J.
  1996, \apjl, 457, L51

\bibitem[{{Hogan} \& {Weymann}(1987)}]{1987MNRAS.225P...1H}
{Hogan}, C.~J. \& {Weymann}, R.~J. 1987, \mnras, 225, 1P

\bibitem[{{Hu} {et~al.}(1991){Hu}, {Songaila}, {Cowie}, \&
  {Stockton}}]{1991ApJ...368...28H}
{Hu}, E.~M., {Songaila}, A., {Cowie}, L.~L., \& {Stockton}, A. 1991, \apj, 368,
  28

\bibitem[{{Humphrey} {et~al.}(2008){Humphrey}, {Villar-Mart{\'\i}n},
  {S{\'a}nchez}, {di Serego Alighieri}, {De Breuck}, {Binette}, {Tadhunter},
  {Vernet}, {Fosbury}, \& {Stasielak}}]{2008MNRAS.390.1505H}
{Humphrey}, A., {Villar-Mart{\'\i}n}, M., {S{\'a}nchez}, S.~F., {et~al.} 2008,
  \mnras, 390, 1505

\bibitem[{Hunter(2007)}]{Hunter2007}
Hunter, J.~D. 2007, Computing in Science {\&} Engineering, 9, 90

\bibitem[{{Ikeuchi} \& {Ostriker}(1986)}]{1986ApJ...301..522I}
{Ikeuchi}, S. \& {Ostriker}, J.~P. 1986, \apj, 301, 522

\bibitem[{{Ivezi{\'c}} {et~al.}(2019){Ivezi{\'c}}, {Kahn}, {Tyson}, {Abel},
  {Acosta}, {Allsman}, {Alonso}, {AlSayyad}, {Anderson}, {Andrew}, {Angel},
  {Angeli}, {Ansari}, {Antilogus}, {Araujo}, {Armstrong}, {Arndt}, {Astier},
  {Aubourg}, {Auza}, {Axelrod}, {Bard}, {Barr}, {Barrau}, {Bartlett}, {Bauer},
  {Bauman}, {Baumont}, {Bechtol}, {Bechtol}, {Becker}, {Becla}, {Beldica},
  {Bellavia}, {Bianco}, {Biswas}, {Blanc}, {Blazek}, {Blandford}, {Bloom},
  {Bogart}, {Bond}, {Booth}, {Borgland}, {Borne}, {Bosch}, {Boutigny},
  {Brackett}, {Bradshaw}, {Brandt}, {Brown}, {Bullock}, {Burchat}, {Burke},
  {Cagnoli}, {Calabrese}, {Callahan}, {Callen}, {Carlin}, {Carlson},
  {Chandrasekharan}, {Charles-Emerson}, {Chesley}, {Cheu}, {Chiang}, {Chiang},
  {Chirino}, {Chow}, {Ciardi}, {Claver}, {Cohen-Tanugi}, {Cockrum}, {Coles},
  {Connolly}, {Cook}, {Cooray}, {Covey}, {Cribbs}, {Cui}, {Cutri}, {Daly},
  {Daniel}, {Daruich}, {Daubard}, {Daues}, {Dawson}, {Delgado}, {Dellapenna},
  {de Peyster}, {de Val-Borro}, {Digel}, {Doherty}, {Dubois},
  {Dubois-Felsmann}, {Durech}, {Economou}, {Eifler}, {Eracleous}, {Emmons},
  {Fausti Neto}, {Ferguson}, {Figueroa}, {Fisher-Levine}, {Focke}, {Foss},
  {Frank}, {Freemon}, {Gangler}, {Gawiser}, {Geary}, {Gee}, {Geha}, {Gessner},
  {Gibson}, {Gilmore}, {Glanzman}, {Glick}, {Goldina}, {Goldstein}, {Goodenow},
  {Graham}, {Gressler}, {Gris}, {Guy}, {Guyonnet}, {Haller}, {Harris},
  {Hascall}, {Haupt}, {Hernandez}, {Herrmann}, {Hileman}, {Hoblitt}, {Hodgson},
  {Hogan}, {Howard}, {Huang}, {Huffer}, {Ingraham}, {Innes}, {Jacoby}, {Jain},
  {Jammes}, {Jee}, {Jenness}, {Jernigan}, {Jevremovi{\'c}}, {Johns}, {Johnson},
  {Johnson}, {Jones}, {Juramy-Gilles}, {Juri{\'c}}, {Kalirai}, {Kallivayalil},
  {Kalmbach}, {Kantor}, {Karst}, {Kasliwal}, {Kelly}, {Kessler}, {Kinnison},
  {Kirkby}, {Knox}, {Kotov}, {Krabbendam}, {Krughoff}, {Kub{\'a}nek},
  {Kuczewski}, {Kulkarni}, {Ku}, {Kurita}, {Lage}, {Lambert}, {Lange},
  {Langton}, {Le Guillou}, {Levine}, {Liang}, {Lim}, {Lintott}, {Long},
  {Lopez}, {Lotz}, {Lupton}, {Lust}, {MacArthur}, {Mahabal}, {Mandelbaum},
  {Markiewicz}, {Marsh}, {Marshall}, {Marshall}, {May}, {McKercher}, {McQueen},
  {Meyers}, {Migliore}, {Miller}, {Mills}, {Miraval}, {Moeyens}, {Moolekamp},
  {Monet}, {Moniez}, {Monkewitz}, {Montgomery}, {Morrison}, {Mueller},
  {Muller}, {Mu{\~n}oz Arancibia}, {Neill}, {Newbry}, {Nief}, {Nomerotski},
  {Nordby}, {O'Connor}, {Oliver}, {Olivier}, {Olsen}, {O'Mullane}, {Ortiz},
  {Osier}, {Owen}, {Pain}, {Palecek}, {Parejko}, {Parsons}, {Pease},
  {Peterson}, {Peterson}, {Petravick}, {Libby Petrick}, {Petry},
  {Pierfederici}, {Pietrowicz}, {Pike}, {Pinto}, {Plante}, {Plate}, {Plutchak},
  {Price}, {Prouza}, {Radeka}, {Rajagopal}, {Rasmussen}, {Regnault}, {Reil},
  {Reiss}, {Reuter}, {Ridgway}, {Riot}, {Ritz}, {Robinson}, {Roby}, {Roodman},
  {Rosing}, {Roucelle}, {Rumore}, {Russo}, {Saha}, {Sassolas}, {Schalk},
  {Schellart}, {Schindler}, {Schmidt}, {Schneider}, {Schneider}, {Schoening},
  {Schumacher}, {Schwamb}, {Sebag}, {Selvy}, {Sembroski}, {Seppala}, {Serio},
  {Serrano}, {Shaw}, {Shipsey}, {Sick}, {Silvestri}, {Slater}, {Smith},
  {Smith}, {Sobhani}, {Soldahl}, {Storrie-Lombardi}, {Stover}, {Strauss},
  {Street}, {Stubbs}, {Sullivan}, {Sweeney}, {Swinbank}, {Szalay}, {Takacs},
  {Tether}, {Thaler}, {Thayer}, {Thomas}, {Thornton}, {Thukral}, {Tice},
  {Trilling}, {Turri}, {Van Berg}, {Vanden Berk}, {Vetter}, {Virieux},
  {Vucina}, {Wahl}, {Walkowicz}, {Walsh}, {Walter}, {Wang}, {Wang}, {Warner},
  {Wiecha}, {Willman}, {Winters}, {Wittman}, {Wolff}, {Wood-Vasey}, {Wu},
  {Xin}, {Yoachim}, \& {Zhan}}]{2019ApJ...873..111I}
{Ivezi{\'c}}, {\v{Z}}., {Kahn}, S.~M., {Tyson}, J.~A., {et~al.} 2019, \apj,
  873, 111

\bibitem[{Jones {et~al.}(2001)Jones, Oliphant, Peterson, {et~al.}}]{Jones2001}
Jones, E., Oliphant, T., Peterson, P., {et~al.} 2001, {SciPy}: Open source
  scientific tools for {Python}, [Online; accessed <today>]

\bibitem[{{Kakuma} {et~al.}(2019){Kakuma}, {Ouchi}, {Harikane}, {Inoue},
  {Komiyama}, {Kusakabe}, {Liu}, {Matsuda}, {Matsuoka}, {Mawatari}, {Momose},
  {Ono}, {Shibuya}, \& {Taniguchi}}]{2019arXiv190600173K}
{Kakuma}, R., {Ouchi}, M., {Harikane}, Y., {et~al.} 2019, ArXiv e-prints
  [\eprint[arXiv]{1906.00173}]

\bibitem[{{Katz} {et~al.}(1996){Katz}, {Weinberg}, \&
  {Hernquist}}]{1996ApJS..105...19K}
{Katz}, N., {Weinberg}, D.~H., \& {Hernquist}, L. 1996, \apjs, 105, 19

\bibitem[{{Keel} {et~al.}(1999){Keel}, {Cohen}, {Windhorst}, \&
  {Waddington}}]{1999AJ....118.2547K}
{Keel}, W.~C., {Cohen}, S.~H., {Windhorst}, R.~A., \& {Waddington}, I. 1999,
  \aj, 118, 2547

\bibitem[{{Khaire} {et~al.}(2019){Khaire}, {Walther}, {Hennawi}, {O{\~n}orbe},
  {Luki{\'c}}, {}, {Prochaska}, {Tripp}, {Burchett}, \&
  {Rodriguez}}]{2019MNRAS.486..769K}
{Khaire}, V., {Walther}, M., {Hennawi}, J.~F., {et~al.} 2019, \mnras, 486, 769

\bibitem[{{Kollmeier} {et~al.}(2010){Kollmeier}, {Zheng}, {Dav{\'e}}, {Gould},
  {Katz}, {Miralda-Escud{\'e}}, \& {Weinberg}}]{2010ApJ...708.1048K}
{Kollmeier}, J.~A., {Zheng}, Z., {Dav{\'e}}, R., {et~al.} 2010, \apj, 708, 1048

\bibitem[{{Kulkarni} {et~al.}(2019{\natexlab{a}}){Kulkarni}, {Keating},
  {Haehnelt}, {Bosman}, {Puchwein}, {Chardin}, \&
  {Aubert}}]{2019MNRAS.485L..24K}
{Kulkarni}, G., {Keating}, L.~C., {Haehnelt}, M.~G., {et~al.}
  2019{\natexlab{a}}, \mnras, 485, L24

\bibitem[{{Kulkarni} {et~al.}(2019{\natexlab{b}}){Kulkarni}, {Worseck}, \&
  {Hennawi}}]{2019MNRAS.488.1035K}
{Kulkarni}, G., {Worseck}, G., \& {Hennawi}, J.~F. 2019{\natexlab{b}}, \mnras,
  488, 1035

\bibitem[{{Kull} \& {B{\"o}hringer}(1999)}]{1999A&A...341...23K}
{Kull}, A. \& {B{\"o}hringer}, H. 1999, \aap, 341, 23

\bibitem[{{Leclercq} {et~al.}(2017){Leclercq}, {Bacon}, {Wisotzki}, {Mitchell},
  {Garel}, {Verhamme}, {Blaizot}, {Hashimoto}, {Herenz}, {Conseil},
  {Cantalupo}, {Inami}, {Contini}, {Richard}, {Maseda}, {Schaye}, {Marino},
  {Akhlaghi}, {Brinchmann}, \& {Carollo}}]{2017A&A...608A...8L}
{Leclercq}, F., {Bacon}, R., {Wisotzki}, L., {et~al.} 2017, \aap, 608, A8

\bibitem[{{Luki{\'c}} {et~al.}(2015){Luki{\'c}}, {Stark}, {Nugent}, {White},
  {Meiksin}, \& {Almgren}}]{2015MNRAS.446.3697L}
{Luki{\'c}}, Z., {Stark}, C.~W., {Nugent}, P., {et~al.} 2015, \mnras, 446, 3697

\bibitem[{{Lusso} {et~al.}(2019){Lusso}, {Fumagalli}, {Fossati}, {Mackenzie},
  {Bielby}, {Arrigoni Battaia}, {Cantalupo}, {Cooke}, {Cristiani}, {Dayal},
  {D'Odorico}, {Haardt}, {Lofthouse}, {Morris}, {Peroux}, {Prichard},
  {Rafelski}, {Simcoe}, {Swinbank}, \& {Theuns}}]{2019MNRAS.485L..62L}
{Lusso}, E., {Fumagalli}, M., {Fossati}, M., {et~al.} 2019, \mnras, 485, L62

\bibitem[{{Martin} {et~al.}(2014){Martin}, {Chang}, {Matuszewski}, {Morrissey},
  {Rahman}, {Moore}, \& {Steidel}}]{2014ApJ...786..106M}
{Martin}, D.~C., {Chang}, D., {Matuszewski}, M., {et~al.} 2014, \apj, 786, 106

\bibitem[{{Matsuda} {et~al.}(2012){Matsuda}, {Yamada}, {Hayashino}, {Yamauchi},
  {Nakamura}, {Morimoto}, {Ouchi}, {Ono}, {Umemura}, \&
  {Mori}}]{2012MNRAS.425..878M}
{Matsuda}, Y., {Yamada}, T., {Hayashino}, T., {et~al.} 2012, \mnras, 425, 878

\bibitem[{{McCarthy} {et~al.}(1990){McCarthy}, {Spinrad}, {van Breugel},
  {Liebert}, {Dickinson}, {Djorgovski}, \& {Eisenhardt}}]{1990ApJ...365..487M}
{McCarthy}, P.~J., {Spinrad}, H., {van Breugel}, W., {et~al.} 1990, \apj, 365,
  487

\bibitem[{{Meiksin} \& {White}(2003)}]{2003MNRAS.342.1205M}
{Meiksin}, A. \& {White}, M. 2003, \mnras, 342, 1205

\bibitem[{{Meiksin}(2009)}]{2009RvMP...81.1405M}
{Meiksin}, A.~A. 2009, Reviews of Modern Physics, 81, 1405

\bibitem[{{Momose} {et~al.}(2014){Momose}, {Ouchi}, {Nakajima}, {Ono},
  {Shibuya}, {Shimasaku}, {Yuma}, {Mori}, \& {Umemura}}]{2014MNRAS.442..110M}
{Momose}, R., {Ouchi}, M., {Nakajima}, K., {et~al.} 2014, \mnras, 442, 110

\bibitem[{{Morrissey} {et~al.}(2018){Morrissey}, {Matuszewski}, {Martin},
  {Neill}, {Epps}, {Fucik}, {Weber}, {Darvish}, {Adkins}, {Allen}, {Bartos},
  {Belicki}, {Cabak}, {Callahan}, {Cowley}, {Crabill}, {Deich}, {Delecroix},
  {Doppman}, {Hilyard}, {James}, {Kaye}, {Kokorowski}, {Kwok}, {Lanclos},
  {Milner}, {Moore}, {O'Sullivan}, {Parihar}, {Park}, {Phillips}, {Rizzi},
  {Rockosi}, {Rodriguez}, {Salaun}, {Seaman}, {Sheikh}, {Weiss}, \&
  {Zarzaca}}]{2018ApJ...864...93M}
{Morrissey}, P., {Matuszewski}, M., {Martin}, D.~C., {et~al.} 2018, \apj, 864,
  93

\bibitem[{{O{\~n}orbe} {et~al.}(2019){O{\~n}orbe}, {Davies}, {Luki{\'c}}, {},
  {Hennawi}, \& {Sorini}}]{2019MNRAS.486.4075O}
{O{\~n}orbe}, J., {Davies}, F.~B., {Luki{\'c}}, {et~al.} 2019, \mnras, 486,
  4075

\bibitem[{{O{\~n}orbe} {et~al.}(2017){O{\~n}orbe}, {Hennawi}, \&
  {Luki{\'c}}}]{2017ApJ...837..106O}
{O{\~n}orbe}, J., {Hennawi}, J.~F., \& {Luki{\'c}}, Z. 2017, \apj, 837, 106

\bibitem[{{Oteo} {et~al.}(2018){Oteo}, {Ivison}, {Dunne}, {Manilla-Robles},
  {Maddox}, {Lewis}, {de Zotti}, {Bremer}, {Clements}, {Cooray}, {Dannerbauer},
  {Eales}, {Greenslade}, {Omont}, {Perez─Fourn{\'o}n}, {Riechers}, {Scott},
  {\VAN{Werf}{Van der}{van der} Werf}, {Weiss}, \&
  {Zhang}}]{2018ApJ...856...72O}
{Oteo}, I., {Ivison}, R.~J., {Dunne}, L., {et~al.} 2018, \apj, 856, 72

\bibitem[{{Ouchi} {et~al.}(2018){Ouchi}, {Harikane}, {Shibuya}, {Shimasaku},
  {Taniguchi}, {Konno}, {Kobayashi}, {Kajisawa}, {Nagao}, {Ono}, {Inoue},
  {Umemura}, {Mori}, {Hasegawa}, {Higuchi}, {Komiyama}, {Matsuda}, {Nakajima},
  {Saito}, \& {Wang}}]{2018PASJ...70S..13O}
{Ouchi}, M., {Harikane}, Y., {Shibuya}, T., {et~al.} 2018, \pasj, 70, S13

\bibitem[{{Partridge} \& {Peebles}(1967)}]{1967ApJ...147..868P}
{Partridge}, R.~B. \& {Peebles}, P.~J.~E. 1967, \apj, 147, 868

\bibitem[{{Planck Collaboration} {et~al.}(2014){Planck Collaboration}, {Ade},
  {Aghanim}, {Armitage-Caplan}, {Arnaud}, {Ashdown}, {Atrio-Barandela},
  {Aumont}, {Baccigalupi}, {Banday}, {Barreiro}, {Bartlett}, {Battaner},
  {Benabed}, {Beno{\^\i}t}, {Benoit-L{\'e}vy}, {Bernard}, {Bersanelli},
  {Bielewicz}, {Bobin}, {Bock}, {Bonaldi}, {Bond}, {Borrill}, {Bouchet},
  {Bridges}, {Bucher}, {Burigana}, {Butler}, {Calabrese}, {Cappellini},
  {Cardoso}, {Catalano}, {Challinor}, {Chamballu}, {Chary}, {Chen}, {Chiang},
  {Chiang}, {Christensen}, {Church}, {Clements}, {Colombi}, {Colombo},
  {Couchot}, {Coulais}, {Crill}, {Curto}, {Cuttaia}, {Danese}, {Davies},
  {Davis}, {de Bernardis}, {de Rosa}, {de Zotti}, {Delabrouille}, {Delouis},
  {D{\'e}sert}, {Dickinson}, {Diego}, {Dolag}, {Dole}, {Donzelli}, {Dor{\'e}},
  {Douspis}, {Dunkley}, {Dupac}, {Efstathiou}, {Elsner}, {En{\ss}lin},
  {Eriksen}, {Finelli}, {Forni}, {Frailis}, {Fraisse}, {Franceschi}, {Gaier},
  {Galeotta}, {Galli}, {Ganga}, {Giard}, {Giardino}, {Giraud-H{\'e}raud},
  {Gjerl{\o}w}, {Gonz{\'a}lez-Nuevo}, {G{\'o}rski}, {Gratton}, {Gregorio},
  {Gruppuso}, {Gudmundsson}, {Haissinski}, {Hamann}, {Hansen}, {Hanson},
  {Harrison}, {Henrot-Versill{\'e}}, {Hern{\'a}ndez-Monteagudo}, {Herranz},
  {Hildebrandt}, {Hivon}, {Hobson}, {Holmes}, {Hornstrup}, {Hou}, {Hovest},
  {Huffenberger}, {Jaffe}, {Jaffe}, {Jewell}, {Jones}, {Juvela},
  {Keih{\"a}nen}, {Keskitalo}, {Kisner}, {Kneissl}, {Knoche}, {Knox}, {Kunz},
  {Kurki-Suonio}, {Lagache}, {L{\"a}hteenm{\"a}ki}, {Lamarre}, {Lasenby},
  {Lattanzi}, {Laureijs}, {Lawrence}, {Leach}, {Leahy}, {Leonardi},
  {Le{\'o}n-Tavares}, {Lesgourgues}, {Lewis}, {Liguori}, {Lilje},
  {Linden-V{\o}rnle}, {L{\'o}pez-Caniego}, {Lubin}, {Mac{\'\i}as-P{\'e}rez},
  {Maffei}, {Maino}, {Mandolesi}, {Maris}, {Marshall}, {Martin},
  {Mart{\'\i}nez-Gonz{\'a}lez}, {Masi}, {Massardi}, {Matarrese}, {Matthai},
  {Mazzotta}, {Meinhold}, {Melchiorri}, {Melin}, {Mendes}, {Menegoni},
  {Mennella}, {Migliaccio}, {Millea}, {Mitra}, {Miville-Desch{\^e}nes},
  {Moneti}, {Montier}, {Morgante}, {Mortlock}, {Moss}, {Munshi}, {Murphy},
  {Naselsky}, {Nati}, {Natoli}, {Netterfield}, {N{\o}rgaard-Nielsen},
  {Noviello}, {Novikov}, {Novikov}, {O'Dwyer}, {Osborne}, {Oxborrow}, {Paci},
  {Pagano}, {Pajot}, {Paladini}, {Paoletti}, {Partridge}, {Pasian},
  {Patanchon}, {Pearson}, {Pearson}, {Peiris}, {Perdereau}, {Perotto},
  {Perrotta}, {Pettorino}, {Piacentini}, {Piat}, {Pierpaoli}, {Pietrobon},
  {Plaszczynski}, {Platania}, {Pointecouteau}, {Polenta}, {Ponthieu}, {Popa},
  {Poutanen}, {Pratt}, {Pr{\'e}zeau}, {Prunet}, {Puget}, {Rachen}, {Reach},
  {Rebolo}, {Reinecke}, {Remazeilles}, {Renault}, {Ricciardi}, {Riller},
  {Ristorcelli}, {Rocha}, {Rosset}, {Roudier}, {Rowan-Robinson},
  {Rubi{\~n}o-Mart{\'\i}n}, {Rusholme}, {Sandri}, {Santos}, {Savelainen},
  {Savini}, {Scott}, {Seiffert}, {Shellard}, {Spencer}, {Starck}, {Stolyarov},
  {Stompor}, {Sudiwala}, {Sunyaev}, {Sureau}, {Sutton}, {Suur-Uski}, {Sygnet},
  {Tauber}, {Tavagnacco}, {Terenzi}, {Toffolatti}, {Tomasi}, {Tristram},
  {Tucci}, {Tuovinen}, {T{\"u}rler}, {Umana}, {Valenziano}, {Valiviita},
  {\VAN{Tent}{Van}{van} Tent}, {Vielva}, {Villa}, {Vittorio}, {Wade},
  {Wandelt}, {Wehus}, {White}, {White}, {Wilkinson}, {Yvon}, {Zacchei}, \&
  {Zonca}}]{2014A&A...571A..16P}
{Planck Collaboration}, {Ade}, P.~A.~R., {Aghanim}, N., {et~al.} 2014, \aap,
  571, A16

\bibitem[{{Prescott} {et~al.}(2013){Prescott}, {Dey}, \&
  {Jannuzi}}]{2013ApJ...762...38P}
{Prescott}, M. K.~M., {Dey}, A., \& {Jannuzi}, B.~T. 2013, \apj, 762, 38

\bibitem[{{Puchwein} {et~al.}(2019){Puchwein}, {Haardt}, {Haehnelt}, \&
  {Madau}}]{2019MNRAS.485...47P}
{Puchwein}, E., {Haardt}, F., {Haehnelt}, M.~G., \& {Madau}, P. 2019, \mnras,
  485, 47 (P19)

\bibitem[{{Rahmati} {et~al.}(2013){Rahmati}, {Pawlik}, {Rai{\v{c}}evi{\'c}}, \&
  {Schaye}}]{2013MNRAS.430.2427R}
{Rahmati}, A., {Pawlik}, A.~H., {Rai{\v{c}}evi{\'c}}, M., \& {Schaye}, J. 2013,
  \mnras, 430, 2427

\bibitem[{{Rauch} {et~al.}(2011){Rauch}, {Becker}, {Haehnelt}, {Gauthier},
  {Ravindranath}, \& {Sargent}}]{2011MNRAS.418.1115R}
{Rauch}, M., {Becker}, G.~D., {Haehnelt}, M.~G., {et~al.} 2011, \mnras, 418,
  1115

\bibitem[{{Rauch} {et~al.}(2013){Rauch}, {Becker}, {Haehnelt}, {Gauthier}, \&
  {Sargent}}]{2013MNRAS.429..429R}
{Rauch}, M., {Becker}, G.~D., {Haehnelt}, M.~G., {Gauthier}, J.-R., \&
  {Sargent}, W. L.~W. 2013, \mnras, 429, 429

\bibitem[{{Rauch} {et~al.}(2008){Rauch}, {Haehnelt}, {Bunker}, {Becker},
  {Marleau}, {Graham}, {Cristiani}, {Jarvis}, {Lacey}, {Morris}, {Peroux},
  {R{\"o}ttgering}, \& {Theuns}}]{2008ApJ...681..856R}
{Rauch}, M., {Haehnelt}, M., {Bunker}, A., {et~al.} 2008, \apj, 681, 856

\bibitem[{{Rauch} {et~al.}(1997){Rauch}, {Miralda-Escud{\'e}}, {Sargent},
  {Barlow}, {Weinberg}, {Hernquist}, {Katz}, {Cen}, \&
  {Ostriker}}]{1997ApJ...489....7R}
{Rauch}, M., {Miralda-Escud{\'e}}, J., {Sargent}, W. L.~W., {et~al.} 1997,
  \apj, 489, 7

\bibitem[{{Richard} {et~al.}(2019){Richard}, {Bacon}, {Blaizot}, {Boissier},
  {Boselli}, {NicolasBouch{\'e}}, {Brinchmann}, {Castro}, {Ciesla}, {Crowther},
  {Daddi}, {Dreizler}, {Duc}, {Elbaz}, {Epinat}, {Evans}, {Fossati},
  {Fumagalli}, {Garcia}, {Garel}, {Hayes}, {Herrero}, {Hugot}, {Humphrey},
  {Jablonka}, {Kamann}, {Kaper}, {Kelz}, {Kneib}, {de Koter}, {Krajnovi{\'c}},
  {Kudritzki}, {Langer}, {Lardo}, {Leclercq}, {Lennon}, {Mahler}, {Martins},
  {Massey}, {Mitchell}, {Monreal-Ibero}, {Najarro}, {Opitom}, {Papaderos},
  {P{\'e}roux}, {Revaz}, {Roth}, {Rousselot}, {Sander}, {Simmonds Wagemann},
  {Smail}, {Swinbank}, {Tramper}, {Urrutia}, {Verhamme}, {Vink}, {Walsh},
  {Weilbacher}, {Wendt}, {Wisotzki}, \& {Yang}}]{2019arXiv190601657R}
{Richard}, J., {Bacon}, R., {Blaizot}, J., {et~al.} 2019, ArXiv e-prints
  [\eprint[arXiv]{1906.01657}]

\bibitem[{{Roche} {et~al.}(2014){Roche}, {Humphrey}, \&
  {Binette}}]{2014MNRAS.443.3795R}
{Roche}, N., {Humphrey}, A., \& {Binette}, L. 2014, \mnras, 443, 3795

\bibitem[{{Rosdahl} \& {Blaizot}(2012)}]{2012MNRAS.423..344R}
{Rosdahl}, J. \& {Blaizot}, J. 2012, \mnras, 423, 344

\bibitem[{{Sachkov} {et~al.}(2018){Sachkov}, {Shustov}, \& {G{\'o}mez de
  Castro}}]{2018SPIE10699E..3GS}
{Sachkov}, M., {Shustov}, B., \& {G{\'o}mez de Castro}, A.~I. 2018, in Society
  of Photo-Optical Instrumentation Engineers (SPIE) Conference Series, Vol.
  10699, Space Telescopes and Instrumentation 2018: Ultraviolet to Gamma Ray,
  ed. J.-W.~A. {den Herder}, S.~{Nikzad}, \& K.~{Nakazawa}, 106993G

\bibitem[{{S{\'a}nchez} \& {Humphrey}(2009)}]{2009A&A...495..471S}
{S{\'a}nchez}, S.~F. \& {Humphrey}, A. 2009, \aap, 495, 471

\bibitem[{{Schaye} {et~al.}(2000){Schaye}, {Theuns}, {Rauch}, {Efstathiou}, \&
  {Sargent}}]{2000MNRAS.318..817S}
{Schaye}, J., {Theuns}, T., {Rauch}, M., {Efstathiou}, G., \& {Sargent}, W.
  L.~W. 2000, \mnras, 318, 817

\bibitem[{{Scholz} \& {Walters}(1991)}]{1991ApJ...380..302S}
{Scholz}, T.~T. \& {Walters}, H.~R.~J. 1991, \apj, 380, 302

\bibitem[{{Scholz} {et~al.}(1990){Scholz}, {Walters}, {Burke}, \&
  {Scott}}]{1990MNRAS.242..692S}
{Scholz}, T.~T., {Walters}, H.~R.~J., {Burke}, P.~J., \& {Scott}, M.~P. 1990,
  \mnras, 242, 692

\bibitem[{{Sharples} {et~al.}(2013){Sharples}, {Bender}, {Agudo Berbel},
  {Bezawada}, {Castillo}, {Cirasuolo}, {Davidson}, {Davies}, {Dubbeldam},
  {Fairley}, {Finger}, {F{\"o}rster Schreiber}, {Gonte}, {Hess}, {Jung},
  {Lewis}, {Lizon}, {Muschielok}, {Pasquini}, {Pirard}, {Popovic}, {Ramsay},
  {Rees}, {Richter}, {Riquelme}, {Rodrigues}, {Saviane}, {Schlichter},
  {Schmidtobreick}, {Segovia}, {Smette}, {Szeifert}, {van Kesteren}, {Wegner},
  \& {Wiezorrek}}]{2013Msngr.151...21S}
{Sharples}, R., {Bender}, R., {Agudo Berbel}, A., {et~al.} 2013, The Messenger,
  151, 21

\bibitem[{{Silva} {et~al.}(2016){Silva}, {Kooistra}, \&
  {Zaroubi}}]{2016MNRAS.462.1961S}
{Silva}, M.~B., {Kooistra}, R., \& {Zaroubi}, S. 2016, \mnras, 462, 1961

\bibitem[{{Silva} {et~al.}(2013){Silva}, {Santos}, {Gong}, {Cooray}, \&
  {Bock}}]{2013ApJ...763..132S}
{Silva}, M.~B., {Santos}, M.~G., {Gong}, Y., {Cooray}, A., \& {Bock}, J. 2013,
  \apj, 763, 132

\bibitem[{{Sobral} {et~al.}(2013){Sobral}, {Smail}, {Best}, {Geach}, {Matsuda},
  {Stott}, {Cirasuolo}, \& {Kurk}}]{2013MNRAS.428.1128S}
{Sobral}, D., {Smail}, I., {Best}, P.~N., {et~al.} 2013, \mnras, 428, 1128

\bibitem[{{Springel}(2005)}]{2005MNRAS.364.1105S}
{Springel}, V. 2005, \mnras, 364, 1105

\bibitem[{{Springel} {et~al.}(2001){Springel}, {Yoshida}, \&
  {White}}]{2001NewA....6...79S}
{Springel}, V., {Yoshida}, N., \& {White}, S. D.~M. 2001, \na, 6, 79

\bibitem[{{Steidel} {et~al.}(2000){Steidel}, {Adelberger}, {Shapley},
  {Pettini}, {Dickinson}, \& {Giavalisco}}]{2000ApJ...532..170S}
{Steidel}, C.~C., {Adelberger}, K.~L., {Shapley}, A.~E., {et~al.} 2000, \apj,
  532, 170

\bibitem[{{Steidel} {et~al.}(2011){Steidel}, {Bogosavljevi{\'c}}, {Shapley},
  {Kollmeier}, {Reddy}, {Erb}, \& {Pettini}}]{2011ApJ...736..160S}
{Steidel}, C.~C., {Bogosavljevi{\'c}}, M., {Shapley}, A.~E., {et~al.} 2011,
  \apj, 736, 160

\bibitem[{{Tanimura} {et~al.}(2019){Tanimura}, {Hinshaw}, {McCarthy},
  {\VAN{Waerbeke}{Van}{van} Waerbeke}, {Aghanim}, {Ma}, {Mead}, {Hojjati}, \&
  {Tr{\"o}ster}}]{2019MNRAS.483..223T}
{Tanimura}, H., {Hinshaw}, G., {McCarthy}, I.~G., {et~al.} 2019, \mnras, 483,
  223

\bibitem[{{Thatte} {et~al.}(2014){Thatte}, {Clarke}, {Bryson}, {Schnetler},
  {Tecza}, {Bacon}, {Remillieux}, {Mediavilla}, {Herreros Linares}, {Arribas},
  {Evans}, {Lunney}, {Fusco}, {O'Brien}, {Tosh}, {Ives}, {Finger}, {Houghton},
  {Davies}, {Lynn}, {Allen}, {Zieleniewski}, {Kendrew}, {Ferraro-Wood},
  {P{\'e}contal-Rousset}, {Kosmalski}, {Richard}, {Jarno}, {Gallie},
  {Montgomery}, {Henry}, {Zins}, {Freeman}, {Garc{\'\i}a-Lorenzo},
  {Rodr{\'\i}guez-Ramos}, {Revuelta}, {Hernandez Suarez}, {Bueno-Bueno},
  {Gigante-Ripoll}, {Garcia}, {Dohlen}, \& {Neichel}}]{2014SPIE.9147E..25T}
{Thatte}, N.~A., {Clarke}, F., {Bryson}, I., {et~al.} 2014, in Society of
  Photo-Optical Instrumentation Engineers (SPIE) Conference Series, Vol. 9147,
  Ground-based and Airborne Instrumentation for Astronomy V, ed. S.~K.
  {Ramsay}, I.~S. {McLean}, \& H.~{Takami}, 914725

\bibitem[{{Tinker} {et~al.}(2008){Tinker}, {Kravtsov}, {Klypin}, {Abazajian},
  {Warren}, {Yepes}, {Gottl{\"o}ber}, \& {Holz}}]{2008ApJ...688..709T}
{Tinker}, J., {Kravtsov}, A.~V., {Klypin}, A., {et~al.} 2008, \apj, 688, 709

\bibitem[{{Toshikawa} {et~al.}(2016){Toshikawa}, {Kashikawa}, {Overzier},
  {Malkan}, {Furusawa}, {Ishikawa}, {Onoue}, {Ota}, {Tanaka}, {Niino}, \&
  {Uchiyama}}]{2016ApJ...826..114T}
{Toshikawa}, J., {Kashikawa}, N., {Overzier}, R., {et~al.} 2016, \apj, 826, 114

\bibitem[{{Toshikawa} {et~al.}(2018){Toshikawa}, {Uchiyama}, {Kashikawa},
  {Ouchi}, {Overzier}, {Ono}, {Harikane}, {Ishikawa}, {Kodama}, {Matsuda},
  {Lin}, {Onoue}, {Tanaka}, {Nagao}, {Akiyama}, {Komiyama}, {Goto}, \&
  {Lee}}]{2018PASJ...70S..12T}
{Toshikawa}, J., {Uchiyama}, H., {Kashikawa}, N., {et~al.} 2018, \pasj, 70, S12

\bibitem[{{Umehata} {et~al.}(2019){Umehata}, {Fumagalli}, {Smail}, {Matsuda},
  {Swinbank}, {Cantalupo}, {Sykes}, {Ivison}, {Steidel}, {Shapley}, {Vernet},
  {Yamada}, {Tamura}, {Kubo}, {Nakanishi}, {Kajisawa}, {Hatsukade}, \&
  {Kohno}}]{2019Sci...366...97U}
{Umehata}, H., {Fumagalli}, M., {Smail}, I., {et~al.} 2019, Science, 366, 97

\bibitem[{{Valls-Gabaud} \& {MESSIER
  Collaboration}(2017)}]{2017IAUS..321..199V}
{Valls-Gabaud}, D. \& {MESSIER Collaboration}. 2017, in Formation and Evolution
  of Galaxy Outskirts, ed. A.~{Gil de Paz}, J.~H. {Knapen}, \& J.~C. {Lee},
  Vol. 321, 199--201

\bibitem[{{Vanzella} {et~al.}(2017){Vanzella}, {Balestra}, {Gronke}, {Karman},
  {Caminha}, {Dijkstra}, {Rosati}, {De Barros}, {Caputi}, {Grillo}, {Tozzi},
  {Meneghetti}, {Mercurio}, \& {Gilli}}]{2017MNRAS.465.3803V}
{Vanzella}, E., {Balestra}, I., {Gronke}, M., {et~al.} 2017, \mnras, 465, 3803

\bibitem[{{Venemans} {et~al.}(2007){Venemans}, {R{\"o}ttgering}, {Miley}, {van
  Breugel}, {de Breuck}, {Kurk}, {Pentericci}, {Stanford}, {Overzier}, {Croft},
  \& {Ford}}]{2007A&A...461..823V}
{Venemans}, B.~P., {R{\"o}ttgering}, H.~J.~A., {Miley}, G.~K., {et~al.} 2007,
  \aap, 461, 823

\bibitem[{{Verhamme} {et~al.}(2006){Verhamme}, {Schaerer}, \&
  {Maselli}}]{2006A&A...460..397V}
{Verhamme}, A., {Schaerer}, D., \& {Maselli}, A. 2006, \aap, 460, 397

\bibitem[{{Viel} {et~al.}(2004){Viel}, {Haehnelt}, \&
  {Springel}}]{2004MNRAS.354..684V}
{Viel}, M., {Haehnelt}, M.~G., \& {Springel}, V. 2004, \mnras, 354, 684

\bibitem[{{Villar-Mart{\'\i}n} {et~al.}(2007){Villar-Mart{\'\i}n},
  {S{\'a}nchez}, {Humphrey}, {Dijkstra}, {di Serego Alighieri}, {De Breuck}, \&
  {Gonz{\'a}lez Delgado}}]{2007MNRAS.378..416V}
{Villar-Mart{\'\i}n}, M., {S{\'a}nchez}, S.~F., {Humphrey}, A., {et~al.} 2007,
  \mnras, 378, 416

\bibitem[{{Walther} {et~al.}(2019){Walther}, {O{\~n}orbe}, {Hennawi}, \&
  {Luki{\'c}}}]{2019ApJ...872...13W}
{Walther}, M., {O{\~n}orbe}, J., {Hennawi}, J.~F., \& {Luki{\'c}}, Z. 2019,
  \apj, 872, 13

\bibitem[{{\VAN{Walt}{Van der}{van der} Walt} {et~al.}(2011){\VAN{Walt}{Van
  der}{van der} Walt}, {Colbert}, \& {Varoquaux}}]{2011CSE....13b..22V}
{\VAN{Walt}{Van der}{van der} Walt}, S., {Colbert}, S.~C., \& {Varoquaux}, G.
  2011, Computing in Science and Engineering, 13, 22

\bibitem[{{Weinberg} \& {et al.}(1999)}]{1999elss.conf..346W}
{Weinberg}, D. \& {et al.} 1999, in Evolution of Large Scale Structure : From
  Recombination to Garching, ed. A.~J. {Banday}, R.~K. {Sheth}, \& L.~N. {da
  Costa}, 346

\bibitem[{{Wisotzki} {et~al.}(2016){Wisotzki}, {Bacon}, {Blaizot},
  {Brinchmann}, {Herenz}, {Schaye}, {Bouch{\'e}}, {Cantalupo}, {Contini},
  {Carollo}, {Caruana}, {Courbot}, {Emsellem}, {Kamann}, {Kerutt}, {Leclercq},
  {Lilly}, {Patr{\'\i}cio}, {Sandin}, {Steinmetz}, {Straka}, {Urrutia},
  {Verhamme}, {Weilbacher}, \& {Wendt}}]{2016A&A...587A..98W}
{Wisotzki}, L., {Bacon}, R., {Blaizot}, J., {et~al.} 2016, \aap, 587, A98

\bibitem[{{Wisotzki} {et~al.}(2018){Wisotzki}, {Bacon}, {Brinchmann},
  {Cantalupo}, {Richter}, {Schaye}, {Schmidt}, {Urrutia}, {Weilbacher},
  {Akhlaghi}, {Bouch{\'e}}, {Contini}, {Guiderdoni}, {Herenz}, {Inami},
  {Kerutt}, {Leclercq}, {Marino}, {Maseda}, {Monreal-Ibero}, {Nanayakkara},
  {Richard}, {Saust}, {Steinmetz}, \& {Wendt}}]{2018Natur.562..229W}
{Wisotzki}, L., {Bacon}, R., {Brinchmann}, J., {et~al.} 2018, \nat, 562, 229

\bibitem[{{Wold} {et~al.}(2017){Wold}, {Finkelstein}, {Barger}, {Cowie}, \&
  {Rosenwasser}}]{2017ApJ...848..108W}
{Wold}, I. G.~B., {Finkelstein}, S.~L., {Barger}, A.~J., {Cowie}, L.~L., \&
  {Rosenwasser}, B. 2017, \apj, 848, 108

\end{thebibliography}

\appendix

\section{Model parameters and fitting functions}
\label{ap:Model parameters}

\subsection{Emission processes}

This section contains the fitting functions for the relevant quantities in the formulae for recombination and collisional excitation emissivity, \cref{eq:Recombination emissivity,eq:Collisional excitation emissivity} (in \cref{ssec:Lya recombination emission,ssec:Lya collisional excitation emission}), which are repeated here for clarity.

\vspace{1.5ex} \noindent Recombination emissivity (\cref{eq:Recombination emissivity}):
\begin{equation}
    \label{eq:AppRecombination emissivity}
    \epsilon_\text{rec}(T) = f_\text{rec, A/B} (T) \, n_\text{e} \, n_\text{HII} \, \alpha_\text{A/B}(T) \, E_\text{\lya}
\end{equation}
\noindent Collisional excitation emissivity (\cref{eq:Collisional excitation emissivity}):
\begin{equation}
	\label{eq:AppCollisional excitation emissivity}
    \epsilon_\text{exc}(T) = \gamma_\text{1s2p} (T) \, n_\text{e} \, n_\text{HI} \, E_\text{\lya}
\end{equation}

\subsubsection{Recombination fitting functions}

The underlying equation governing \lya\ emission due to recombination in the IGM is given in \cref{eq:AppRecombination emissivity}. The recombination fraction $f_\text{rec, A/B}$ gives the number of recombinations that ultimately result in the emission of a \lya\ photon. It can be modelled using the relations given in \citet{2008ApJ...672...48C, 2014PASA...31...40D} -- this can be summarised as follows:
\begin{equation*}
	f_\text{rec, A/B} = \left\{
	\begin{array}{r}
    	\multicolumn{1}{l}{0.41 - 0.165 \log_{10} \left( \frac{T}{10^4 \, \mathrm{K}} \right)} \\
        
        \qquad \qquad \qquad - 0.015 \left( \frac{T}{10^4 \, \mathrm{K}} \right)^{-0.44}, \, \text{case-A} \\
        
        \multicolumn{1}{l}{0.686 - 0.106 \log_{10} \left( \frac{T}{10^4 \, \mathrm{K}} \right)} \\
        
        \qquad \qquad \qquad - 0.009 \left( \frac{T}{10^4 \, \mathrm{K}} \right)^{-0.44}, \, \text{case-B} \\
	\end{array}
	\right.
\end{equation*}

\noindent The recombination coefficient, $\alpha_\text{A/B}$, is given in the work of \citet{2011piim.book.....D}:
\begin{equation*}
	\alpha_\text{A/B} = \left\{
	\begin{array}{r}
    	\multicolumn{1}{l}{4.13 \cdot 10^{-13} \left( \frac{T}{10^4 \, \mathrm{K}} \right)^{-0.7131-0.0115 \log_{10} \left( \frac{T}{10^4 \, \mathrm{K}} \right)} \, \mathrm{cm^{3} \, s^{-1}},} \\
        
        \text{case-A} \\
        
    	\multicolumn{1}{l}{2.54 \cdot 10^{-13} \left( \frac{T}{10^4 \, \mathrm{K}} \right)^{-0.8163-0.0208 \log_{10} \left( \frac{T}{10^4 \, \mathrm{K}} \right)} \, \mathrm{cm^{3} \, s^{-1}},} \\
        
        \text{case-B} \\
	\end{array}
	\right.
\end{equation*}

\subsubsection{Collisional excitation fitting functions}

For collisional excitation, the \lya\ luminosity density is given by \cref{eq:AppCollisional excitation emissivity}. The function $\gamma_\text{1s2p}$ in this formula is given by
\begin{equation}
	\gamma_\text{1s2p} (T) = \Gamma(T) \exp \left( -\frac{E_\text{\lya}}{k_\text{B}T} \right),
\end{equation}

\noindent with $k_\text{B}$ the Boltzmann constant. The function $\Gamma(T)$ is characterised in \citet{1990MNRAS.242..692S, 1991ApJ...380..302S} as follows:
\begin{equation}
	\label{eq:Gamma}
	\Gamma (T) = \exp \left( \sum_{i=0}^{5} c_i \left( \ln T \right)^i \right),
\end{equation}

\noindent where the coefficients $c_i$ found by \citet{1990MNRAS.242..692S, 1991ApJ...380..302S} are dependent on the temperature regime, and are shown in \cref{tab:Coefficients}. As noted in \cref{sssec:Collisional excitation emissivity}, the rates are not identical to those applied in the cosmological hydrodynamical simulation, but in the relevant temperature regime deviate so little that the \lya\ emission would not be appreciably changed.
\begin{table}
	\centering
	\caption[Coefficients $c_i$ in \cref{eq:Gamma}]{
	    Coefficients $c_i$ in \cref{eq:Gamma}, and their corresponding temperature regimes.
    }
	\begin{tabular}{c c l l l}
		
		& & \textbf{Regime 1}	& \textbf{Regime 2}	& \textbf{Regime 3} \\
		
		$c_0$ &	& $-1.630155 \cdot 10^{2}$	& $5.279996 \cdot 10^{2}$	& $-2.8133632 \cdot 10^{3}$ \\
		
		$c_1$ &	& $8.795711 \cdot 10^{1}$	& $-1.939399 \cdot 10^{2}$	& $8.1509685 \cdot 10^{2}$ \\
		
		$c_2$ &	& $-2.057117 \cdot 10^{1}$	& $2.718982 \cdot 10^{1}$	& $-9.4418414 \cdot 10^{1}$ \\
		
		$c_3$ &	& $2.359573$				& $-1.883399$				& $5.4280565$ \\
		
		$c_4$ &	& $-1.339059 \cdot 10^{-1}$	& $6.462462 \cdot 10^{-2}$	& $-1.5467120 \cdot 10^{-1}$ \\
		
		$c_5$ &	& $3.021507 \cdot 10^{-3}$	& $-8.811076 \cdot 10^{-4}$	& $1.7439112 \cdot 10^{-3}$ \\
	\end{tabular}
	
	\vspace{0.5ex}
	
	\begin{tabular}{c c}
		
		\textbf{Regimes}	& \textbf{Temperature values} \\
		
		Regime 1	& $2 \cdot 10^3 \, \mathrm{K} \leq T < 6 \cdot 10^4 \, \mathrm{K}$ \\
		
		Regime 2	& $6 \cdot 10^4 \, \mathrm{K} \leq T < 6 \cdot 10^6 \, \mathrm{K}$ \\
		
		Regime 3	& $6 \cdot 10^6 \, \mathrm{K} \leq T \leq 1 \cdot 10^8 \, \mathrm{K}$ \\
	\end{tabular}
	\label{tab:Coefficients}
\end{table}

\begin{figure}
	\centering
	\includegraphics[width=\columnwidth]{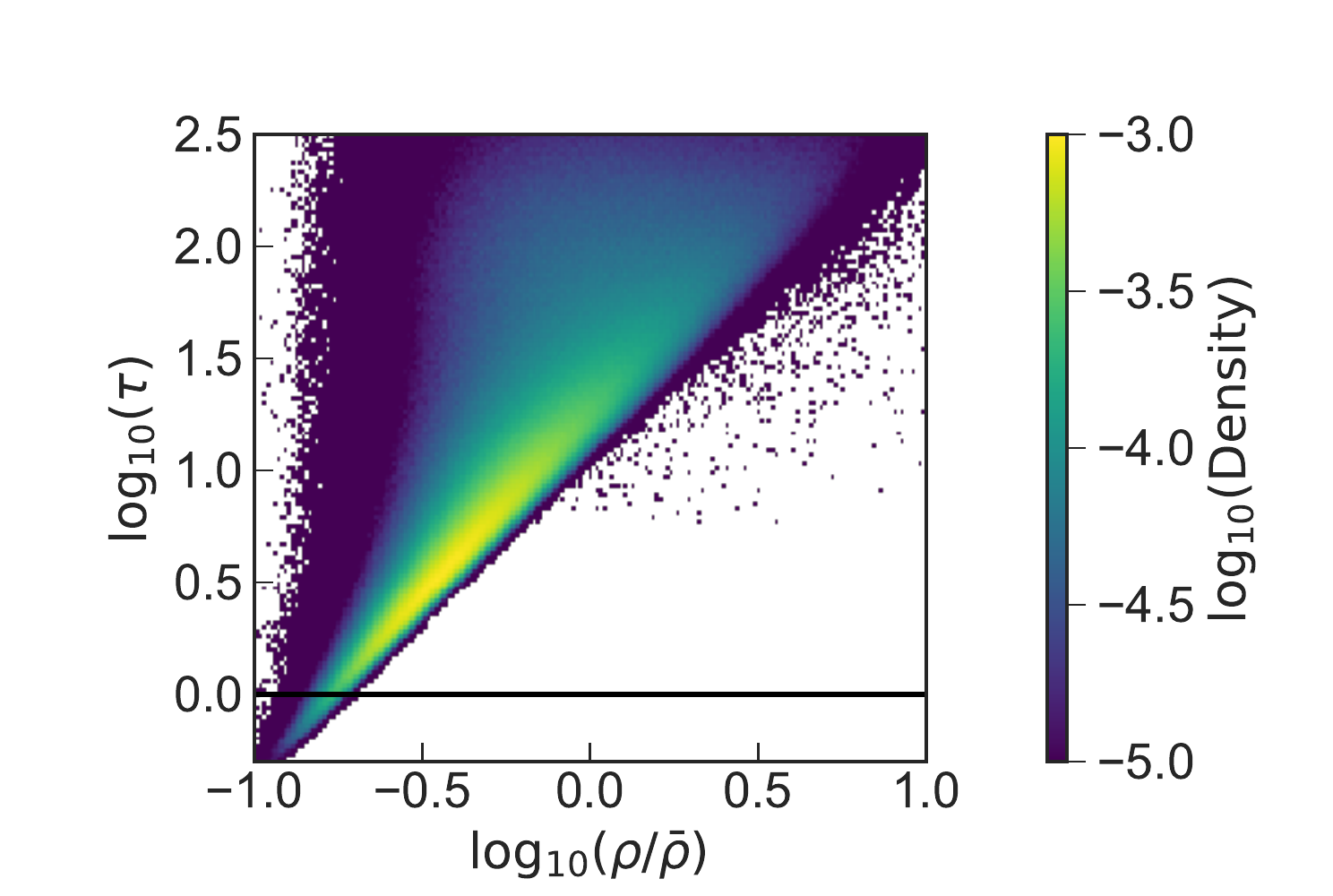}
	\caption[2D histogram of \lya\ optical depth $\tau$ and overdensity $\rho/\bar{\rho}$ at $z=4.8$]{Two-dimensional density histogram for each of $2048$ pixels in spectra along $5000$ (randomly selected) lines of sight at $z=4.8$, as a function of both the \lya\ optical depth $\tau$ and overdensity $\rho/\bar{\rho}$ in the sightline, both measured at line centre (the optical depth having been divided by $2$ to account just for the hydrogen between the source and the observer, see text).}
	\label{fig:2D scatter}
\end{figure}
\begin{figure*}
	\centering
	\includegraphics[width=\linewidth]{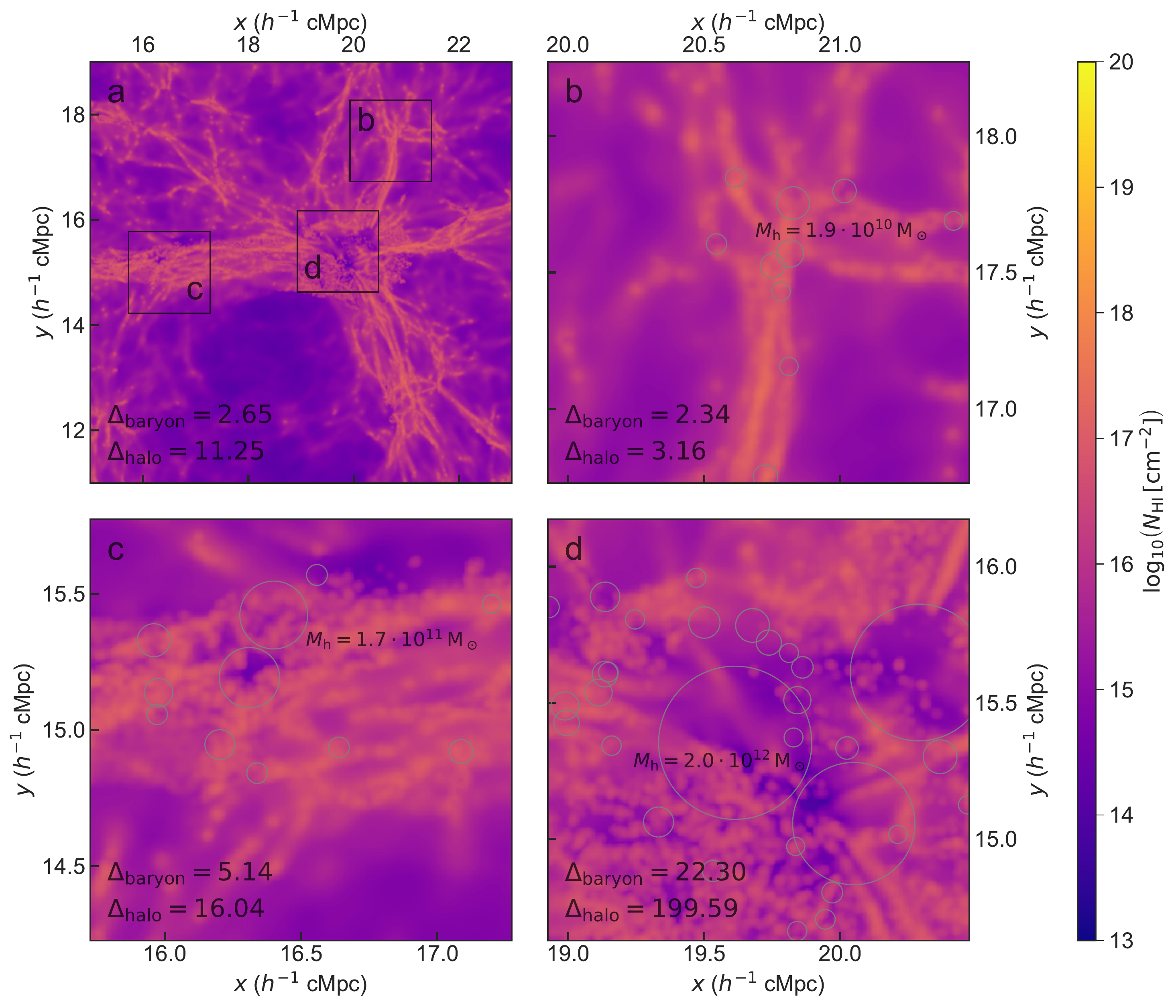}
	\caption[Observed column density at $z=4.8$]
	{Simulated column density of neutral hydrogen, $N_\text{HI}$, in a simulation snapshot at $z=4.8$. The same regions as in \cref{fig:4nsobs_ov} are shown. Moreover, the same density thresholds used for the collisional excitation component are applied, i.e. only gas below half the critical self-shielding density is shown, meaning this is the column density that would correspond to a narrowband image of the low-density gas with $\Delta \lambda_\text{obs} = 3.75 \, \text{\AA}$ ($\ssim 1.19 \, h^{-1} \, \mathrm{cMpc}$). Panel~\textbf{a}, an overview of part of the simulation snapshot, corresponding to region~1 in \cref{fig:SB} (centred on the same comoving coordinates both spatially and spectrally, but now less extended in wavelength range). This panel shows a region of $8 \times 8 \, h^{-2} \, \mathrm{cMpc}^2$ ($5.2 \times 5.2 \, \mathrm{arcmin}^2$) on a pixel grid of $1024 \times 1024$. Panels~\textbf{b}-\textbf{d}, column density maps of neutral hydrogen the size of the MUSE FOV consisting of $300 \times 300$ pixels. The areas covered by these maps are indicated by the black squares in the overview panel~a. In the bottom left corner of each panel, two different measures of the region's overdensity are shown (see \cref{ssec:Simulated observations} for more details). In panels~b-d, halos with halo mass of $M_\mathrm{h} > 10^{9.5} \, \mathrm{M_\odot}$ are shown as circles, their size indicating their projected virial radius (see \cref{ssec:Simulated observations}). The most massive halo in each panel is annotated.}
	\label{fig:NHI}
\end{figure*}
\begin{figure*}
    \centering
    \includegraphics[width=\linewidth]{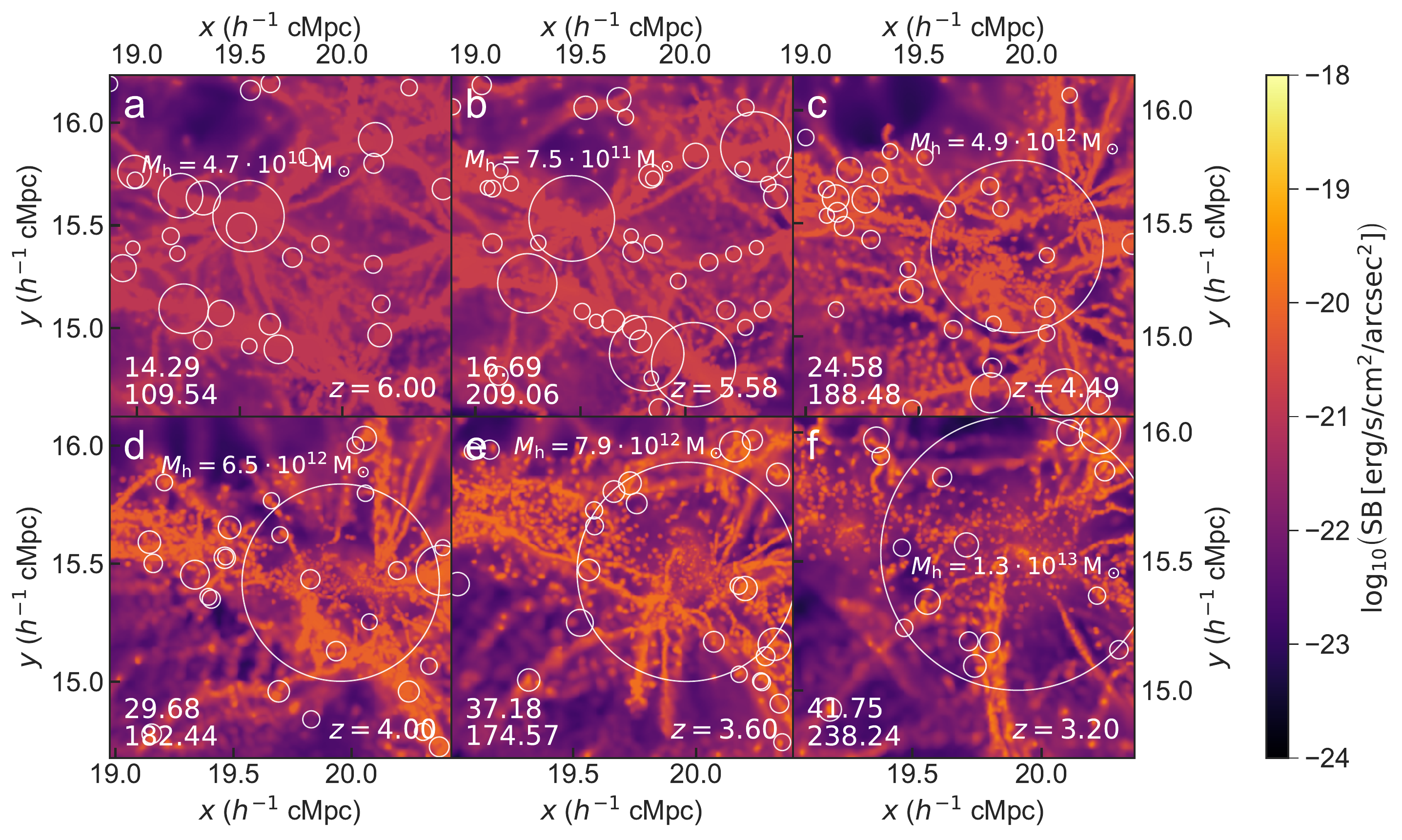}
    \caption[Observed \lya\ surface brightness at different redshifts]
    {\lya\ surface brightness for a combination of recombination emission (of all gas in the simulation) below the mirror limit, and collisional excitation of gas below half the critical self-shielding density (note that both thresholds evolve as a function of redshift, see \cref{fig:UVB_limits}), from simulation snapshots at redshifts of $z=6.00$, $z=5.58$, $z=4.49$, $z=4.00$, $z=3.60$, $z=3.20$ for a narrowband with $\Delta \lambda_\text{obs} = 3.75 \, \text{\AA}$; at $z=5.76$, this corresponds to $\ssim 1.19 \, h^{-1} \, \mathrm{cMpc}$, but this again changes with redshift. They all display a pixel grid of $300 \times 300$ and the angular size of the MUSE FOV ($1 \times 1 \, \mathrm{arcmin}^2$), which translates to different physical sizes at each corresponding redshift. The regions are all centred at the same comoving transverse coordinates as panel~d in \cref{fig:4nsobs_ov} and \cref{fig:4nsobs_ov150} -- however, the narrowband centre (the coordinate along the line of sight) has been chosen to coincide with the most massive halo in each panel, to ensure the entire filament is captured in each panel. The two numbers in the bottom left corner show the same two different measures of the region's overdensity, $\Delta_\mathrm{baryon}$ and $\Delta_\mathrm{halo}$ (see \cref{sssec:Cosmic variance and narrowband widths} for more details). Halos with halo mass of $M_\mathrm{h} > 10^{9.5} \, \mathrm{M_\odot}$ are shown as circles, their size indicating their projected virial radius. The most massive halo in each panel is annotated. Note that in this figure the scale varies between different panels, since the angular size is kept constant across all redshifts.}
    \label{fig:4nsobs_mos}
\end{figure*}

\section{\texorpdfstring{\lya}{\lyatext} optical depth}
\label{ap:Lya optical depth}

This work does not contain treatment of \lya\ line radiative transfer effects (\cref{sssec:Radiative transfer effects}). For our purposes, the treatment without radiative transfer will be able to give us valuable insights about the lower-density IGM filaments on large, cosmological scales, without having to resort to implementing computationally expensive radiative transfer methods that are difficult to accurately model, since e.g. the effects of dust are poorly constrained.

In \cref{fig:2D scatter}, a two-dimensional density histogram for each of $2048$ pixels in mock \lya\ absorption spectra along $5000$ lines of sight at $z=4.8$ is shown as a function of both the \lya\ optical depth $\tau$ and overdensity $\rho/\bar{\rho}$ in the relevant pixel. These spectra are extracted on-the-fly at redshift intervals $\Delta z = 0.1$ and are constructed from the gas density and neutral fraction, temperature, and peculiar velocity of neutral hydrogen along these lines of sight \citep[for details, see][ where they are studied in the context of the \lya\ forest]{2017MNRAS.464..897B}. The peculiar velocity of a given pixel's density has been used to translate its position to redshift space where optical depth is determined, and therefore both density and optical depth are effectively measured at line centre. The optical depth has been divided by a factor of $2$ to account for the fact that on average only half of the matter will be in between the source and the observer -- the other half is located behind the source.\footnote{Note that the division by $2$ is necessary as the \lya\ optical depths were originally extracted for studying \lya\ forest absorption in the spectra of background sources in which case all the gas that affects a pixel in redshift space is in front of the source in real space.} From this figure, it is clear that at mean density optical depths of order $10$ are reached, indicating that radiative transfer will have an effect on most regions. However, effectively this plot is still showing an overestimated measure of optical depth. Since it uses a measure of optical depth at line centre, this does not mean that physically no \lya\ emission will be detected in the optically thick regime ($\tau > 1$). Many \lya\ photons may actually be able to escape, as an initial scattering does not only change the direction of propagation of photons, but also shifts their frequency, and the optical depth decreases quickly when moving away from line centre -- an example of this effect is the \lya\ radiation from galaxies, where densities are high enough to have optical depths of the order of $10^6$, but escape away from line centre is still possible. The optical depth thus mostly informs the expected degree of scattering, i.e. both spatial and spectral broadening of the line profile.

Additionally, the neutral hydrogen (\ion{H}{I}) column density at $z=4.8$ is shown in \cref{fig:NHI}, for precisely the same simulation region (and density limits used for collisional excitation) as in \cref{fig:4nsobs_ov}, with the same narrowband width of $\Delta \lambda_\text{obs} = 3.75 \, \text{\AA}$ (equivalent to $\ssim 1.19 \, h^{-1} \, \mathrm{cMpc}$), and pixel grids of pixel grid of $1024 \times 1024$ (panel~a) and $300 \times 300$ (panels~b-d). The overview map (panel~a) shows that all areas have column densities of at least $N_\text{HI} \sim 10^{15} \, \mathrm{cm^{-2}}$. The most extreme features of the low-density gas show column densities of $10^{17}$-$10^{18} \, \mathrm{cm^{-2}}$, the range of Lyman-limit systems. Simulations that are except for the self-shielding prescription very similar to the one used here have been found to match observational \ion{H}{I} column density distributions well at lower redshifts (where data is more abundant), at least up to $N_\text{HI} \sim 3 \cdot 10^{16} \, \mathrm{cm^{-2}}$, where self-shielding is expected to have a negligible effect \citep{2017MNRAS.464..897B}. At higher column densities, the self-shielding prescription that we use \citep{2013MNRAS.430.2427R} was calibrated to yield realistic column density distributions. At the highest column densities, our simulation will certainly be affected by our simplistic galaxy formation model. These high densities are, however, not the focus of this study.

As with \cref{fig:2D scatter}, it has to be taken into account that this is the column density projected for the entire narrowband. Emitting structures seen within this slice will always lie between the boundaries of this region, and so part of the column density that is projected here may be behind the emitting region, as seen from the observer's perspective. This means that, on average, the actual values of column densities photons travels through is about half of what is displayed.

As discussed in \cref{sssec:Radiative transfer effects}, it is expected that the precise way in which these scattering processes affect the perceived surface brightness images will be the result of a competition between two underlying effects: either the photons emerging from the filamentary structure might be spread out, causing the signal to become fainter, or the filament signal might be enhanced by \lya\ radiation coming from nearby dense structures (where additional radiation is likely to be produced in galaxies) that is scattered in the filament, causing the filaments to appear brighter. As mentioned, similar simulations including radiative transfer indeed show a mixture of these two effects, where the surface brightness of filaments generally is not affected much, or is even boosted \citetext{private communication, Weinberger, 2019}.

\section{Redshift evolution}
\label{ap:Redshift evolution}

The region extensively discussed in \cref{ssec:Simulated observations}, shown in panel~d of \cref{fig:4nsobs_ov} and all panels in \ref{fig:4nsobs_ov150}, is shown at different redshifts in \cref{fig:4nsobs_mos}, again displaying the combination of recombination emission of all gas in the simulation below the mirror limit, and collisional excitation of gas below half the critical self-shielding density. The panels shown are centred at the same transverse comoving coordinates as panel~d in \cref{fig:4nsobs_ov} and all panels in \ref{fig:4nsobs_ov150}, but the narrowband centre (the coordinate along the line of sight) has now been chosen to coincide with the most massive halo in each panel, to ensure the same structure is captured in each panel. Each panel covers the angular size of the MUSE FOV, the physical extent of which varies at different redshifts.

Following the redshift evolution from high to low (going from panel~a to panel~f), we note that the comoving size of the observed region shrinks (although the angular size of the FOV is kept at $1 \times 1 \, \mathrm{arcmin}^2$), roughly from $\ssim 1.5 \times 1.5 \, h^{-2} \, \mathrm{cMpc}^2$ to just over $\ssim 1 \times 1 \, h^{-2} \, \mathrm{cMpc}^2$. The appearance of new massive ($M_\mathrm{h} > 10^{9.5} \, \mathrm{M_\odot}$) halos, and their evolution -- both in relative movement and mass accretion, indicated by the increase of the virial radii -- can also be traced between the different panels. Panel~e, at $z=3.60$, has the same redshift as shown in the bottom two panels of \cref{fig:4nsobs_ov150}.

With these conservative limits that exclude emission from the dense (and complicated) central regions of halos, \lya\ emission appears brighter at low redshift, where the mirror limit is less affected by surface brightness dimming and self-shielding effects only start to play a role at higher overdensities, as discussed in \cref{sssec:Sensitivity analysis}. Panels~a and b appear particularly homogeneous, with large portions being impacted by the mirror limit ($24.6\%$ and $22.1\%$ of pixels exceeding the mirror limit). We note that at low redshift, on the other hand, there is less low-density gas that is luminous in \lya, especially within the large, central halo -- the gas there is likely denser and hotter, and thus less effective at emitting \lya\ radiation (at least within the low-density regime that we are considering; cf. \cref{fig:z_evolution_lum,fig:Luminosity phase space} and their discussion in the text).

\end{document}